\documentclass[11pt]{article}
\pdfoutput=1 
\usepackage{shorthand}
\usepackage{mathtools}
\usepackage{booktabs}
\usepackage[english]{babel}
\usepackage{amsmath,amssymb,amsbsy,amstext, amsthm, simplewick, amsfonts,braket}
\usepackage{graphicx}
\usepackage[small]{caption}
\usepackage{siunitx}
\usepackage{upgreek}
\usepackage{framed}
\usepackage{wrapfig}
\usepackage{multirow}
\usepackage{bbm}
\usepackage[numbers,sort&compress]{natbib}
\usepackage[svgnames,dvipsnames,x11names]{xcolor}
\usepackage[utf8x]{inputenc}
\usepackage{selinput}
\usepackage{bm}
\usepackage{float}
\usepackage{dsfont}
\usepackage{caption}
\usepackage{subcaption}
\usepackage{sidecap}
\usepackage{longtable}
\usepackage{anyfontsize}

\usepackage{epstopdf}
\usepackage{cancel}
\usepackage{tcolorbox}
\usepackage{latexsym,amsmath,amssymb,epsfig}
\usepackage{braket}
\usepackage{tensor}
\usepackage{tocloft}

\usepackage{tikz-cd}
\usepackage{ytableau}
\ytableausetup{centertableaux,boxsize=.8em}
\usepackage{genyoungtabtikz}
\Yboxdim{14pt} 
\Ylinethick{.9pt}

\usepackage{physics}

\usepackage{tcolorbox}
\definecolor{greyish2}{rgb}{.96,.96,.96}

\def\xyma{\xymatrix@M.7em}
\def\xymas{\xymatrix@M.1em}

\newcommand{\Comment}[1]{{}}
\definecolor{darkblue}{rgb}{0.15,0.35,0.55}
\definecolor{reddish}{rgb}{0.65, 0.2, 0.2}
\definecolor{darkgreen}{RGB}{50,150,0}
\definecolor{greyish2}{rgb}{.96,.96,.96}
\usepackage[linktocpage=true]{hyperref}
\hypersetup{
colorlinks=true,
citecolor=darkblue,
linkcolor=reddish,
urlcolor=darkblue,
pdfauthor={},
pdftitle={},
pdfsubject={}
}

\DeclareFontFamily{OT1}{rsfs10}{}
\DeclareFontShape{OT1}{rsfs10}{m}{n}{ <-> rsfs10 }{}
\DeclareMathAlphabet{\mathscript}{OT1}{rsfs10}{m}{n}

\def\gsim{ \lower .75ex \hbox{$\sim$} \llap{\raise .27ex \hbox{$>$}} }
\def\lsim{ \lower .75ex \hbox{$\sim$} \llap{\raise .27ex \hbox{$<$}} }
\def\be{\begin{equation}}
\def\ee{\end{equation}}
\def\bea{\begin{eqnarray}}
\def\eea{\end{eqnarray}}

\newcommand{\baaa}{\begin{eqnarray}}
\newcommand{\eaaa}{\end{eqnarray}}

\newcommand{\eq}[1]{\begin{equation}#1\end{equation}}

\newcommand{\spl}[1]{\begin{split} #1 \end{split}}
\newcommand{\fg}[1]{\begin{figure}[tbp]\centering #1 \end{figure}}

\DeclareMathOperator{\E}{e}

\newcommand{\D}{{\rm d}}
\renewcommand{\d}{{\rm d}}

\newcommand{\Mpl}{M_{\rm Pl}} 

\newcommand{\overbar}[1]{\mkern 2.5mu\overline{\mkern-2.5mu#1\mkern-2.5mu}\mkern 2.5mu}
\renewcommand \bar [1] {\overbar{#1}}

\usepackage[letterpaper,margin=1in]{geometry}

\usepackage{tikz}
\usetikzlibrary{decorations}
\pgfdeclaredecoration{complete sines}{initial}
{
    \state{initial}[
        width=+0pt,
        next state=upsine,
        persistent precomputation={\pgfmathsetmacro\matchinglength{
            \pgfdecoratedinputsegmentlength / int(\pgfdecoratedinputsegmentlength/\pgfdecorationsegmentlength)}
            \setlength{\pgfdecorationsegmentlength}{\matchinglength pt}
        }] {}
    \state{upsine}[width=\pgfdecorationsegmentlength,next state=downsine]{
        \pgfpathsine{\pgfpoint{0.25\pgfdecorationsegmentlength}{0.5\pgfdecorationsegmentamplitude}}
        \pgfpathcosine{\pgfpoint{0.25\pgfdecorationsegmentlength}{-0.5\pgfdecorationsegmentamplitude}}
    }
    \state{downsine}[width=\pgfdecorationsegmentlength,next state=upsine]{
        \pgfpathsine{\pgfpoint{0.25\pgfdecorationsegmentlength}{-0.5\pgfdecorationsegmentamplitude}}
        \pgfpathcosine{\pgfpoint{0.25\pgfdecorationsegmentlength}{0.5\pgfdecorationsegmentamplitude}}
}
    \state{final}{}
}

\definecolor{greyish}{rgb}{.90,.90,.90}
\definecolor{greyish2}{rgb}{.96,.96,.96}
\usepackage{xcolor,colortbl}
\usepackage{tcolorbox}

\usepackage[all]{xy}


\numberwithin{equation}{section}

\begin{document}

%
\renewcommand{\thefootnote}{\fnsymbol{footnote}}
\vspace{0truecm}
\thispagestyle{empty}

\begin{center}
{\fontsize{21.5}{18} \bf{Toward the CompactHedron:} \\ [14pt]
{\fontsize{16}{18} \bf   Bounds on Neutron Stars from Microscopic Principles}}
\end{center} 

\vspace{.3truecm}

\begin{center}
{\fontsize{13.25}{18}\selectfont
Thomas Apostolidis,${}^{\rm a}$
Alessandro Longo,${}^{\rm a}$
Borna Salehian,${}^{\rm b}$
and Luca Santoni${}^{\rm a}$
}
\end{center}
\vspace{.5truecm}

\begin{small}

\centerline{{\it ${}^{\rm a}$Universit\'e Paris Cit\'e, CNRS, Astroparticule et Cosmologie,}}
\centerline{{\it 10 Rue Alice Domon et L\'eonie Duquet, F-75013 Paris, France}} 

 \vspace{.3cm}

\centerline{{\it ${}^{\rm b}$Institut f\"ur Theoretische Physik, ETH Z\"urich,}}
\centerline{{\it Wolfgang-Pauli-Strasse 27, 8093 Z\"urich, Switzerland}} 

\end{small}

\vspace{.6cm}
\begin{abstract}
\noindent
We investigate the implications of causality and positivity for the tidal Love numbers of gravitating objects. We work within the framework of point-particle effective field theory, where the Love numbers correspond to the Wilson coefficients of operators constructed from the curvature tensor and its derivatives. For simplicity, we focus on the quadrupolar gravito-electric sector, which provides the leading contribution to tidal deformability. We show that requiring the tidal response to be retarded and to decay at high frequencies, together with the assumption of passivity, leads to nontrivial bounds on the Love numbers. These bounds delineate an allowed region in the parameter space of the effective theory and apply to any localized object.
In particular, we study their application to neutron stars. We compute the dynamical Love numbers through fourth order in the frequency expansion for a representative set of polytropic and tabulated neutron-star equations of state and show that, in the low-compactness regime, the bounds translate into constraints on the first oscillation mode of the star. Importantly, these constraints are conceptually independent of the stability and thermodynamic conditions commonly imposed in the literature. While our focus is primarily on conservative systems, we also discuss how the bounds are modified in the presence of dissipation. Our results lay the ground for a systematic framework to derive bounds on compact objects from fundamental principles.
\end{abstract}

\newpage

\setcounter{page}{2}
\setcounter{tocdepth}{2}
\tableofcontents
\newpage
\renewcommand*{\thefootnote}{\arabic{footnote}}
\setcounter{footnote}{0}

\section{Introduction}

The idea that general principles can constrain the dynamics of quantum field theories and yield sharp predictions for physical observables is a long-standing one, finding application in fields as varied as particle physics, conformal field theory, hydrodynamics, holography, gravity, and cosmology. Within particle physics, this approach traces back to the $S$-matrix program of the 1960s and is grounded in the well-established link between microcausality, unitarity, and the analytic structure of scattering amplitudes \cite{Eden:1966dnq,Mizera:2023tfe}. In more recent years, perhaps one of the most notable incarnations of this reasoning has been the requirement that effective field theory operator coefficients satisfy certain inequalities and positivity bounds, following from the assumption of a Lorentz-invariant, local, and unitary ultraviolet completion \cite{Pham:1985cr,Adams:2006sv}. Over the last two decades, this approach has driven major advances in high-energy physics (see, e.g., Refs.~\cite{Correia:2020xtr,Kruczenski:2022lot,deRham:2022hpx} for some reviews).

More recently, similar consistency-based approaches have been extended to hydrodynamics and other many-body systems, where causality, stability, and symmetry have been exploited to derive nontrivial constraints on transport coefficients, collective-mode dispersion relations, and the parameter space of effective theories; see,  e.g., Refs.~\cite{Grozdanov:2020koi,Heller:2022ejw,Heller:2023jtd,Gavassino:2023myj,Delacretaz:2025ifh}. These developments are currently part of a broader effort to understand how fundamental principles constrain the dynamics of systems in which Lorentz invariance is broken by the underlying state or background. Such systems arise naturally in condensed matter physics, finite-temperature and finite-density matter, and cosmology, where the presence of a preferred frame nonlinearly realizes the Poincaré group~\cite{Creminelli:2022onn,Creminelli:2023kze,Hui:2023pxc,Green:2023ids,Creminelli:2024lhd,Hui:2025aja,Creminelli:2025rxj,Lee:2025kgs,deRham:2025mjh,Baumann:2022jpr}. This line of research has its roots in the seminal work of Kronig and Kramers on the electromagnetic response of media \cite{kronig1926theory,kramers1927diffusion}. Extending bootstrap techniques to these settings offers a promising route to establishing universal, model-independent bounds on their low-energy dynamics.

In this work and a companion paper~\cite{unpub}, we aim to develop a similar program for astrophysical compact objects. Working within the framework of the point-particle effective field theory (EFT)~\cite{Goldberger:2004jt,Goldberger:2005cd,Porto:2005ac}, we address a central question: \textit{What constraints do fundamental principles impose on the Wilson coefficients of the point-particle effective theory?}

The point-particle EFT builds on a simple physical intuition: when observed from sufficiently large distances, an extended object appears, to leading order, as a point particle. Consider, for instance, an inspiraling binary system. When the separation between the two bodies is much larger than their characteristic sizes, their dynamics can be described, at leading order, in terms of point particles. As the inspiral progresses and the separation decreases, finite-size effects become increasingly important. Within the EFT framework, these corrections are systematically incorporated through higher-dimensional operators localized on the worldlines of the two bodies and organized in a derivative expansion. Their Wilson coefficients encode the finite-size properties of the objects, including their response to external gravitational fields.
At first sight, once the relevant symmetries---spacetime diffeomorphism invariance and worldline reparametrization invariance---are imposed, these coefficients might appear to be arbitrary parameters, subject only to the standard EFT power-counting rules. As we demonstrate below, this is not generally the case. Under suitable assumptions concerning causality, the high-frequency behavior of the response function, and passivity, the Wilson coefficients governing the tidal response---commonly referred to as tidal Love numbers~\cite{Rodriguez:2026iot,Chakraborty:2026qru}---must satisfy positivity bounds and a hierarchy of nontrivial inequalities. 

Our approach is deliberately agnostic about the origin and validity of the underlying assumptions. Rather than attempting to establish them, we investigate their consequences and derive the constraints they impose on the effective theory. In the low-compactness regime, where these assumptions are more readily justified, we show that the resulting bounds on the Love numbers translate into nontrivial constraints on the internal dynamics of neutron stars when applied to concrete models of their interiors. In particular, we find that these inequalities impose an upper bound on the EFT cutoff and, consequently, on the frequency of the star's fundamental oscillation mode.  Moreover, we obtain a measure of how well a single-mode description approximates the linear response. Remarkably, these bounds are qualitatively distinct from the conventional requirements imposed on physically viable equations of state, such as the subluminality of the sound speed and thermodynamic stability. To the best of our knowledge, constraints of this kind have not previously been identified. 

The situation becomes more subtle in the high-compactness regime, where the validity of the assumptions underlying our bounds requires closer scrutiny. Indeed, the vanishing of the static Love numbers of black holes in four-dimensional general relativity~\cite{Fang:2005qq,Damour:2009vw,Binnington:2009bb,Kol:2011vg,Gurlebeck:2015xpa,Hui:2020xxx,Rai:2024lho,Rodriguez:2026iot,Chakraborty:2026qru} already highlights potential tensions with a naive application of the dispersion relation~\cite{Goldberger:2020fot,Porto:2016pyg}. Resolving these tensions and establishing the conditions under which our framework can be consistently extended to strongly gravitating compact objects are important questions that we leave for future work.

As part of this analysis, we extend the calculation of neutron-star dynamical Love numbers to quartic order in frequency, going beyond previous results obtained up to quadratic order~\cite{Chakrabarti:2013lua,Steinhoff:2016rfi,Pitre:2023xsr,HegadeKR:2024agt,HegadeKR:2025qwj,HegadeKR:2026iou,Counsell:2024pua,Saketh:2024juq,Jakobsen:2023pvx,Mandal:2023hqa,Apostolidis:2026qsg,Jarequi:2026cyp,Saketh:2026trm} (see Refs.~\cite{Rodriguez:2026iot,Chakraborty:2026qru} for a more comprehensive list of relevant works).
We compute these coefficients for a representative set of neutron-star equations of state, including both polytropic models~\cite{Hinderer:2007mb,Binnington:2009bb,Damour:2009vw} and more realistic descriptions of dense nuclear matter~\cite{Hinderer:2009ca,Oertel:2016bki}. For simplicity, we restrict our analysis to the leading quadrupolar Love numbers. However, both the calculation and the resulting bounds can be straightforwardly extended to higher orders in the EFT multipole expansion.
We stress that, although we apply the bounds here to neutron stars, they are general statements within the point-particle EFT and, as such, apply to any localized self-gravitating object.

For most of this paper, we neglect dissipation in the system. Dissipative effects will be  considered separately in Section~\ref{diss}, where we show that allowing for nonzero dissipative coefficients dramatically alters the structure of the final results, effectively eliminating the bounds. Nontrivial bounds can nevertheless be recovered by imposing additional information on the dissipative sector.

Taken together, our results represent a first step toward establishing what, in analogy with Refs.~\cite{Arkani-Hamed:2020blm,Heller:2023jtd}, we dub a ``CompactHedron'': a systematic framework for deriving  general bounds on compact objects in the strong-gravity regime from fundamental principles.

The paper is organized as follows. In Section~\ref{sec:ppEFT}, we briefly review the main ingredients of the point-particle EFT and set up the notation. In Section~\ref{sec:arc_bounds}, we review on general grounds the assumptions underlying the bounds and their derivation using the arc variables of Ref.~\cite{Bellazzini:2020cot}, and apply them to the response function of the point-particle EFT. In Section~\ref{sec:NeutronStar}, we explicitly compute the dynamical Love numbers of neutron stars modeled as perfect fluids by solving the Einstein equations for a set of polytropic and physically motivated equations of state and then matching the results to the EFT. In Section~\ref{sec:eos}, we apply the bounds to the dynamical Love numbers and discuss their implications for neutron-star physics. Finally, Section~\ref{diss} is devoted to discussing how the results are modified in the presence of dissipation. Some technical details are collected in Appendices~\ref{app:ppEFT} and \ref{app:GRcomputation}.

\paragraph{Conventions:} We adopt the mostly-plus metric signature, $(-,+,\cdots,+)$, and work in natural units with $\hbar = c = 1$. The reduced Planck mass is defined as $\Mpl=  (8 \pi G)^{-\frac1{D-2}}$, with $G$ Newton's gravitational constant and $D$ the number of spacetime dimensions. 
For an asymptotically flat spacetime, the Schwarzschild radius is related to the mass  $M$ and the Newton's constant $G$ through
\begin{equation}
GM = \frac{(D-2)}{16\pi} \frac{2\pi^\frac{D-1}{2}}{\Gamma\left(\frac{D-1}{2}\right)}r_s^{D-3},
\end{equation}
which yields $r_s=2GM$ in $D=4$. Throughout this work, we will interchangeably use $r_s$ and $2GM$.
Spacetime indices are denoted by Greek letters $\mu,\nu,\cdots$, while spatial indices are denoted by Latin letters $i, j, \cdots$. The notation $( {\cdots} )_T$ indicates the trace-subtracted symmetrization of the enclosed indices.

\newpage

\section{Point-particle effective field theory}
\label{sec:ppEFT}

We begin by summarizing the basic structure of the point-particle EFT framework. Further details can be found in the original references~\cite{Goldberger:2004jt,Goldberger:2005cd,Porto:2005ac}, as well as in a series of comprehensive reviews~\cite{Goldberger:2006bd,Foffa:2013qca,Rothstein:2014sra,Porto:2016pyg,Levi:2018nxp,Goldberger:2022ebt,Goldberger:2022rqf,Rodriguez:2026iot,Chakraborty:2026qru}.
In this work, we focus on non-rotating objects, leaving the study of spinning particles for future work. 

The EFT action for a non-rotating  object can be decomposed as
\begin{equation}
    S=S_{\mathrm{pp}}+S_{\text {bulk}}+S_{\text{int}} \,,
\label{eq:ppEFT0}
\end{equation}
where $S_{\mathrm{pp}}$ describes the point-particle motion, $S_{\mathrm{bulk}}$ governs the dynamics of the gravitational field in the bulk, and $S_{\mathrm{int}}$ encodes the interactions between the object and the gravitational field.

The point-particle action is given by the standard worldline action
\begin{equation}
	S_{\rm pp}  = -M \int \D \tau = 
	-M \int \D \sigma 
	\sqrt{-g_{\mu\nu}(X)\frac{\D X^\mu}{\d \sigma}\frac{\d X^\nu}{\d \sigma}} \; ,
	\label{eq:Spp}
\end{equation}
where $\tau$ denotes the proper time along the object's worldline, $M$ is its mass, $X^\mu(\sigma)$ specifies its spacetime trajectory, and $\sigma$ is an affine parameter that parametrizes the worldline.

The bulk action is described by the Einstein--Hilbert term,
\begin{equation}
    S_{\mathrm{bulk}} = \frac{M_{\mathrm{Pl}}^{2}}{2} \int \mathrm{d}^4 x \, \sqrt{-g} \, R .
\end{equation}
The point-particle description is supplemented by a tower of higher-dimensional operators in $S_{\mathrm{int}}$, which encode the finite-size structure of the object. These operators account for the fact that the  object is not fundamentally pointlike and capture its response to external gravitational fields. Beyond the intrinsic multipole couplings, the interactions between the worldline degrees of freedom and the bulk gravitational field can be written as~\cite{Goldberger:2004jt,Goldberger:2005cd,Chakrabarti:2013lua,Goldberger:2020fot,Saketh:2023bul,Combaluzier--Szteinsznaider:2025eoc}: 
\begin{equation}
    S_{\mathrm{int}}  =  \int \mathrm{d}\tau \,
    Q_E^{ij}(\tau) \, E_{ij} 
 + \text{magnetic} + \text{higher multipoles} \,,
\label{SintE}
\end{equation}
where $E_{ij}$ denotes the electric component of the Weyl tensor $C_{\mu\nu\rho\sigma}$,
\begin{equation}
E_{ij}\equiv C_{0i0j},.
\end{equation}
In Eq.~\eqref{SintE}, we displayed only the leading gravito-electric quadrupolar interaction, which provides the dominant finite-size contribution and is the one we will focus on throughout the rest of the work. Note that the external tidal gravitational field is assumed to vary over scales much larger than the size of the object, for the EFT to be consistent.

The operator $Q_E$ schematically represents the induced (electric-type) multipole moments and encodes information about the object's finite-size structure, including both conservative and, when present, dissipative effects. In general, $Q_E$ is a composite operator whose dynamics is determined by the object's underlying microscopic physics. Since the details of the internal dynamics are generally inaccessible, the standard EFT strategy is to integrate out the corresponding microscopic degrees of freedom and determine $Q_E$ using response theory~\cite{Goldberger:2004jt,Goldberger:2005cd,Goldberger:2020fot,Ivanov:2022hlo,Saketh:2022xjb,Saketh:2023bul,Glazer:2024eyi,Combaluzier--Szteinsznaider:2025eoc,Apostolidis:2026qsg}. In the presence of dissipation, this procedure is naturally formulated within the Schwinger--Keldysh framework (see Appendix~\ref{app:ppEFT} for a summary).

Ultimately, within linear response theory, the induced quadrupole $Q_E^{ij}$ can be expressed in terms of the gravito-electric field $E_{ij}$ as
\begin{equation}
    \langle Q_{E,I}^{ij}(\tau)\rangle=\int \d \tau'\,K_{IJ}^{(E)}{}^{ij\vert i' j'}(\tau-\tau'){E}^J_{i' j'}(\tau')\,,
\label{eq:lQrmain}
\end{equation}
where $I,J=\{+,-\}$ are Keldysh indices,  $K_{IJ}^{(E)ij\vert i'j'}$ is the response function, and the brackets $\langle{\cdots}\rangle$ denote the expectation value.  In particular, $K_{+-}^{(E)ij\vert i'j'}$ is related to the retarded Green's function of the $Q_E^{ij}$ operators  (see, e.g., Refs.~\cite{Goldberger:2020fot,Saketh:2022xjb,Saketh:2023bul,Glazer:2024eyi,Combaluzier--Szteinsznaider:2025eoc,Rodriguez:2026iot}):
\begin{equation}
K^{(E)ij\vert i'j'}_{+-}({\tau}_1-{\tau}_2)
= i \langle [Q_{{E},+}^{ij}(\tau_1), Q_{{E},-}^{i'j'}(\tau_2)] \rangle\theta(\tau_1-\tau_2) .
\label{eq:KpmGR}
\end{equation}

The retarded correlator $K_{+-}^{(E)ij\vert i'j'}$ is the quantity that enters the calculation of the object's response (see Eq.~\eqref{E_opf} below). In particular, it is the central object whose analytic structure we will exploit to derive constraints on the object's physical response, as discussed in the following section.

Note that spherical symmetry of the background fixes the spatial tensorial structure of $K_{+-}^{(E)ij\vert i'j'}$, namely,
\begin{equation}
{K_{+-}^{(E)ij}}_{i'j'}(\tau) \equiv  {K_{+-}^{(E)}} (\tau) \, \delta^{(  i}_{ ( i'}\delta^{j)_T}_{j')_T} ,
\label{ap:KpmcEs}
\end{equation}
where we have introduced the index-free kernel $K_{+-}^{(E)}$, and where $({\cdots})_T$ denotes the trace-free symmetrization of the enclosed indices.
At low energies, we can  expand the  Fourier transform of the response function ${K_{+-}^{(E)}}$ in powers of the frequency  as 
\begin{equation}
{K_{+-}^{(E)}(\omega)} = \frac{R_\star^5}{G}c_E+i \omega M R_\star^5 \nu_E+ \omega^2   \frac{R_\star^8}{G^2M}c_{\dot E}+i \omega^3  
    \frac{R_\star^8}{G} \nu_{\dot{E}}+ \omega^4   \frac{R_\star^{11}}{G^3M^2}c_{\ddot E}+\cdots ,
\label{ap:KpmcEs}
\end{equation}
where we have introduced the dimensionless couplings $c_E$, $\nu_E$, $c_{\dot E}$, $\nu_{\dot E}$, and $c_{\ddot E}$. The parameters $R_\star$ and $M$, which denote the radius and mass of the object, respectively, have been factored out according to standard power-counting arguments~\cite{Rodriguez:2026iot}. In particular, we have identified the characteristic dynamical timescale of the object as $(R_\star^3/(GM))^{1/2}$, whose inverse provides a dimensional estimate of the energy cutoff in the EFT.  
As a consequence of retardation, the frequency-space response function ${K_{+-}^{(E)}(\omega)}$ is analytic in the upper half of the complex $\omega$ plane.

In the following, we will mainly work under the assumption that the object's low-frequency response is dominated by conservative effects and neglect dissipation, which we will include in Section~\ref{diss}. In this limit, using the solution \eqref{eq:lQrmain}, the effective interaction action boils down to  (see also Eq.~\eqref{eq:gammainineft})
\begin{equation}
    S_{\mathrm{int}}  = \frac{1}{2} \int \mathrm{d}\tau 
    \left( \frac{R_\star^5}{G}c_E E_{ij}E^{ij} + \frac{R_\star^8}{G^2M}c_{\dot{E}} \partial_t E_{ij}\partial_t E^{ij}+ \frac{R_\star^{11}}{G^3M^2}c_{\ddot{E}} \partial_t^2 E_{ij}\partial_t^2 E^{ij}  +\cdots\right),
\label{eq:EFTconsint}
\end{equation}
up to quartic order in the time-derivative expansion.

\section{Arc bounds for Love numbers}
\label{sec:arc_bounds}

We have seen that the set of tidal Love number coefficients of a compact object can be viewed as the low-energy expansion of a retarded Green’s function, as defined in Eq.~\eqref{eq:KpmGR}. This perspective allows us to derive generic relations among these coefficients from the general properties of the Green’s function, without relying on the details of the underlying system. The basic idea is closely related to the one underlying the Kramers--Kronig relations \cite{kronig1926theory,kramers1927diffusion}, which, for example, imply that the electric permittivity of a dielectric is larger than that of the vacuum (see also Ref.~\cite{Creminelli:2024lhd}). Here we extend this approach to higher-derivative operators and show that, under the assumption of passivity (the system can only absorb energy from an external source), provided the Fourier-transformed response function decays at large $|\omega|$, analyticity implies a tower of positivity constraints on the Wilson coefficients associated with finite size effects operators. More specifically, we follow the approach of Ref.~\cite{Bellazzini:2020cot}, which recasts analogous questions in the context of the $S$-matrix as a moment problem, a well-studied  problem in mathematics.

However, one difference with respect to the $S$-matrix case is the presence of dissipation in the system (see also Ref.~\cite{Creminelli:2024lhd}). As anticipated, we will mainly neglect dissipation in what follows. We will comment on the possibility of deriving bounds in the presence of dissipation in Section~\ref{diss}, where we discuss the differences with purely conservative systems, while leaving a thorough analysis to future work.
A more fundamental question is whether a Green’s function with the properties we require (analyticity, positivity, and suitable behavior at large frequency, to be discussed below) can be defined for gravitational response. In this paper, we take these properties as assumptions and study their consequences, leaving a deeper understanding of their origin to a separate work~\cite{unpub}.

In the rest of this section, we first review the assumptions we make about retarded Green’s functions and then derive the resulting bounds on the Love numbers following the approach of Ref.~\cite{Bellazzini:2020cot}. We conclude with a brief discussion of these bounds in a toy model. More realistic compact objects will be considered in the following section.

\subsection{Retarded Green's function}\label{retG}

The retarded Green’s function encodes the response of the system to an external perturbation at linear order, and is therefore also referred to as the linear response function. To remain general and avoid cluttering the notation, we derive here the consequences of causality for a generic Green’s function, which we denote by $G_R(t)$. When needed, this function will be identified with $K^{(E)}_{+-}(\tau)$ introduced in the previous section.

Causality requires the response to vanish before the external perturbation is applied,
\begin{equation}
    G_R(t)=0\qquad \text{for}\qquad t<0\,,
\end{equation}
a property usually referred to as retardedness. This condition is closely related to the analyticity of $G_R(\omega)$, the Fourier-space representation of the linear response, defined as
\begin{equation}
    G_R(\omega)=\int \dd{t} \, \E^{i\omega t}G_R(t)\,.
    \label{FouGR}
\end{equation}
For complex $\omega$ in the upper half-plane (UHP), i.e.~${\rm UHP}=\{\omega \in \mathbb{C}\,|\,{\rm Im}[\omega]>0\}$, the exponential factor in Eq.~\eqref{FouGR} improves the convergence of the integral: since $t\geq0$ by causality, we have $\E^{i\omega t}=\E^{i{\rm Re}[\omega]t}\E^{-{\rm Im}[\omega]t}$, and the second factor provides exponential suppression at large $t$.\footnote{\label{Bogoliubov}More mathematically, the retarded Green’s function is assumed to be a tempered distribution. It can then be shown that a tempered distribution is retarded if and only if its Fourier transform is analytic in the UHP and satisfies a bound of the form $|G_R(\omega)|\leq A(1+|\omega|)^n(1+|{\rm Im}[\omega]|^{-m})$, for suitable constants $A$, $n$, and $m$. See Refs.~\cite{Bogolubov:1990ask,Creminelli:2025rxj}.} 

The analyticity of the retarded Green’s function allows one to derive the Kramers–Kronig relations, also known as dispersion relations, which relate the real and imaginary parts of $G_R(\omega)$ in the limit ${\rm Im}[\omega]\to 0^+$. Applying Cauchy’s theorem, one finds
\begin{equation}\label{DispRelUnsub}
    {\rm Re}[G_R](\omega)=\frac{1}{\pi}{\rm PV}\int \dd{\omega'} \frac{{\rm Im}[G_R](\omega')}{\omega'-\omega}+\mathcal{C}_\infty\,,
\end{equation}
where ${\rm PV}$ denotes the Cauchy principal value, which is needed to regulate the singularity at $\omega'=\omega$, and $\mathcal{C}_\infty$ denotes a possible contribution from the contour at infinity. In what follows, we will neglect $\mathcal{C}_\infty$. A sufficient condition for this contribution to vanish is that the Green’s function falls off at large frequencies, which we will take as one of our assumptions and discuss it in more detail below. We will use a slightly different, though closely related, dispersion relation for the arc variables. We will introduce these relations below and discuss their properties in more detail.

The imaginary part of the retarded Green’s function, which appears on the right-hand side of Eq.~\eqref{DispRelUnsub}, has a direct physical interpretation: it determines the energy absorbed by the system from the external source. At leading order, the energy transferred to the system is quadratic in the external source and is given by\footnote{Here, we assume that the external source is switched on and off sufficiently rapidly in the far past and far future, respectively, such that the integral is convergent. Otherwise, one can consider the rate of energy exchange, which is also determined by the imaginary part.}
\begin{equation}\label{enab}
    \Delta H=\int \frac{\dd{\omega}}{2\pi}\omega\, {\rm Im}\,\big[K^{(E)}_{+-}\big](\omega)E^{ij}(\omega)E_{ij}^*(\omega)\,,
\end{equation}
where $E_{ij}(\omega)$ is the Fourier transform of the external tidal field, and where we have replaced $G_R$ by $K^{(E)}_{+-}$.

We assume that the system is passive, meaning that the net energy transferred from the source to the system is non-negative for any source profile. Equivalently, a passive system cannot be used to extract net work through an external perturbation: it may emit energy, but the net energy exchange is always an absorption. Since the source profile is arbitrary, this condition requires the kernel multiplying $E^{ij}(\omega)E_{ij}^*(\omega)$ in Eq.~\eqref{enab} to be non-negative,
\begin{equation}
    \omega \, {\rm Im}[G_R](\omega)\geq 0 \, , \quad\omega \in \mathbb{R}\,.
    \label{posim}
\end{equation}
It follows that ${\rm Im}[G_R](\omega)$ is non-negative for positive $\omega$ and non-positive for negative $\omega$.\footnote{One can show that, assuming a suitable asymptotic behavior, the sign-definiteness of ${\rm Im}[G_R](\omega)$ can in fact be extended to the entire region of analyticity. Defining $F_R(\omega)\equiv \omega\,G_R(\omega)$, one can conclude that ${\rm Im}[F_R](\omega)\geq 0$ for $\omega\in$ UHP, in which case $F_R$ is a Herglotz--Nevanlinna function. See, for instance, Ref.~\cite{Creminelli:2025rxj}.}     

We note that the retarded Green’s function is real in the time domain, independently of whether the system is passive. This follows either from Eq.~\eqref{eq:lQrmain}, since both the source and the induced multipole are real, or directly from Eq.~\eqref{eq:KpmGR}, since the operator is Hermitian. Consequently, the Fourier-space Green’s function satisfies the reality condition
\begin{equation}
G_R(\omega)^*=G_R(-\omega^*)\,.
\label{reality}
\end{equation}
In particular, ${\rm Im}[G_R](\omega)$ is odd in $\omega$ for $\omega\in\mathbb{R}$. Using this property, we can rewrite the dispersion relation \eqref{DispRelUnsub} in a form that makes the sign of the imaginary part more manifest:
\begin{equation}\label{DispRelUnsub2}
    {\rm Re}[G_R](\omega)=\frac{2}{\pi}{\rm PV}\int_0^{\infty} \dd{\omega'} \frac{\omega'\,{\rm Im}[G_R](\omega')}{\omega'^2-\omega^2}\,,
\end{equation}
where we have dropped the contribution $\mathcal{C}_\infty$. For a passive system, the numerator in the integrand is non-negative by Eq.~\eqref{posim}. In the limit $\omega\to0$, the denominator is also non-negative, and we therefore obtain
\begin{equation}
{\rm Re}[G_R](\omega=0)\geq0\,,
\label{KKpos}
\end{equation}
with equality if and only if ${\rm Im}[G_R](\omega)=0$ for all frequencies. Note that $G_R(0)={\rm Re}[G_R](0)$, since the imaginary part vanishes for $\omega=0$. This is the standard positivity constraint on the static ($\omega=0$) response that follows from the Kramers–Kronig relations.

One can try taking derivatives of Eq.~\eqref{DispRelUnsub2} with respect to $\omega$ and then taking the limit $\omega\to0$ to obtain analogous relations for the higher-order terms in the low-frequency expansion of ${\rm Re}[G_R](\omega)$, i.e.~the dynamical Love numbers~\cite{Rodriguez:2026iot,Chakraborty:2026qru}. However, in the presence of dissipation, one has to be careful when regularizing the pole at $\omega'=\omega$ that appears after taking the derivatives. See Refs.~\cite{Creminelli:2024lhd,Creminelli:2025rxj} for a discussion. An equivalent approach is to work directly with the arc variables, which we develop in the following subsection. 

Before moving on to the arc variables, let us summarize the assumptions that went into deriving Eq.~\eqref{KKpos}, as well as those underlying the analogous bounds below, and assess their validity for the gravitational response of compact objects. We have assumed that:
\begin{enumerate}
\item there exists a real causal function $G_R(t)$;
\item $G_R(\omega)$ decays at high frequencies; 
\item $G_R(\omega)$ satisfies the positivity property.
\end{enumerate}
Let us unpack these assumptions one by one.

\paragraph{Existence:} As described in Eq.~\eqref{eq:lQrmain}, the retarded Green’s function is initially defined within the worldline EFT, whose validity is limited to frequencies below $(GM/R_\star^3)^{1/2}$. In deriving the bounds, we are \emph{assuming} that this response function can be extended beyond the regime of validity of the EFT. In the full theory, one can indeed define a retarded Green’s function $G_R(t,x,x')$ describing the linear response, which also accounts for the spatial propagation of the source and therefore resolves the internal spatial structure of the compact object. What is less obvious is whether, in gravity, one can construct from this full Green’s function a real causal function $G_R(t)$ whose low-frequency expansion consistently captures the tidal Love numbers, including the static, dissipative, and dynamical contributions. It can be shown that this construction can be carried out in a linear theory, such as electromagnetism or weak gravity~\cite{unpub}. We leave the corresponding question in full general relativity for future work. See also Ref.~\cite{Correia:2025enx} for a related discussion on the $S$-matrix in black hole background.

\paragraph{Asymptotic behavior:} In addition, we assume that the Green's function decays at high (complex) frequencies, so that the contribution from the large semicircle $\mathcal{C}_\infty$ can be neglected. Unlike the $S$-matrix, whose asymptotic behavior is constrained by unitarity (at least in the presence of a mass gap), there is no general principle that fixes the high-frequency behavior of a retarded Green's function. In fact, depending on the operator, it can grow as any polynomial, $G_R(\omega) \sim \omega^n$; see footnote~\ref{Bogoliubov}. Nevertheless, decay at high frequencies is a reasonable assumption when a weakly coupled description is available. Physically, when the external perturbation oscillates on time scales much shorter than the characteristic response time of the system, the system cannot develop a coherent response. For example, in the electromagnetic response of a medium, the high-frequency behavior is governed by the approximately free motion of the microscopic degrees of freedom, such as electrons. In this regime, the induced multipole moments vanish at high frequency \cite{Creminelli:2024lhd}, leading to high-frequency transparency, or plasma behavior. A similar picture applies to weakly gravitating systems, where the characteristic frequency is set by the fundamental mode of the star. This is consistent with modeling the stellar response, to first approximation, in the Newtonian limit as a sum over normal modes, as we will discuss below (see also Refs.~\cite{Rodriguez:2026iot,Chakraborty:2026qru} and references therein). Relativistic corrections and strong-field effects can modify this simple picture, but for small compactness, they are expected to give only subleading corrections (see, e.g., Refs.~\cite{Pitre:2023xsr,Saes:2025jvr,HegadeKR:2026kku,Apostolidis:2026qsg} for more quantitative statements).\footnote{If the decay condition is not satisfied, one can instead use a subtracted dispersion relation. In practice, one divides $G_R(\omega)$ by a suitable polynomial $P(\omega)$ such that $G_R(\omega)/P(\omega)$ has sufficiently soft asymptotic behavior, and then writes the dispersion relation for this ratio.}

\paragraph{Positivity:} Finally, we assume that ${\rm Im}[G_R](\omega)$ satisfies the positivity condition~\eqref{posim}, motivated by the passivity of the object. At low frequencies, where the worldline EFT description is valid, the argument is essentially the same as the one given above. See Refs.~\cite{Goldberger:2005cd,Goldberger:2020fot,Biggs:2024dgp} for explicit computations in this regime. Beyond the EFT regime, however, it is not obvious in general relativity that the net energy absorbed by the object is directly related to the imaginary part of the $G_R(\omega)$ defined above. Consequently, passivity alone may not necessarily imply the positivity condition \eqref{posim} at arbitrary frequencies~\cite{unpub}.

\vspace{1cm}

As stated above, our goal in the present work is not so much to justify these assumptions---which we elaborate on further in Ref.~\cite{unpub}---as to take them as a starting point and explore their consequences. We expect them to provide a reasonable description of weakly gravitating compact objects. It is important to emphasize, however, that when applied to black holes, the vanishing of the static Love number is in sharp tension with the positivity condition \eqref{KKpos}.  As we will see in the following sections, these assumptions in fact fail even for neutron stars at high compactness.

\subsection{The arc variables and the moments}

\begin{figure*}[t]
\centering
\includegraphics[width=.5\linewidth]{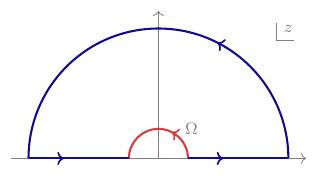}
\caption{The integration contour used in the definition of the arc variables. The defining contour in Eq.~\eqref{ArcEven1} is shown in red, and is subsequently deformed into the blue contour.}
\label{thearc}
\end{figure*}

Given the assumptions of the previous subsection (analyticity, decay at infinity, and positivity of $G_R(\omega)$) we now want to extend the positivity bound in Eq.~\eqref{KKpos} to higher-order terms. For this purpose, it is convenient to first introduce the arc variables, following Ref.~\cite{Bellazzini:2020cot}, and formulate the problem as a moment problem. We define
\begin{equation}
\label{ArcEven1}
a_{2n}(\Omega) \equiv \frac{1}{i \pi} \int_{\curvearrowleft_{\Omega}} \dd{z}\, \frac{G_R(z)}{z^{2 n+1}}, \quad n=0,1,2,\dots\,,
\end{equation}
where the integration contour $\curvearrowleft_{\Omega}$ is a semicircle of radius $\Omega$ in the ${\rm UHP}$ centered at the origin, as shown in Figure~\ref{thearc}.

There are several differences between our construction and the arc variables introduced in Ref.~\cite{Bellazzini:2020cot}. There, the arc variables are defined for a $2\to2$ scattering amplitude, which is an analytic function of the invariant energy squared, namely the Mandelstam variable $s$. Here, we construct analogous variables for the linear response function $G_R(\omega)$, which is a function of frequency. Moreover, in the $S$-matrix case, the asymptotic behavior is constrained by unitarity, requiring a kernel with a sufficiently high power of $s$, $1/s^{2n+3}$. In our case, by assuming that $G_R(\omega)$ vanishes at infinity, the weaker kernel $1/\omega^{2n+1}$ is sufficient. Another difference concerns the location of the contour. In the $S$-matrix case, the contour is centered at the crossing-symmetric point, $s=2m^2$ for scattering particles of mass $m$. In our case, the reality condition \eqref{reality} plays a role analogous to crossing symmetry, and the contours are therefore centered at the origin. Finally, our notation differs slightly from that of Ref.~\cite{Bellazzini:2020cot}: we denote the arcs by $a_{2n}$ rather than $a_n$.

One motivation for introducing the arc variables in Ref.~\cite{Bellazzini:2020cot} is to deal with amplitudes beyond tree level, where loop corrections generate an imaginary part even at low energies. A retarded Green's function generically has a nonzero imaginary part at low frequencies, as is the case, for example, for dissipative coefficients. This provides a natural motivation for adopting the same language here. We emphasize, however, that one could instead work directly with the dispersion relation \eqref{DispRelUnsub2}, as discussed in Ref.~\cite{Creminelli:2024lhd}.

Exploiting the analyticity of $G_R(\omega)$ in the UHP, we can use Cauchy's theorem to deform the integration contour in Eq.~\eqref{ArcEven1} into a contour along the real axis and a large semicircle at complex infinity (see Figure~\ref{thearc}). By our assumption on the asymptotic behavior of $G_R(\omega)$, the contribution from the latter vanishes, giving
\begin{equation}
\begin{aligned}
\label{ArcEven2}
a_{2n}(\Omega)=\frac{1}{i \pi} \int_{\Omega}^{+\infty} \dd{z} \,\left(\frac{G_R(z)}{z^{2 n+1}}-\frac{G_R(z)^{*}}{z^{2 n+1}}\right) =\frac{2}{\pi} \int_{\Omega}^{+\infty} \dd{z}\, \frac{{\rm Im}[G_R](z)}{z^{2 n+1}}\,,
\end{aligned}
\end{equation}
where in the first equality we used the reality condition \eqref{reality}, and for the second equality we used the definition of the imaginary part. Positivity of ${\rm Im}[G_R](\omega)$ then implies that all arcs are positive,
\begin{equation}
\label{ArcEvenPos}
a_{2n}(\Omega)\geq0\,,\quad n=0,1,2,\dots.
\end{equation}

Notice that deriving Eq.~\eqref{ArcEvenPos} required all of the assumptions discussed in the previous subsection. As emphasized in Ref.~\cite{Bellazzini:2020cot}, however, the positivity of the individual arcs does not exhaust the constraints that follow from their interpretation as moments of a positive measure. Indeed, after changing variables to $x=\Omega^{2}/z^{2}$ in Eq.~\eqref{ArcEven2}, we can write
\begin{equation}
\label{A3}
\Omega^{2 n} a_{2n}(\Omega)=\int_{0}^{1} \dd{\mu(x)}\, x^{n}\equiv m_n\,,\quad n=0,1,2,\dots\,, 
\end{equation}
where the positive measure is $\dd{\mu(x)} \equiv \frac{1}{\pi} \frac{\dd{x}}{x} {\rm Im}[G_R](\Omega / \sqrt{x})$ on the interval $x\in[0,1]$, and $m_n$ denotes its $n$-th moment. Thus, the sequence $\{a_0,\Omega^2 a_2,\Omega^4 a_4,\dots\}$ forms a moment sequence associated with a positive measure supported on $[0,1]$.

The necessary and sufficient conditions for an infinite sequence to be a sequence of moments on a given interval, and more generally on other spaces, have been extensively studied in mathematics; see, for instance, Refs.~\cite{shohat1943problem,lasserre2009moments,schmuedgen2017moment}. Here, we are interested in the truncated moment problem: given a finite sequence $\{m_0,m_1,\dots,m_{2N}\}$,\footnote{The criteria differ slightly depending on whether the last term in the sequence is an even or odd moment. Here, we focus only on the even case.} what conditions must it satisfy in order to admit an extension to a sequence of moments? For the finite interval $[0,1]$, the necessary and sufficient conditions can be summarized as follows:
\begin{equation}
\label{ArcHankel}
H_{N}^{0}(m) \succeq 0\,,\quad\text{and} \quad H_{N-1}^{1}(m)-H_{N-1}^{2}(m)\succeq0\,,
\end{equation}
in terms of Hankel matrices $H_{n}^{\ell}(m)\in\mathbb{R}^{(n+1)\times(n+1)}$ defined as
\begin{equation}
\left(H_{n}^{\ell}(m)\right)_{i j}\equiv m_{i+j+\ell}\,, \quad i,j=0,\dots,n\,.
\end{equation}
In matrix notation, Eq.~\eqref{ArcHankel} reduces to
\begin{equation}
\label{ArcHankel-matrix}
\begin{bmatrix}
m_0&m_1&\dots&m_N\\
m_1&m_2&\dots&m_{N+1}\\
\vdots&\vdots&\dots&\vdots\\
m_N&m_{N+1}&\dots&m_{2N}
\end{bmatrix}\succeq 0\,,\quad\text{and} \quad
\begin{bmatrix}
m_1-m_2&\dots&m_N-m_{N+1}\\
\vdots&\ddots&\vdots\\
m_N-m_{N+1}&\dots&m_{2N-1}-m_{2N}
\end{bmatrix}\succeq 0\,.
\end{equation}
The symbol $\succeq$ in Eqs.~\eqref{ArcHankel} and \eqref{ArcHankel-matrix} denotes that the symmetric matrix is positive semidefinite. The origin of these conditions can be understood as follows. For a positive measure on $[0,1]$, the integral of any non-negative function over the interval must itself be non-negative. In particular, we can consider the following family of polynomials:
\begin{equation}
\int_{0}^{1} \dd{\mu(x)}\,P_N(x)^2\geq0\,,\quad \int_{0}^{1} \dd{\mu(x)}\,x(1-x)P_{N-1}(x)^2\geq0\,,
\label{polH}
\end{equation}
where $P_n(x)$ is any polynomial of degree $n$, not necessarily positive. Parametrizing it as $P_n(x)=\sum_{i=0}^n v_i x^i$ gives
\begin{equation}
\sum_{i,j=0}^N\,v_i\,\big(m_{i+j}\big)\,v_j\geq0\,,\quad \sum_{i,j=0}^{N-1}\,v_i\,\big(m_{i+j+1}-m_{i+j+2}\big)\,v_j\geq0\,,   
\end{equation}
which are precisely the conditions in Eq.~\eqref{ArcHankel}. The nontrivial part is that these conditions are also sufficient: Eq.~\eqref{ArcHankel} guarantees that $\{m_0,m_1,\dots,m_{2N}\}$ can be extended to a sequence of moments associated with a positive measure on $[0,1]$. We refer the reader to the references above for the proof and further details.

\begin{figure*}[t]
\centering
\includegraphics[width=.44\linewidth]{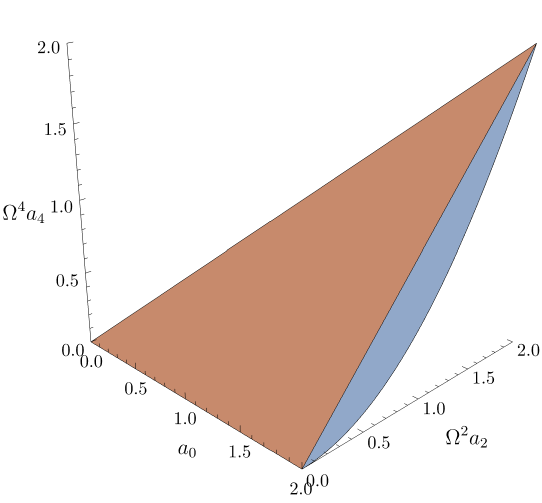}
\hspace{4mm}
\includegraphics[width=.44\linewidth]{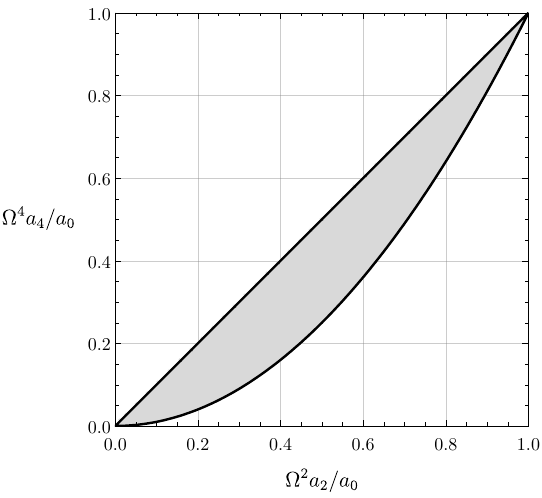}
\caption{The allowed region for the arcs from Eq.~\eqref{arcoptcons}. \textit{Left:} it can be shown that the allowed region is a convex cone; it is the conic hull of vectors $(1,x,x^2)$ for $x\in[0,1]$. \textit{Right:} An equivalent representation, by normalizing with $a_0>0$.}
\label{arcbound}
\end{figure*}

In what follows, we will mainly be interested in the case $N=1$, corresponding to the sequence $\{m_0,m_1,m_2\}$. It is straightforward to check that Eq.~\eqref{ArcHankel} then implies
$m_0\geq m_1\geq m_2\geq0$, and $m_0m_2-m_1^2\geq0$. In terms of the arc variables, these conditions become
\begin{equation}
a_0(\Omega)\geq \Omega^2a_2(\Omega)\geq \Omega^4a_4(\Omega)\geq0\,,\qquad
a_0(\Omega)a_4(\Omega)-a_2(\Omega)^2\geq0\,.
\label{arcoptcons}
\end{equation}
Clearly, Eq.~\eqref{arcoptcons} is stronger than the positivity of the individual arcs in Eq.~\eqref{ArcEvenPos}. The allowed region is shown in Figure~\ref{arcbound}. In the following subsection, we express the arc variables in terms of the low-energy expansion of the retarded Green's function, thereby translating these moment constraints into bounds on the EFT coefficients.

\subsection{Arc bounds in the EFT}

If $\Omega$ is not too large and lies within the regime of validity of the low-energy description, we can evaluate the arcs from their defining expression in Eq.~\eqref{ArcEven1} using the EFT. We can expand the retarded Green's function as
\begin{equation}
\label{AL1}
G_R(\omega)=\sum_{j=0}^{+\infty} \lambda_{j} \omega^{j}=\lambda_0+i\tilde{\lambda}_1\omega+\lambda_2\omega^2+\dots\,, 
\end{equation}
for $|\omega|<\Lambda$, where $\Lambda$ denotes the EFT cutoff. In the context of the gravitational response of compact objects, where $G_R$ is identified with the kernel $K_{+-}^{(E)}$, the coefficients $\lambda_j$ are related to the tidal Love numbers. The reality condition \eqref{reality} implies that the even coefficients are real, while the odd coefficients are purely imaginary. It is therefore convenient to write $\lambda_{2k+1}\equiv i\tilde{\lambda}_{2k+1}$. Comparing with Eq.~\eqref{ap:KpmcEs}, we find the following relation between the $\lambda_j$ and the dimensionless tidal Love number coefficients:
\begin{equation}
\lambda_0=\frac{R_{\star}^5}{G}c_E\,,\quad\tilde{\lambda}_1=MR_{\star}^5 \nu_E\,,\quad \lambda_2=\frac{R_{\star}^8}{G^2M}c_{\dot{E}}\,,\quad\tilde{\lambda}_3=\frac{R_{\star}^8}{G}\nu_{\dot{E}}\,,\quad\lambda_4=\frac{R_{\star}^{11}}{G^3M^2}c_{\ddot{E}}\,, \quad \cdots \, .
\label{eq:ctolambda}
\end{equation}
As mentioned below Eq.~\eqref{ap:KpmcEs}, the EFT cutoff is expected to be set by the characteristic frequency of the object, i.e.,
\begin{equation}
\Lambda\approx\Lambda_\star\equiv\left(\frac{GM}{R_{\star}^3}\right)^{1/2}\,.
\label{cutoffLstar}
\end{equation}

Choosing $\Omega<\Lambda$ in Eq.~\eqref{ArcEven1}, we can evaluate the arcs in terms of the coefficients in the low-energy expansion \eqref{AL1}:
\begin{equation}
\label{AL2}
a_{2 n}(\Omega)=\sum_{j=0}^{+\infty} \frac{\lambda_{j} \Omega^{j-2 n}}{\pi} \int_{0}^{\pi} \dd{\theta} \, \E^{i(j-2 n) \theta} =\lambda_{2 n}-\frac{2}{\pi} \sum_{k=0}^{+\infty} \frac{\tilde{\lambda}_{2 k+1} \Omega^{2(k-n)+1}}{2(k-n)+1}\,. 
\end{equation}
We used the fact that the integral in the first equality is $\pi$ for $j=2n$, vanishes when $j-2n$ is nonzero and even, and equals $2i/(2k'+1)$ when $j-2n=2k'+1$. In addition, in the final summation, we made the substitution $k'=k-n$. More explicitly, the first three arcs are 
\begin{align}\label{eq:arcsdiss}
& a_{0}(\Omega)=\lambda_{0}-\frac{2}{\pi}\left(\tilde{\lambda}_{1} \Omega+\frac{1}{3} \tilde{\lambda}_{3} \Omega^{3}+\frac{1}{5} \tilde{\lambda}_{5} \Omega^{5}+\ldots\right)\,,  \\
& a_{2}(\Omega)=\lambda_{2}-\frac{2}{\pi}\left(-\frac{\tilde{\lambda}_{1}}{\Omega}+\tilde{\lambda}_{3} \Omega+\frac{1}{3} \tilde{\lambda}_{5} \Omega^{3}+\ldots\right)\,,  \\
& a_{4}(\Omega)=\lambda_{4}-\frac{2}{\pi}\left(-\frac{\tilde{\lambda}_{1}}{3 \Omega^{3}}-\frac{\tilde{\lambda}_{3}}{\Omega}+\tilde{\lambda}_{5} \Omega+\ldots\right)\,.\label{eq:arcsdiss2} 
\end{align}
It is important to note a crucial difference from the tree-level expression for the arcs in the context of the $S$-matrix \cite{Bellazzini:2020cot}. At tree level, each arc is associated with a single coefficient (see below). In contrast, Eq.~\eqref{AL2} involves an infinite series of coefficients, arising from the odd powers of $\omega$, i.e.~the dissipative  numbers. A similar structure appears in the $S$-matrix beyond tree level, where loop corrections generate an imaginary part of the amplitude \cite{Bellazzini:2020cot}. In that case, however, perturbative control of the EFT implies a hierarchy between tree-level and loop contributions, allowing the resulting infinite series to be truncated consistently. See Ref.~\cite{Bellazzini:2020cot} for a more detailed discussion.\footnote{Beyond tree level, the $S$-matrix expansion is more complicated than Eq.~\eqref{AL2}, primarily due to the presence of non-analytic terms, such as $s^a\log(s)^b$, which modify both the real and imaginary parts of the low-energy amplitude.} In the present context, this hierarchy is generally absent: the dissipative terms can already contribute at the same order as the conservative terms.\footnote{The worldline EFT describing astrophysical objects is a classical effective theory, and therefore quantum loop corrections are suppressed. Classical loop corrections, on the other hand, are present and lead to the running of the low-energy coefficients (see the following sections), without invalidating the low-energy expansion in Eq.~\eqref{AL1}. This is in contrast to, for example, the dissipative EFT of fluids, where a more intricate low-energy structure has been shown to arise \cite{Grozdanov:2024fle}.} Consequently, the infinite series in Eq.~\eqref{AL2} cannot, in general, be consistently truncated over the entire range $\Omega<\Lambda$ without additional assumptions.  

The infinite number of terms in the expression for the arcs presents an obstruction to translating the arc bounds directly into bounds on the EFT coefficients. In the limit $\Omega\to0$, however, higher-order terms are increasingly suppressed, and it is therefore consistent to truncate the series. Keeping only the leading-order term, we have
\begin{equation}
a_0\approx \lambda_0\,,\quad\text{and}\quad a_{2n}\approx\frac{2}{\pi}\frac{\tilde{\lambda}_1}{(2n-1)\Omega^{2n-1}}\,,\quad n=1,2,\dots\,.
\end{equation}
It is then straightforward to see that the resulting arc bounds do not provide any information beyond what is already known: $\lambda_0\geq0$ and $\tilde{\lambda}_1\geq0$, which follow, respectively, from the Kramers--Kronig relation and the positivity assumption. For more interesting bounds we should look at finite $\Omega<\Lambda$. 

One can show that the arc bounds become stronger as $\Omega$ is increased. For the subset \eqref{arcoptcons}, this can be seen directly by writing the bounds in the form $B\geq0$, where $B$ is a function of the arc: the $\Omega$-derivative of the left-hand side is always negative, so increasing $\Omega$ moves the bound towards a more restrictive region. For example, for the nonlinear bound, we find
\begin{equation}
\partial_\Omega\bigg(a_0(\Omega)a_4(\Omega)-a_2(\Omega)^2\bigg)=-\frac{4}{\pi}\frac{{\rm Im}[G_R](\Omega)}{\Omega^5}\left(\frac{a_0+\Omega^4a_4}{2}-\Omega^2a_2\right)\leq0\,,
\end{equation}
where we have used Eq.~\eqref{ArcEven2} to evaluate the $\Omega$-derivatives of the arcs. The last inequality follows from
$\Omega^2a_2\leq\sqrt{a_0(\Omega^4a_4)}\leq(a_0+\Omega^4a_4)/2$.
The same argument applies to the other bounds in Eq.~\eqref{arcoptcons}. A more general argument establishing this property for all the bounds can be found in Ref.~\cite{Bellazzini:2020cot}. On the other hand, for finite $\Omega$, the infinite sum in Eq.~\eqref{AL2} cannot be truncated without extra assumptions on the EFT coefficients. 

We defer dealing with dissipation to Section \ref{diss}, and instead, concentrate on conservative systems for which the effect of low-energy dissipation can be neglected. In this case, as for the tree-level $S$-matrix case, the relation between the arcs and the EFT coefficients is simplest:
\begin{equation}
a_{2n}(\Omega)=\lambda_{2n}\,,\qquad(\text{no dissipation})
\label{nodiss}
\end{equation}
for any $n=0,1,2,\dots$. In the context of the gravitational response of compact objects, this holds when it is possible to model, for example, a neutron star as a perfect fluid. This will be the topic of sections \ref{sec:NeutronStar} and \ref{sec:eos}. Note that this assumption does not hold for black holes, for which the dissipative  coefficients are of the same order, in units of the cut off, as the conservative ones~\cite{Chakrabarti:2013lua,Ivanov:2022hlo,Pitre:2023xsr,Perry:2023wmm,Saketh:2023bul,Ivanov:2024sds,Katagiri:2024wbg,Chakraborty:2025wvs,Caron-Huot:2025tlq,Combaluzier--Szteinsznaider:2025eoc,Kobayashi:2025vgl,Correia:2026utp}.\footnote{See the reviews~\cite{Rodriguez:2026iot,Chakraborty:2026qru} for a comprehensive list of references on dynamical tidal effects for black holes.} In the absence of dissipation, using the bounds \eqref{arcoptcons} for the first three arcs, we obtain
\noindent
\begin{tcolorbox}[colframe=white,arc=0pt,colback=greyish2]
\vspace{-0.2cm}\begin{equation}
\lambda_0\geq\Omega^2 \lambda_2 \geq\Omega^4\lambda_4\geq0\,,\qquad\lambda_0 \lambda_4-\lambda_2^2\geq0\,.
\label{EFTbounds}
\end{equation}
\end{tcolorbox}
In the absence of dissipation, with the assumptions of Section~\ref{retG}, the conservative tidal Love numbers are not only positive but also satisfy the relations \eqref{EFTbounds}. Notice the difference between the bounds in Eq.~\eqref{EFTbounds}: the first three inequalities involve the size of the arc $\Omega$, while the last is independent of it.     

\subsection{The boundary and the mode sum}\label{subsec:modes}

When are the arc bounds  saturated? Equivalently, what happens on the boundary of the allowed region? We have seen that the allowed region is described by Eq.~\eqref{ArcHankel}, i.e.~suitable combination of Hankel matrices to be positive semidefinite. The boundary is then described as either of the matrices in Eq.~\eqref{ArcHankel} becomes singular, i.e.~there exists a real nonzero vector $v_i$ such that $(H_N^0)_{ij}v^iv^j=0$ or $(H_{N-1}^1-H_{N-1}^2)_{ij}v^iv^j=0$. As argued for Eq.~\eqref{polH}, this condition can be translated into the following integrals
\begin{equation}
\int_{0}^{1} \dd{\mu(x)}\,P_v(x)^2=0\,,\quad\text{or}\quad \int_{0}^{1} \dd{\mu(x)}\,x(1-x)P_v(x)^2=0\,, 
\label{polpoint}
\end{equation}
in which $P_v(x)=\sum_iv_ix^i$, is the polynomial associated to $v^i$. Since the integrand is non-negative, this is possible only when $\dd{\mu(x)}$ is supported only on the zeros of $P_v$, and for the second condition, possibly at the end points $x=0,1$. As a result, the measure must be of the form
\begin{equation}
\dd{\mu(x)}=\sum_k\rho_k\delta(x-x_k)\dd{x}\,,
\label{pointmu}
\end{equation}
where $x_k$'s are zeros of $P_v$ or the end points, and $\rho_k\geq0$ the corresponding magnitude. The nontrivial part is that, for every point on the boundary there is a \emph{unique} measure of the form \eqref{pointmu}. In contrast, for the points in the interior of the allowed region, there are infinitely many compatible measures which can be both supported on points, as in Eq.~\eqref{pointmu}, or more general. It is only for the boundary points that the measure has to be as in Eq.~\eqref{pointmu} and is uniquely determined. For more details see Ref.~\cite{schmuedgen2017moment}. The magnitudes $\rho_k$ can be determined from the first few moments using Eq.~\eqref{pointmu},
\begin{equation}
    m_n=\sum_k\rho_kx_k^n\,,
    \label{rhos}
\end{equation}
and then solving the system for the $\rho_k$'s. With the measure specified, one is able to compute all of the moment sequence, not just a truncation.  

It is straightforward to rewrite the measure in Eq.~\eqref{pointmu} in terms of the imaginary part of the retarded Green's function. From the definition, given below Eq.~\eqref{A3}, we conclude that
\begin{equation}
{\rm Im}[G_R](\omega)=\pi\,{\rm sgn}(\omega)\sum_k\rho_k\,\omega_k^2\,\delta\big(\omega^2-\omega_k^2\big)\,,\quad\text{for}\quad|\omega|\geq\Omega\,,
\label{ImGpoint}
\end{equation}
where $\omega_k\equiv\Omega/\sqrt{x_k}\geq\Omega$. Note that both $\omega_k$ and $\rho_k$ can depend on $\Omega$. Eq.~\eqref{ImGpoint} gives ${\rm Im}[G_R](\omega)$ only for $|\omega|\geq\Omega$. However, it is not difficult to see that Eq.~\eqref{ImGpoint} is in fact the expression for the imaginary part over the whole real line.\footnote{Since $G_R(\omega)$ is analytic in the UHP, all of its singularities must lie in the lower half-plane (LHP). By assumption in Eq.~\eqref{AL1}, $G_R(\omega)$ is analytic in a disc with radius $\Lambda\geq\Omega$ around the origin. From Eq.~\eqref{ImGpoint}, it has a series of poles at $\omega=\pm\omega_k$ on the real line. However, it cannot have any other singularity strictly in the LHP, i.e.~at finite distance from the real line. The contribution of such singularity, is necessarily a real-analytic function for $\omega\in\mathbb{R}$, and therefore, has a smooth imaginary part over the whole real line. This is in contradiction with Eq.~\eqref{ImGpoint}. Therefore, all singularities of $G_R(\omega)$ are on the real line, and Eq.~\eqref{ImGpoint} is valid everywhere.} Then, we can use Cauchy's theorem to construct the retarded Green's function\footnote{Similar holds in the context of the $S$-matrix, where the boundary corresponds to a theory for distinct massive resonances \cite{Bellazzini:2020cot}.}
\begin{equation}
G_R(\omega)=\frac{1}{\pi}\int_{-\infty}^{+\infty}\frac{\dd{z}}{z-\omega-i\epsilon}{\rm Im}[G_R](z)=\sum_k\frac{\rho_k\omega_k^2}{-(\omega+i\epsilon)^2+\omega_k^2}\,,
\label{GRmod}
\end{equation}
where the $i\epsilon$ is compatible with the retarded prescription. We conclude that, in order to saturate the bounds, the Green's function is necessarily a sum of normal modes with frequency $\omega_k$. In particular, this proves that the boundary of the allowed region necessarily corresponds to a conservative system, where all dissipative terms vanish. 

Let us then study boundaries of the allowed region for the subset $\{m_0,m_1,m_2\}=\{a_0,\Omega^2a_2,\Omega^4a_4\}$, in Eq.~\eqref{arcoptcons}. See also Figure~\ref{arcbound}. The linear boundary is $m_1=m_2$, corresponding to $H_0^1-H_0^2=0$. The associated polynomial is $P_v=1$. The measure only can have support on the end points $x=0$ and $x=1$, with magnitudes $\rho_0=m_0-m_1$, and $\rho_1=m_1$ respectively. However, this boundary depends on $\Omega$. The maximum value of $\Omega$ for which the bounds are satisfied gives an upper bound on the EFT cutoff, as will be discussed in section~\ref{sec:eos}. 

The nonlinear boundary, $m_0m_2=m_1^2$ for $m_0\neq0$, corresponds to $H_1^0(m)$ becoming singular. It is straightforward to check that the null (eigen-)vector is $v=(1,-m_0/m_1)$, with the associated $P_v(x)=1-\frac{m_0}{m_1}x$. Therefore, there is only a single mode in Eq.~\eqref{pointmu}, corresponding to the zero of $P_v$, located at  
\begin{equation}
\omega_\star=\frac{\Omega}{\sqrt{m_1/m_0}}=\sqrt{\frac{\lambda_0}{\lambda_2}}=\sqrt{\frac{\lambda_2}{\lambda_4}}\,,
\end{equation}
where the second equality comes from $\lambda_0\lambda_4=\lambda_2^2$. Notice that this is independent of $\Omega$. From Eq.~\eqref{rhos}, the corresponding magnitude is found to be $\rho_\star=m_0=\lambda_0$. The full retarded Green's function is as in \eqref{GRmod} with only a single term. The EFT cutoff is $\Lambda=\omega_\star$, and for $|\omega|<\omega_\star$, all the coefficients are fixed in terms of $\lambda_0$ and $\lambda_2$:
\begin{equation}
G_R(\omega)=\sum_{n=0}^\infty\frac{\lambda_2^n}{\lambda_0^{n-1}}\omega^{2n}=\lambda_0+\lambda_2\omega^2+\frac{\lambda_2^2}{\lambda_0}\omega^4+\frac{\lambda_2^3}{\lambda_0^2}\omega^6+\dots\,.
\end{equation}

As previously stated, in the Newtonian regime, the response of a star can be robustly modeled as a sum of collective modes with different normal frequencies, as in Eq.~\eqref{GRmod} (see, for instance, Refs.~\cite{Chakrabarti:2013xza,Andersson:2019ahb}, and Ref.~\cite{Lipparini_1989} in the context of nuclear physics). 
Besides providing intuition for the large-$|\omega|$ behavior of the response, this picture gives a simple toy model in which the arc bounds \eqref{EFTbounds} can be checked explicitly. Indeed it is straightforward to see that from Eq.~\eqref{GRmod}  
\eq{
\lambda_0=\sum_k\rho_k\,,\quad\lambda_2=\sum_k\frac{\rho_k}{\omega_k^2}=\lambda_0\,\mathbf{E}\bigg[\frac{1}{\omega_k^2}\bigg]\,,\quad\lambda_4=\sum_k\frac{\rho_k}{\omega_k^4}=\lambda_0\,\mathbf{E}\bigg[\frac{1}{\omega_k^4}\bigg]\,,
}
where the expectation value $\mathbf{E}[\cdot]$ is taken with respect to the distribution of the modes,~$\rho_k/\sum_i\rho_i$. The bound in Eq.~\eqref{EFTbounds} is then proportional to the variance
\begin{equation}
\lambda_0\lambda_4-\lambda_2^2=\lambda_0^2\,\mathbf{Var}\bigg[\frac{1}{\omega_k^2}\bigg]\geq0\,,
\end{equation}
where saturation occurs only if there is only a single mode in the response. As a result, the strict positivity of the nonlinear bound in Eq.~\eqref{EFTbounds} is a measure of how well the response function can be approximated by a single mode. See also Section~\ref{sec:eos}. Note that validity of the bounds in Eq.~\eqref{EFTbounds} is beyond approximating the response in terms of a sum over the normal modes, and only relies on our assumptions stated above.

\newpage

\section{Dynamical Love numbers of neutron stars}
\label{sec:NeutronStar}

In this section, we apply the Love-number bounds derived in the previous section to gravitating objects. In particular, we consider neutron stars modeled as perfect fluids. This allows us to neglect dissipative effects and directly apply the bounds \eqref{EFTbounds}.
Although the assumptions discussed in Section~\ref{sec:arc_bounds} underlying Eqs.~\eqref{EFTbounds} are well motivated in the low-compactness regime, here we do not impose any restriction on the compactness of the star. We compute instead the response coefficients for generic compactness by solving the background and perturbation equations both in the interior and exterior of the star in full general relativity, without imposing any restriction on its size or mass. This will allow us to investigate possible subtleties that arise in the strong-gravity regime and to highlight the differences from the weak-gravity regime.

To this end, we match the EFT description \eqref{eq:EFTconsint} to explicit stellar models and determine the Wilson coefficients $c_E$, $c_{\dot{E}}$, and $c_{\ddot{E}}$ in terms of the properties and dynamics of the stellar interior. In the following, we summarize the main ingredients of stellar perturbation theory, obtain the solution numerically through fourth order in frequency, and then discuss the matching procedure and the bounds.

\subsection{Interior solution and stellar perturbation theory}
\label{sec:maininterior}

\subsubsection{Background solution}

We take the unperturbed configuration of the star to be a static, spherically symmetric spacetime described by the background metric $\bar{g}_{\mu\nu}$. In Schwarzschild-like coordinates, $(t,r,\theta,\varphi)$, its line element is
\begin{equation}  
    \d s^2 = \bar{g}_{\mu\nu}\d x^\mu \d x^\nu = -\E^{\nu(r)} \d t^2+\E^{\lambda(r)} \d r^2 + r^2\left(\d\theta^2+\sin^2\theta \d\varphi^2\right),
\label{eq:BGmetric}
\end{equation}
where $\nu(r)$ and $\lambda(r)$ are functions of the radial coordinate only.\footnote{We use $\lambda$ to parametrize the $r$--$r$ component of the background metric. This should not be confused with the EFT coefficients introduced in the previous section.}
We assume that the matter content of the star is described by a perfect fluid, with energy-momentum tensor given by
\begin{equation}
    T_{\mu \nu}=\left(\epsilon+p\right)u_\mu u_\nu +p \bar g_{\mu \nu},
\label{eq:BGfluid}
\end{equation}
where $\epsilon$ and $p$ denote the total energy density and pressure, respectively, while $u^\mu=(\E^{-\nu/2},0,0,0)$ is the fluid four-velocity in the fluid rest frame. As a timelike vector, it is normalized according to $\bar g_{\mu\nu}u^\mu u^\nu=-1$.

For the unperturbed metric and energy-momentum tensor, the Einstein equations reduce to the standard Tolman--Oppenheimer--Volkoff (TOV) equations~\cite{Tolman:1939jz,Oppenheimer:1939ne}:
\begin{subequations}
\label{eq:TOVeqs}
\begin{align}
    m'(r)& =4\pi r^2 \epsilon(r),\\
    \nu'(r)& =G\frac{2m(r)+8\pi r^3 p(r)}{r\left(r-2Gm(r)\right)},\\
    p'(r)& =-G\left(\epsilon(r)+p(r)\right)\frac{m(r)+4\pi r^3p(r)}{r\left(r-2Gm(r)\right)},
\end{align}
\end{subequations}
where the prime denotes differentiation with respect to the radial coordinate $r$, and where we have defined $m(r)\equiv\frac{r}{2G}(1-\E^{-\lambda})$, which corresponds to the mass enclosed within a sphere of  radius $r$.

The radius of the star, $R_\star$, is determined by the condition
\begin{equation}
    p(R_\star)=0 .
\end{equation}
The  total mass $M$ of the star is, by definition,
\begin{equation}
    m(R_\star)\equiv M .
\end{equation}

The system \eqref{eq:TOVeqs} contains four unknown functions but only three independent equations and therefore requires an equation of state to be closed. In the following, we will assume the fluid to be barotropic, i.e.~$p=p(\epsilon)$. In addition, it will sometimes be useful to introduce the speed of sound, defined as
\begin{equation}
    c_s^2=\frac{p'(r)}{\epsilon'(r)}\,.
\label{eq:soundspeed}
\end{equation}

\subsubsection{Dynamical perturbation equations}
\label{sec:interiorperturbations}

We are interested in solving the coupled dynamics of the fluid and gravitational fluctuations, perturbatively in frequency. To this end, we perturb the background metric $\bar g_{\mu\nu}$ and the fluid energy-momentum tensor $T_{\mu\nu}$ according to $\bar g_{\mu\nu}\to\bar g_{\mu\nu}+\delta g_{\mu\nu}$  and $T_{\mu\nu}\to T_{\mu\nu}+\delta T_{\mu\nu}$. 
The procedure is standard~\cite{1967ApJ...149..591T,Lindblom:1983ps,Detweiler:1985zz,Kojima:1992ie} and has been used previously to study the dynamical tidal response of neutron stars up to second order in frequency (see, e.g., Refs.~\cite{Pitre:2023xsr,Apostolidis:2026qsg} and references therein, and Refs.~\cite{Rodriguez:2026iot,Chakraborty:2026qru} for a comprehensive list of relevant works). Extending the calculation to fourth order is computationally straightforward. We therefore keep the discussion concise, focusing on the main ingredients and referring to previous work for the technical details.

We decompose the parity-even metric perturbations into spherical harmonics as
\begin{multline}
\delta g_{\mu\nu}\d x^\mu \d x^\nu=
- r^\ell \int \frac{\D \omega}{2\pi}\left[\E^\nu H^{\ell m}_0(r) \d t^2-2i \omega r H^{\ell m}_1(r) \d t\d r \right.
\\
\left.+\E^\lambda H_2^{\ell m}(r) \d r^2 +r^2 K^{\ell m}(r)(\d\theta^2+\sin^2\theta \d\varphi^2) \right]Y_{\ell m}(\theta,\varphi)\E^{-i\omega t}\,,\label{eq:PerturbedMetric}
\end{multline}
where $Y_{\ell m}(\theta,\varphi)$ denote the scalar spherical harmonics, while $H_0^{\ell m}$, $H_1^{\ell m}$, $H_2^{\ell m}$, and $K^{\ell m}$ are radial functions describing the corresponding metric perturbations. In Eq.~\eqref{eq:PerturbedMetric} we fixed the Regge--Wheeler gauge~\cite{Regge:1957td}.
Similarly, we decompose the spatial components of the fluid four-velocity perturbation as $\delta u^i=u^0 \partial_t \xi^i =-i\omega \E^{-\nu/2}\xi^i$, with~\cite{Lindblom:1983ps,Detweiler:1985zz} 
\begin{subequations}
\begin{align}
    \xi^r& =r^{\ell-1}\E^{-\lambda/2}W^{\ell m}Y_{\ell m}\E^{-i\omega t}\,,\label{eq:expW}
\\
    \xi_A& =-r^\ell V^{\ell m} {\cal  Y}_A^{\ell m} \E^{-i\omega t}\,,\label{eq:expV}
\end{align}
\end{subequations}
where ${\cal Y}_A^{\ell m}=(\partial_\theta Y_{\ell m},\partial_\varphi Y_{\ell m})$ are the even-parity vector spherical harmonics, and $W^{\ell m}(r)$ and $V^{\ell m}(r)$ are radial functions parametrizing the fluid velocity perturbations. The time component $\delta u^0$ is instead fixed by the normalization condition $(\bar g_{\mu\nu}+\delta g_{\mu\nu})(u^\mu+\delta u^\mu)
(u^\nu+\delta u^\nu)=-1$ and is given by $\delta u^0=-\frac{1}{2}\E^{-\nu/2}r^\ell H_0Y_{\ell m}\E^{-i\omega t}$.

Plugging the decompositions $\bar g_{\mu\nu}+\delta g_{\mu\nu}$, $p+\delta p$, $\epsilon+\delta\epsilon$, and $u^\mu+\delta u^\mu$ into the linearized Einstein equations, we obtain the equations governing the coupled dynamics of the fluid and gravitational degrees of freedom. We follow here the conventions and procedure adopted in Ref.~\cite{Apostolidis:2026qsg}, previously introduced by Ref.~\cite{Katagiri:2025qze}. After some algebraic manipulations, the linearized Einstein equations can be recast as a system of three differential equations~\cite{Katagiri:2025qze}:
\begin{subequations}
\label{eq:eqforppHpWpV}
\begin{align}
    {\cal H}''&=\alpha_{\cal H, H'}{\cal H}'+\alpha_{\cal H,H}{\cal H}+\alpha_{{\cal H},W} W+\alpha_{{\cal H},V}V,~\label{eq:eqforppH}\\
    W'&=\alpha_{W,{\cal H}'}{\cal H}'+\alpha_{W,{\cal H}}{\cal H}+\alpha_{W,W}W+\alpha_{W,V}V,\label{eq:eqforpW}\\
    V'&=\alpha_{V,{\cal H}'}{\cal H}'+\alpha_{V,{\cal H}}{\cal H}+\alpha_{V,W}W+\alpha_{V,V}V,\label{eq:eqforpV}
\end{align}
\end{subequations}
where we defined ${\cal H}\equiv-r^\ell H_0$, while the coefficients $\alpha_{I,J}$ are functions of the background quantities and the frequency $\omega$; see Refs.~\cite{Apostolidis:2026qsg,Katagiri:2025qze} and references therein for their explicit expressions.
Note that, although the fluid and gravitational sectors each propagate a single parity-even degree of freedom, it is convenient to formulate the dynamics as a system of three differential equations. In particular, the equation \eqref{eq:eqforppH} for the metric perturbation ${\cal H}$ is second order, while the two equations \eqref{eq:eqforpW}--\eqref{eq:eqforpV}  are first order.

As in Ref.~\cite{Apostolidis:2026qsg}, we solve Eqs.~\eqref{eq:eqforppHpWpV} order by order in the small-frequency expansion. The only difference is that, here, we extend the calculation to fourth order in $\omega$.
We start by expanding ${\cal H}$, $V$, and $W$ as 
\begin{subequations}
\label{eq:PerturbativeHVW}
\begin{align}
    {\cal H}&={\cal H}^{(0)}+(r_s\omega)^2 {\cal H}^{(2)}+(r_s\omega)^4 {\cal H}^{(4)} +\mathcal{O}(\omega^6),\\
    V&=V^{(0)}+(r_s\omega)^2 V^{(2)}+(r_s\omega)^4 V^{(4)}+\mathcal{O}(\omega^6),\\
    W&=W^{(0)}+(r_s\omega)^2 W^{(2)}+(r_s\omega)^4 W^{(4)}+\mathcal{O}(\omega^6).
\end{align}
\end{subequations}
The advantage of this procedure is that, as can be checked explicitly, the coefficients $\alpha_{{\cal H},W}$ and $\alpha_{{\cal H},V}$ begin at order $\omega^2$ upon using the background equations of motion. As a result, the equation for ${\cal H}^{(2n)}$ decouples from $W^{(2n)}$ and $V^{(2n)}$, and depends only on quantities determined at lower orders. Consequently, determining ${\cal H}^{(4)}$---which is what we need to match to the EFT and extract the gravito-electric dynamical response at fourth order---requires solving the full system only up to second order in $\omega$.

At each order in the $r_s\omega$ expansion, the equation for ${\cal H}^{(2n)}$ is thus an inhomogeneous second-order ordinary differential equation. Solving it requires specifying two boundary conditions. The first one is given by regularity at the center of the star, $r=0$.\footnote{Note that regularity at the star's center fixes the interior solution up to an overall constant. This normalization is immaterial and can be absorbed into the amplitude of the tidal field at infinity~\cite{Apostolidis:2026qsg}.} The second one is obtained by imposing the vanishing of the Lagrangian perturbation of the pressure at the stellar surface, $r=R_\star$~\cite{Apostolidis:2026qsg}. These conditions fix the integration constants and select the physical solution.

Note that such a solution cannot in general be found analytically in closed form, except for very special choices of the equation of state. We therefore proceed by solving Eqs.~\eqref{eq:eqforppHpWpV} numerically. Once the metric and fluid perturbations in the interior of the star have been obtained, we can match ${\cal H}$ to the exterior solution. This procedure is outlined in the following section. The results for different choices of the equation of state are instead presented in Section~\ref{sec:eos}.

\subsection{Exterior solution}

The exterior solution to the vacuum Einstein equations can be conveniently obtained by solving the Zerilli equation. The Zerilli equation governs the dynamics of vacuum, parity-even gravitational perturbations, and takes the form~\cite{Zerilli:1970se,Zerilli:1970wzz}
\begin{equation}
\label{eq:Zerillimain}
    \frac{\d^2}{\d r_*^2} \Psi_{\mathrm{Z}}(r)+\left[\omega^2-f(r)\left(\frac{r_s}{r^3}+\frac{2 n}{3 r^2}+\frac{8 n^2(2 n+3)}{3\left(2 n r+3 r_s\right)^2}\right)\right] \Psi_{\mathrm{Z}}(r)=0 .
\end{equation}
where  $f(r)=1-r_s/r$ and $n\equiv(\ell+2)(\ell-1)/2$.\footnote{The notation for $n$ is standard and should not be confused with the polytropic index appearing in Section~\ref{sec:eos}.}
The Zerilli field $\Psi_{\mathrm{Z}}$ is related to the metric perturbations defined in Eq.~\eqref{eq:PerturbedMetric} via (see, e.g., Refs.~\cite{Berti:2009kk,Rodriguez:2026iot})
\begin{align}
\label{eq:ZtoK}
    K & =  - r^{-\ell}f(r) \frac{\mathrm{d} \Psi_{\mathrm{Z}}}{\mathrm{d} r}- \frac{3 r_s^2+3 n r_s r+2(n+1) n r^2}{r^{2+\ell}\left(3 r_s+2 n r\right)} \Psi_{\mathrm{Z}}, 
\\
      H_1 & =-r^{-\ell-1}f(r) \frac{r^2}{2(r-r_s)} \frac{\mathrm{~d} \Psi_{\mathrm{Z}}}{\mathrm{d} r}+\frac{r^{-\ell-1}(3 r_s^2+6 n r_s r-4 n r^2)}{4\left(r-r_s\right)\left(3 r_s+2 n r\right)} \Psi_{\mathrm{Z}}\,,
\end{align}
and
\begin{equation}
\label{eq:algebraic}
    \left(2 n r+3 r_s\right)H_0+2\left[r_s(n+1)-2 r^3 \omega ^2\right]H_1 -\frac{4 n r (r-r_s)-4 r^4 \omega ^2+2 r r_s-3 r_s^2}{2 (r-r_s)}K =0,
\end{equation}
where we set $H_2=H_0$, which follows from the $\theta$--$\varphi$ component of the Einstein equations applied to the metric \eqref{eq:PerturbedMetric}.

The equation \eqref{eq:Zerillimain} can be  solved perturbatively in $\omega$ as
\begin{equation}
\Psi_{\mathrm{Z}}=\Psi_{\mathrm{Z}}^{(0)}+(r_s\omega)^2 \Psi_{\mathrm{Z}}^{(2)}+(r_s\omega)^4 \Psi_{\mathrm{Z}}^{(4)}+\cdots ,
\end{equation}
similarly to Eqs.~\eqref{eq:PerturbativeHVW}.
At each order in $r_s\omega$, the functions $\Psi_{\mathrm{Z}}^{(0)}$, $\Psi_{\mathrm{Z}}^{(2)}$, and $\Psi_{\mathrm{Z}}^{(4)}$ admit closed-form expressions, which can  be obtained using standard Green's function methods. The analytical solution is obtained in terms of harmonic polylogarithm functions \cite{Remiddi:1999ew}. The procedure and the details are presented in Appendix~\ref{app:GRcomputation}~\cite{Combaluzier--Szteinsznaider:2025eoc,Apostolidis:2026qsg}.

For illustration, we give here the expression of the large-$r$ expansion of the $\ell=2$ exterior solution for the Zerilli field (see also Eq.~\eqref{Zer_sol_inf_GR}):\footnote{Note that, compared to Eq.~\eqref{Zer_sol_inf_GR}, we are setting the coefficients of the terms odd in frequency to zero. These terms break time-reversal invariance and are associated with dissipative effects. Since we model the star as a perfect fluid, its dynamics is purely conservative. Consequently, the integration constants multiplying the odd-in-frequency terms in $\Psi_{\mathrm{Z}}$ vanish upon matching the exterior solution to the interior solution obtained in Section~\ref{sec:maininterior}.}
\begin{align}
     \Psi_{\mathrm{Z}}^{\ell=2}(r) \xrightarrow{r\rightarrow\infty}
      &\,\, a_0\frac{ r^3}{r_s^3}+\frac{r_s^2}{r^2}\left(\frac{1 }{5}b_0-\frac{189  }{1024}a_0\right) \nonumber 
      \\
      &+\omega ^2 r_s^2\left[\frac{r^3}{r_s^3} \left(a_2+a_0\frac{107 }{210 }\log \left(\frac{r_s}{r}\right)\right)+\frac{r_s^2}{r^2}\left(b_2-\left(a_0\frac{3011 }{10240}+b_0\frac{107}{1050}\right) \log \left(\frac{r_s}{r}\right)\right)\right] \nonumber 
      \\
      &+\omega ^4 r_s^4\Bigg[\frac{r^3}{r_s^3} \left(a_4+\left(a_0\frac{1695233 }{9261000}+a_2\frac{107}{210}\right) \log \left(\frac{r_s}{r}\right)+a_0\frac{11449}{88200}\log ^2\left(\frac{r_s}{r}\right)\right) \nonumber 
      \\
      &\qquad\quad+\frac{r_s^2}{r^2}\left(b_4- \left(a_0\frac{2634215 }{7225344}+a_2\frac{1987 }{5120}+b_0\frac{1695233 }{46305000}+b_2\frac{107 }{210}\right)\log \left(\frac{r_s}{r}\right)\right. \nonumber 
      \\
      &\qquad\quad\left.-\frac{11449 (945 a_0-1024 b_0) }{451584000}\log ^2\left(\frac{r_s}{r}\right)\right)\Bigg]+\cdots,
\label{Zer_sol_inf_GR_main}
\end{align}
where only the terms corresponding to the $r^3$ and $r^{-2}$ falloffs are shown. 
In Eq.~\eqref{Zer_sol_inf_GR_main}, we denoted with $a_n$ and $b_n$ the two integration constants that arise when solving the Zerilli equation perturbatively at each order $n=0,1,2,\cdots$.

Once the solution for $\Psi_{\mathrm{Z}}$ is obtained, Eqs.~\eqref{eq:ZtoK}--\eqref{eq:algebraic} can be used to construct the variable ${\cal H}\equiv-r^\ell H_0$, which can then be matched to the interior numerical solution for ${\cal H}$ obtained in Section~\ref{sec:interiorperturbations}. Denoting by ${\cal H}^{\rm ext}$ and ${\cal H}^{\rm int}$ the exterior and interior solutions for ${\cal H}$, respectively, the matching is performed by imposing continuity of the field and its first derivative at the radius $R_\star$ of the object, i.e.,
\begin{equation}
\label{surface_cond}
    \mathcal{H}^{\text{ext}}(R_{\star})=\mathcal{H}^{\text{int}}(R_{\star})\,,\qquad\partial_r\mathcal{H}^{\text{ext}}(R_{\star})=\partial_r\mathcal{H}^{\text{int}}(R_{\star})\,.
\end{equation}
The two conditions \eqref{surface_cond} fix two of the integration constants of the boundary-value problem. Finally, requiring that the exterior solution approaches a tidal field at large distances completely fixes the remaining freedom, up to an overall constant amplitude. It is sometimes convenient to eliminate the dependence on this amplitude by defining the ratio~\cite{Hinderer:2007mb}
\begin{equation}
\label{eq:yH}
y\equiv\frac{R_{\star}\partial_r\mathcal{H}(R_{\star})}{\mathcal{H}(R_{\star})}\,.
\end{equation}

\subsection{Matching to point-particle EFT}

Determining the Love number couplings in terms of the neutron star's properties and equation of state---which are implicitly encoded in the coefficients $a_n$ and $b_n$ appearing in Eq.~\eqref{Zer_sol_inf_GR_main}---requires matching the large-$r$ exterior solution \eqref{Zer_sol_inf_GR_main} obtained in the full theory to the corresponding quantity computed within the EFT described by the action \eqref{eq:ppEFT0}. To correctly capture gravitational nonlinearities, and in particular the running behavior manifest in Eq.~\eqref{Zer_sol_inf_GR_main}, the EFT calculation must be performed at loop level. The perturbative order to which the EFT calculation of the Love numbers must be carried out is determined by the angular momentum number $\ell$ of interest and the power $n$ of the frequency $\omega$, and scales as $\mathcal{O}(G^{2\ell+1+n}\omega^n)$. The procedure and power counting are detailed in Refs.~\cite{Combaluzier--Szteinsznaider:2025eoc,Apostolidis:2026qsg} and summarized in Appendix~\ref{app:ppEFT} below.  Here we report the final result and refer to Refs.~\cite{Combaluzier--Szteinsznaider:2025eoc,Apostolidis:2026qsg} (see also Ref.~\cite{Caron-Huot:2025tlq}) and to the appendix for a thorough discussion.

Matching the Zerilli field computed within the EFT \eqref{eq:ppEFT0} to the relativistic result \eqref{Zer_sol_inf_GR_main} yields the following expressions for the renormalized Wilson couplings of the point-particle EFT:\footnote{The expressions \eqref{eq:ccoeffsmain} can be obtained by comparing Eq.~\eqref{ap:KpmcEs} with Eq.~\eqref{K_pm_ab}, where we have set the coefficients of the terms odd in frequency to zero, as dictated by the perfect-fluid description of the stellar interior.}
\begin{tcolorbox}[colframe=white,arc=0pt,colback=greyish2]
\begin{subequations}
\label{eq:ccoeffsmain}
\begin{align}
    c_E & = \frac{8G^5M^5}{45R_\star^5}\frac{b_0}{a_0},
\\
    c_{\dot E} & = \frac{32G^8M^8}{45R_\star^8} \left[- \left(1+\frac{107}{105}\frac{b_0}{a_0}\right)\log (\mu r_s) \right.
\nonumber\\
    &\qquad \qquad\qquad \left. -\frac{67981 }{176400}\frac{ b_0}{ a_0}-\frac{b_0a_2}{a_0^2}+ \frac{945 a_2}{1024 a_0}+\frac{5 b_2}{a_0}-\frac{1015283}{1032192}
    \right],
\\
    c_{\ddot E} & = \frac{128G^{11}M^{11}}{45R_\star^{11}}
    \left[\frac{107}{210}\left(1+\frac{107}{105}\frac{ b_0}{a_0}\right) \log ^2(\mu  r_s) 
    \right.
\nonumber\\
    &\qquad \qquad\qquad
    -\frac{107}{105}\log (\mu  r_s) \left(-\frac{a_2 b_0}{a_0^2}+\frac{945 a_2}{1024 a_0}-\frac{32869 }{1258320 }\frac{b_0}{a_0}+\frac{5 b_2}{a_0}+\frac{76383379}{61358080}\right)
\nonumber\\
    &\qquad \qquad\qquad
    +\frac{a_2^2 b_0}{a_0^3}-\frac{945 }{1024 }\frac{a_2^2}{a_0^2}+\frac{67981 }{176400 }\frac{a_2 b_0}{a_0^2}
    -\frac{5 a_2 b_2}{a_0^2}-\frac{a_4 b_0}{a_0^2}-\frac{203943 }{573440 }\frac{a_2}{a_0}
\nonumber\\
    &\qquad \qquad\qquad \left. +\frac{945 }{1024 }\frac{a_4}{a_0}-\frac{21388060129 }{124467840000 }\frac{b_0}{a_0}-\frac{67981 }{35280 }\frac{b_2}{a_0}+\frac{5 b_4}{a_0} -\frac{6744009765173}{3641573376000} \right],
\end{align}
\end{subequations}
\end{tcolorbox}
\noindent where $\mu$ is the renormalization scale and we have used dimensional regularization (details are in Appendix~\ref{app:ppEFT}). 

In the following, we compute explicitly the coefficients $a_0$, $a_2$, $a_4$, $b_0$, $b_2$, and $b_4$ for a range of standard and physically motivated equations of state, and discuss the implications of the bounds~\eqref{EFTbounds}.

\newpage

\section{Applications to neutron-star equations of state}
\label{sec:eos}
 
We begin by rewriting the bounds~\eqref{EFTbounds} in the notation of Eq.~\eqref{ap:KpmcEs} (see Eq.~\eqref{eq:ctolambda}):
\noindent
\begin{tcolorbox}[colframe=white,arc=0pt,colback=greyish2]
\vspace{-0.45cm}\begin{equation}
B_1(\Omega)\equiv c_E-\Omega^{2} \frac{R_{\star}^3}{GM}c_{\dot{E}}\geq0 \,,\quad B_2(\Omega)\equiv c_{\dot{E}}-\Omega^{2} \frac{R_{\star}^3}{GM}c_{\ddot{E}}\geq0 \, , \quad B_3\equiv c_E c_{\ddot{E}}-c^{2}_{\dot{E}}\geq0\,,
\label{CE_bounds2}
\end{equation}
\end{tcolorbox}
\noindent
in addition to $c_E\geq0$~\cite{unpub}.\footnote{Note that the conditions $c_{\dot E}\geq0$ and $c_{\ddot E}\geq0$ follow from combining $c_E\geq0$ with the bounds \eqref{CE_bounds2}.} Note the appearance of the combination $GM/R_\star^3$, which corresponds to the square of the characteristic energy scale $\Lambda_\star$ defined in Eq.~\eqref{cutoffLstar}.
Since the Love numbers $c_E$, $c_{\dot E}$ and $c_{\ddot E}$ depend implicitly on the interior solution, and hence on the equation of state of the star, the bounds~\eqref{CE_bounds2} could in principle yield nontrivial constraints on the modeling of neutron-star interiors. 

\subsection{Polytropic equation of state}
\label{sec:polytropic}

To illustrate the application of the arc bounds~\eqref{CE_bounds2}, we begin with a simple model for the neutron-star equation of state, namely, a polytropic equation of state defined by~\cite{Hinderer:2007mb,Binnington:2009bb,Damour:2009vw}\footnote{Note that Eq.~\eqref{eq:poly_eos} is equivalent to the relation $p=K \rho^{1+1/n}$, where $\rho$ is the fluid's particle mass density.}
\begin{equation}
\label{eq:poly_eos}
    \epsilon=\left(\frac{p}{K}\right)^{\frac{n}{n+1}}+np\,,
\end{equation}
where $n$ is the polytropic index and $K$ a generic constant parameter. 

\begin{figure*}[p]
\subfloat{%
    {\includegraphics[width=.49\linewidth]{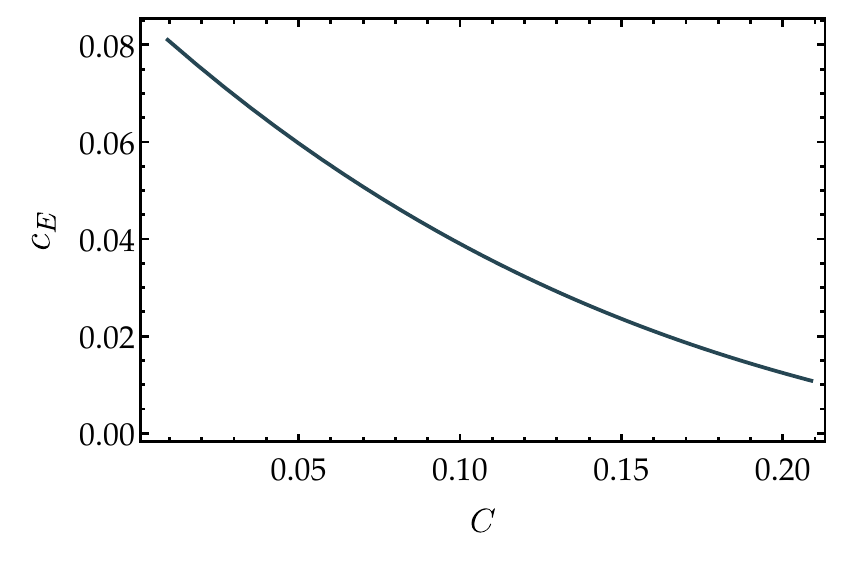}}
            \label{}%
        }
        \subfloat{%
        \includegraphics[width=.49\linewidth]{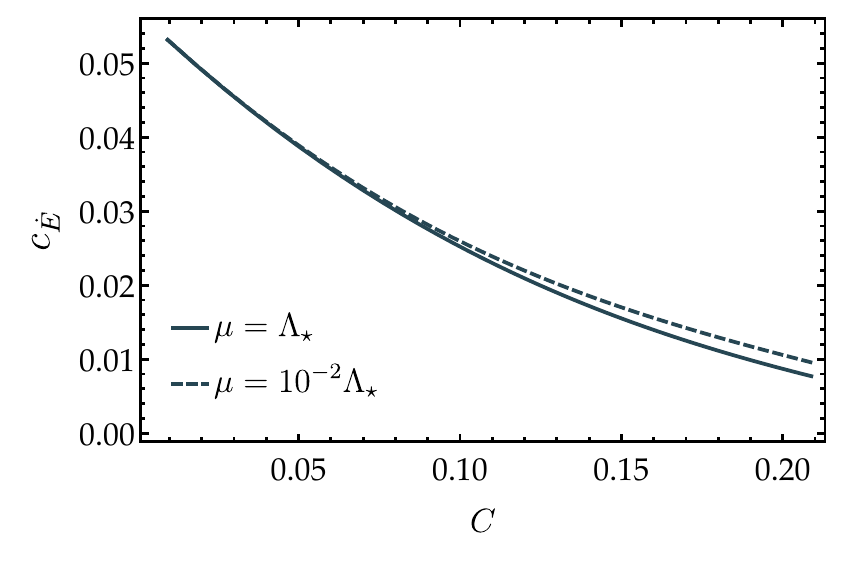}%
            \label{}%
        }\\
        \subfloat{%
    {\includegraphics[width=.49\linewidth]{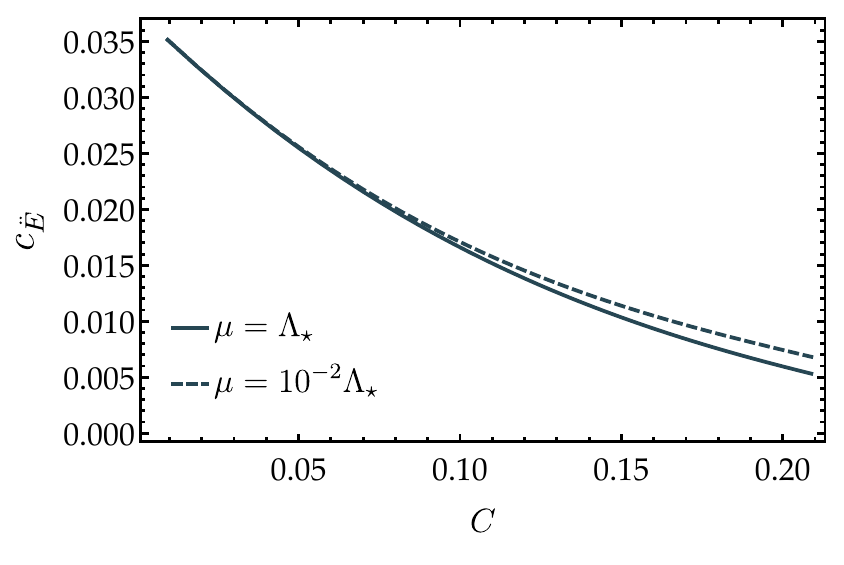}}
            \label{}%
        }
        \subfloat{%
        \includegraphics[width=.49\linewidth]{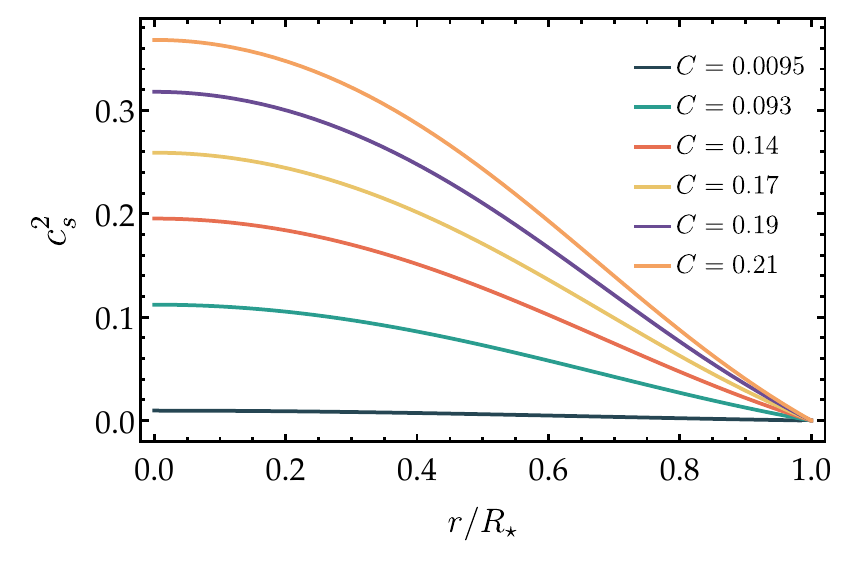}%
            \label{}%
        }\\
        \subfloat{%
        \includegraphics[width=.49\linewidth]{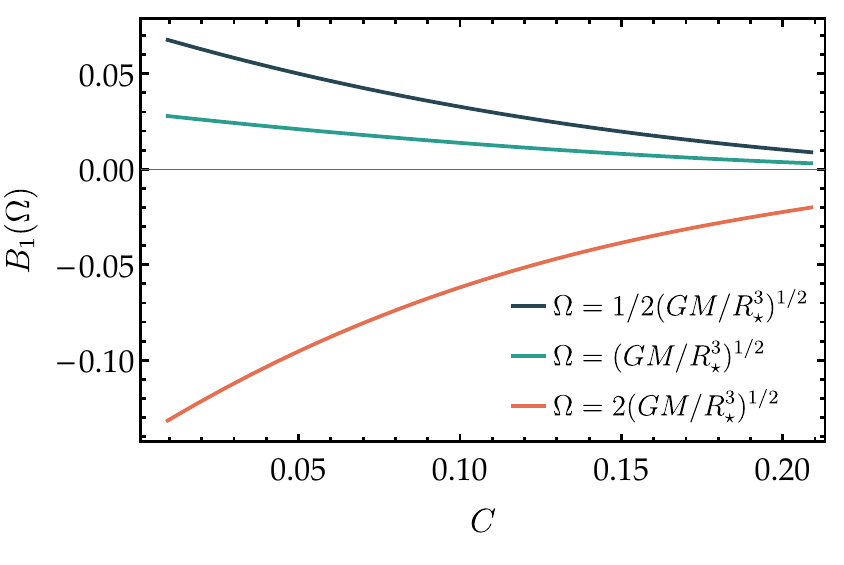}%
            \label{}%
        }\hfill
        \subfloat{%
        \includegraphics[width=.49\linewidth]{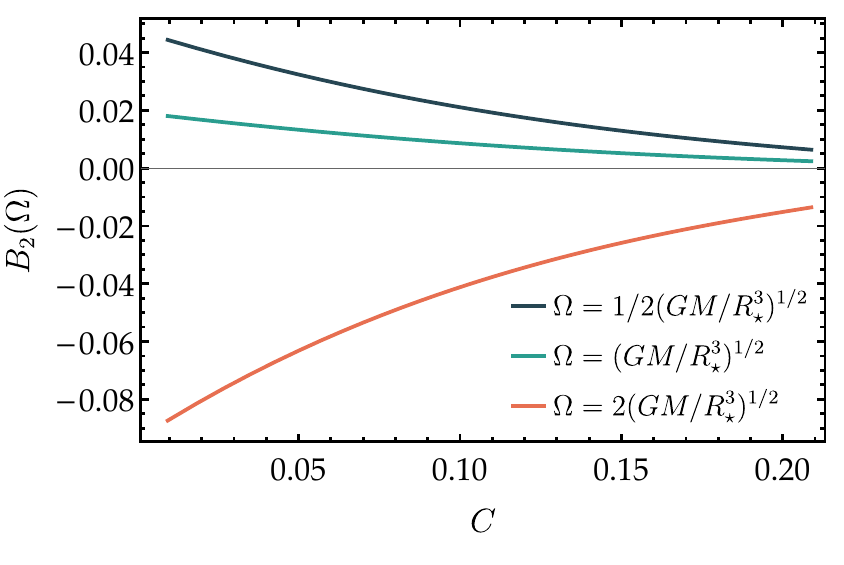}
            \label{}%
        }
         \caption{Static and dynamical Love numbers, squared sound speed, and bounds $B_1$ and $B_2$ defined in Eq.~\eqref{CE_bounds2}. The Love numbers and bounds are plotted as functions of the compactness $C$, while the squared sound speed is shown as a function of the radial coordinate for several values of $C$. Throughout, we adopt a polytropic equation of state with $n=1$ and $K=10^2 G^3 M_\odot^2$. The dynamical Love numbers exhibit renormalization-group running, leading to a dependence on the renormalization scale $\mu$ in $c_{\dot E}$, $c_{\ddot E}$, $B_1$, and $B_2$. To illustrate this dependence, we show $c_{\dot E}$ and $c_{\ddot E}$ for two representative values of $\mu$, while fixing $\mu=\Lambda_\star$ in the plots of $B_1$ and $B_2$.}
   \label{n1_plots}
\end{figure*}

We begin with the case $n=1$, which provides a simple model capturing some of the qualitative features of neutron-star matter. We will consider other values of $n$, as well as more physically motivated equations of state, below. After integrating the stellar structure equations in the interior, we evaluate the inequalities \eqref{CE_bounds2} as a function of the compactness, defined by
\begin{equation}
    C\equiv \frac{GM}{R_{\star}} ,
\end{equation}
considering configurations up to the maximum mass supported by the equation of state. We show the resulting bounds for different values of the arc length $\Omega$ in Figure~\ref{n1_plots}.

A few comments are in order. First, the plots for $B_1$ and $B_2$ in Figure~\ref{n1_plots} show that the bounds are violated as $\Omega$ is increased. Let us call $\Omega_{\rm max}$ the  maximum value of $\Omega$ allowed by the bounds. It is  conveniently expressed in the units of $\Lambda_\star=(GM/R_\star^3)^{1/2}$, as previously defined in Eq.~\eqref{cutoffLstar}.

The violation of the bounds at values larger than $\Omega_{\rm max}$, suggests a breakdown of the EFT~\cite{Bellazzini:2020cot}. In other words, $\Omega_{\rm max}$ becomes a proxy for an upper bound on the EFT cutoff $\Lambda$ (see also Eq.~\eqref{cutoffLstar}). We stress that the values for $\Omega_{\rm max}$ that we will find below need not be the optimal ones. Stronger constraints, beyond the ones that we explored here, exist and might make the upper limit on $\Omega$ more stringent.

\begin{figure*}[t]
\centering
        \subfloat{%
        \includegraphics[width=.45\linewidth]{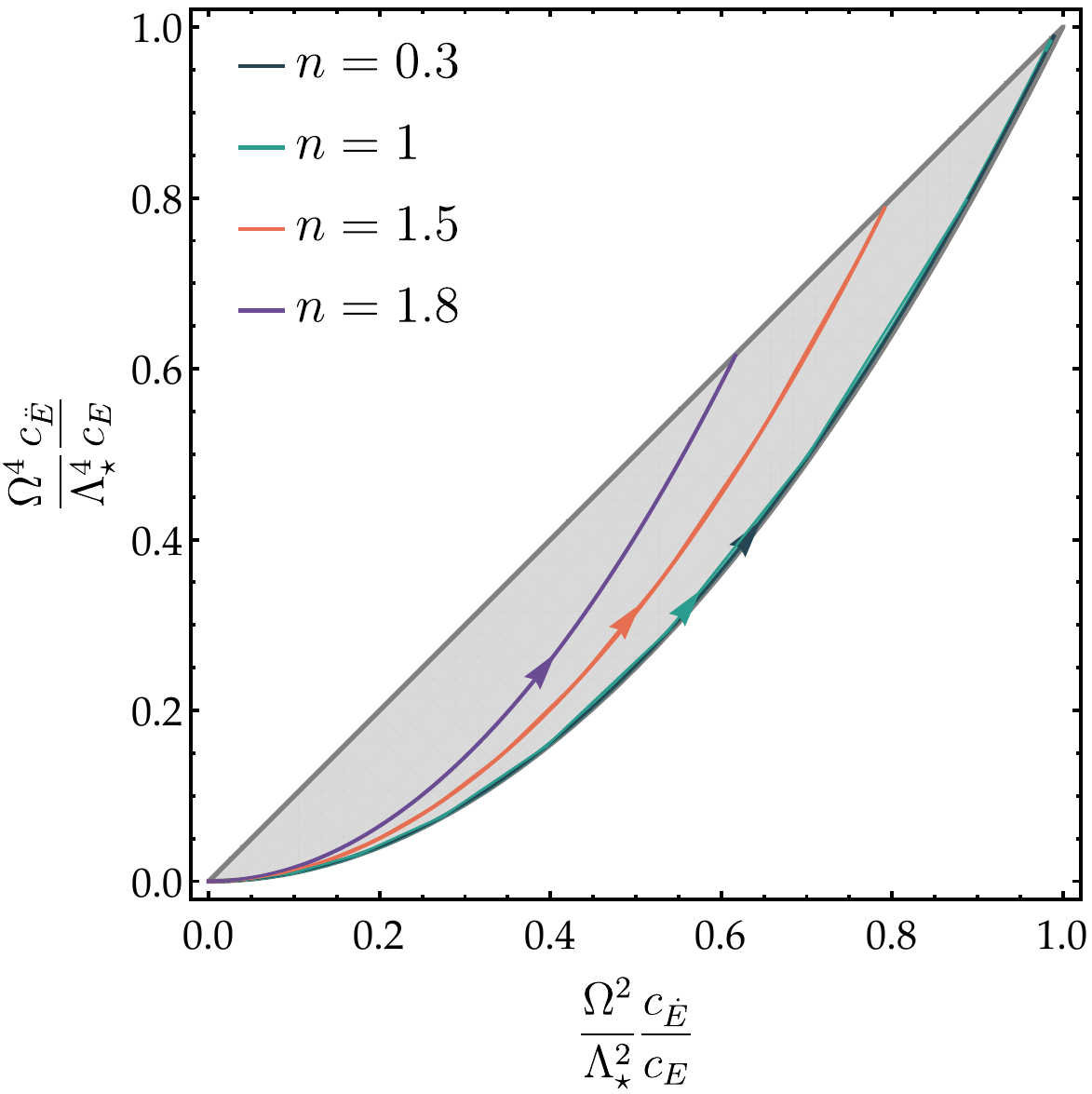}%
            \label{}%
        }
        \subfloat{%
    \raisebox{0.6cm}{\includegraphics[width=.55\linewidth]{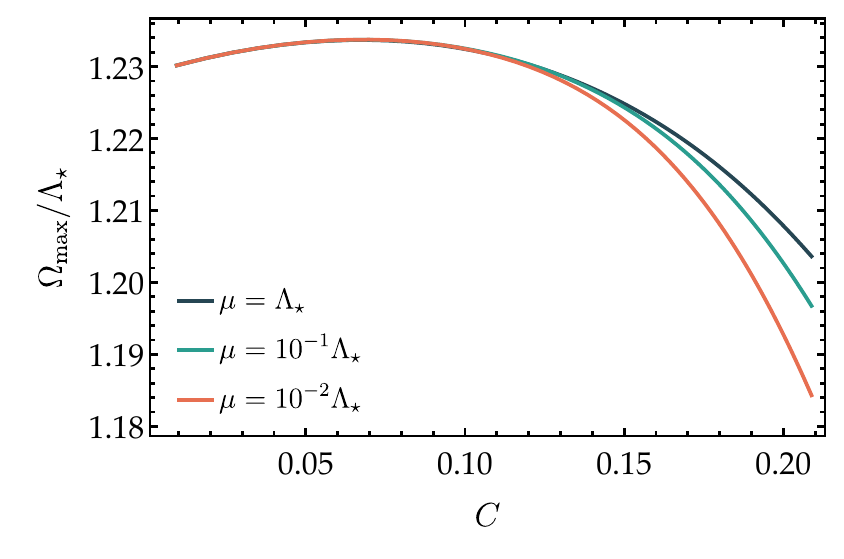}}
            \label{}%
        }
         \caption{\textit{Left:} Curves obtained by increasing $\Omega$ from zero up to the maximum value allowed by $B_2$, for different polytropic indices. In all cases, we fixed $C=0.05$, and  $\mu=\Lambda_\star$. The shaded region corresponds to the parameter space allowed by the arc bounds~\eqref{CE_bounds2}. \textit{Right:} Maximum allowed value of $\Omega$ as a function of compactness for an $n=1$ polytropic equation of state, as determined by $B_2$.}
   \label{n1_amax}
\end{figure*}

For completeness, Figure~\ref{n1_plots} also shows the static and dynamical Love numbers, as well as the squared sound speed, $c_s^2$, as a function of the radial coordinate for different values of the compactness $C$. The sound speed remains subluminal throughout the star for all configurations considered. For $c_{\dot E}$ and $c_{\ddot E}$, we show results for two representative values of the renormalization scale $\mu$.\footnote{By contrast, the static Love number is known not to run when the bulk classical dynamics is governed by general relativity, a property that can be understood in terms of the underlying symmetries~\cite{Hui:2021vcv,Hui:2022vbh,Charalambous:2021kcz,Charalambous:2022rre,Combaluzier-Szteinsznaider:2024sgb,Parra-Martinez:2025bcu}. Quantum effects, which we neglect in this work, generically induce a renormalization-group running for the static coefficients as well~\cite{Elkhidir:2026lgp}.} The dependence on $\mu$ becomes more pronounced as $C$ increases, whereas the dynamical Love numbers become independent of $\mu$ in the weak-gravity limit. (See also the right panel of Figure~\ref{n1_amax}, where we show how $\Omega_{\rm max}$ changes with $\mu$.) This is precisely the regime in which the assumptions underlying our bounds are better justified and their implications can be established with greater confidence. 

In all the plots of Figure~\ref{n1_plots}, we fixed the parameter $K$ in Eq.~\eqref{eq:poly_eos} to the reference value $K=10^2G^3M_\odot^2$, with $M_\odot$ the solar mass. As we show in Table~\ref{K_bounds}, varying $K$  has only a mild effect on $\Omega_{\rm max}$. This weak dependence in the relativistic regime has already been noted in the literature~\cite{Hinderer:2007mb,Postnikov:2010yn}. 
\begin{table}[t]
    \centering
    \begin{tabular}{ccc}
        \toprule
        $K/G^3M_\odot^2$ & $\Omega_{\max}^{(B_1)}/\Lambda_{\star}$ & $\Omega_{\max}^{(B_2)}/\Lambda_{\star}$ \\
        \midrule
        $10^3$ & 1.289 & 1.248 \\
        $10^2$ & 1.241 & 1.233 \\
        $10$ & 1.237 & 1.232 \\
        $1$ & 1.236 & 1.232 \\
        \bottomrule
    \end{tabular}
\caption{Maximum values $\Omega_{\text{max}}$ in units of $\Lambda_{\star}$, as determined by the bounds $B_1$ (second column) and $B_2$ (third column), for different values of the polytropic constant $K$, measured in units of $G^3M_\odot^2$,  with $M_\odot$ the solar mass. In all cases, we fixed $n=1$, $C=0.05$, and  $\mu=\Lambda_\star$.}
\label{K_bounds}
\end{table}

In Table~\ref{Polytrope_bounds}, we explore several values of the polytropic index $n$ and, for each one, compute the maximum value of $\Omega$ allowed by the bounds $B_1$ and $B_2$. In all cases, the compactness is fixed to $C=0.05$ to remain close to the weak-gravity regime, where the dependence on the renormalization scale $\mu$ is suppressed. Notice that, for fixed $n$, the bound on $\Omega$ implied by the inequality $B_2$ is always stronger than the one implied by $B_1$. This can be understood simply as a consequence of the $\Omega$-independent bound $c_E c_{\ddot{E}}-c_{\dot{E}}^2\geq0$. Indeed, dividing this last inequality by $c_{\dot{E}}c_{\ddot{E}}$ yields
\begin{equation}
\Omega_{{\rm max}}^{(B_2)}\leq\Omega_{{\rm max}}^{(B_1)}\,,
\end{equation}
where, by definition,  $\Omega_{{\rm max}}^{(B_1)}=\Lambda_\star\sqrt{c_E/c_{\dot E}}$, and $\Omega_{{\rm max}}^{(B_2)}=\Lambda_\star\sqrt{c_{\dot E}/c_{\ddot E}}$.

We remind that the EFT cutoff, $\Lambda$, is defined by the location of the closest singularity of the response function, corresponding to the frequency of the leading mode in the system. As a result, $\Omega_{\rm max}$ gives an upper bound on the leading mode.
The upper bounds on $\Omega$ reported in Tables~\ref{K_bounds} and \ref{Polytrope_bounds} are compatible with available calculations of the leading-mode frequencies in the Newtonian limit; see, e.g., Refs.~\cite{Kokkotas:1995xe,Kokkotas:1999bd,Andersson:2019ahb,Passamonti:2022yqp,Boston:2023fey,Pnigouras:2025muo}. For illustration, the values of $\Omega_{\rm max}/\Lambda_\star$ can be compared with the dimensionless quadrupolar $f$-mode frequency $\omega_{0,\ell=2}/\Lambda_\star$ in Newtonian gravity. In particular, one finds $\omega_{0,\ell=2}/\Lambda_\star\simeq 1.227$ for $n=1$ and $\omega_{0,\ell=2}/\Lambda_\star\simeq 1.456$ for $n=1.5$~\cite{Pitre:2023xsr}. Moreover, we observe a trend in Table~\ref{Polytrope_bounds}: as $n$ increases, the ratio $\Omega_{\rm{max}}/\Lambda_{\star}$ increases as well. This is consistent with a similar behavior observed for the fundamental mode as a function of $n$; see, for instance, Ref.~\cite{Chan:2014kua}.

\begin{table}[t]
    \centering
    \label{tab:alpha_max}
    \begin{tabular}{c c c}
        \toprule
        $n$ & $\Omega_{\max}^{(B_1)}/\Lambda_{\star}$ & $\Omega_{\max}^{(B_2)}/\Lambda_{\star}$\\
        \midrule
        0.3 & 0.995 & 0.990 \\
        0.5 & 1.051 & 1.047 \\
        1.0 & 1.241 & 1.233  \\
        1.2 & 1.343 & 1.324 \\
        1.5 & 1.506 & 1.474  \\
        1.8 & 2.199 & 1.729  \\
        \bottomrule
    \end{tabular}
    \caption{Maximum values $\Omega_{\text{max}}$, as determined by the bounds $B_1$ (second column) and $B_2$ (third column) for different values of the polytropic index. In all cases, we fixed  $C=0.05$, $\mu=\Lambda_\star$, and $K=10^2G^3M_\odot^2$.}
\label{Polytrope_bounds}
\end{table}

Finally, in Figure~\ref{n1_amax}, left panel, we plot curves obtained by increasing $\Omega$ from zero up to the maximum value allowed by $B_2$, with $C$ and $\mu$ kept fixed, for different polytropic indices. The curves lie within the shaded region allowed by the arc bounds~\eqref{CE_bounds2} (see also the right panel of Figure~\ref{arcbound}), and exit the allowed region at $c_{\dot E}(\Omega^{(B_2)}_{\rm max})^2/c_{E}\Lambda_\star^2=c_{\dot E}^2/c_Ec_{\ddot E}\leq 1$. Therefore, the exit point is a measure of how well the bound $B_3$ is satisfied. Figure~\ref{n1_amax} shows that, as $n$ increases, the curve lies further away from the $B_3$ boundary, implying that the single-mode approximation gets worse (see Section~\ref{subsec:modes}). This behavior is consistent with previous analyses and estimates of the validity of the $f$-mode approximation to the tidal response~\cite{Pitre:2023xsr,Kokkotas:1995xe}. Conversely, in the limit $n\to 0$, the lower boundary of the positivity region is approached. In this case, the star interior becomes an infinitely stiff perfect fluid with constant density. As explained in Ref.~\cite{Ogilvie_2016}, the density perturbation induced by the tidal field is entirely localized on the surface of the star, and the response of each multipolar sector is described by a single oscillator (Kelvin mode) with normal frequency 
\begin{equation}
    \omega_\ell^2=\frac{2\ell(\ell-1)}{2\ell+1}\frac{GM}{R^3_\star}\,.
\end{equation} 

\subsection{More realistic equations of state}

In this section, we move beyond polytropic models and consider tabulated equations of state based on nuclear physics. These models span a broader range of neutron-star tidal deformabilities and support maximum masses exceeding $2M_\odot$, consistent with constraints from pulsar-timing observations (see, e.g., Refs.~\cite{Antoniadis:2013pzd,NANOGrav:2019jur}). 
In particular, we consider three representative equations of state: AP4~\cite{Akmal:1998cf}, 
MS1~\cite{Mueller:1996pm}, and
SLy4~\cite{Douchin:2001sv}.  The corresponding mass-radius diagram can be found, for instance, in Figure~1 of Ref.~\cite{Apostolidis:2026qsg}.

In the literature, many other equations of state have been developed that incorporate additional physical ingredients or improve upon various aspects of the modeling, including nuclear interactions, crust-core consistency, observational constraints, and additional degrees of freedom. Our goal here is not to provide an exhaustive survey of these models, but rather to use AP4, MS1, and SLy4, which have been widely employed in the tidal-deformability literature, as representative benchmarks spanning a range of stiffnesses. We refer the reader to the CompOSE reference manual~\cite{CompOSECoreTeam:2022ddl} and the associated online repository for an extensive collection of tabulated equations of state available in the literature. 

\begin{figure*}[p]
        \subfloat{%
    {\includegraphics[width=.49\linewidth]{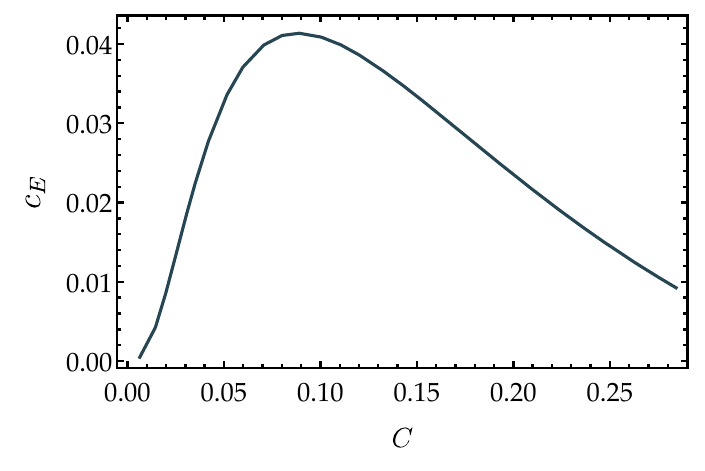}}
            \label{}%
        }
        \subfloat{%
        \includegraphics[width=.49\linewidth]{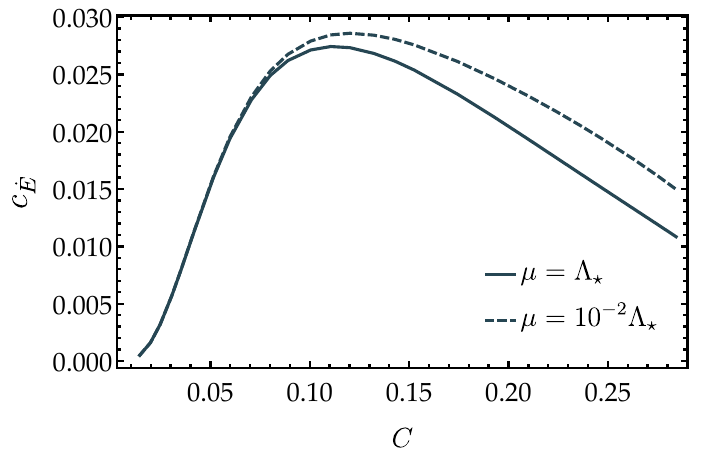}%
            \label{}%
        }\\
        \subfloat{%
    {\includegraphics[width=.49\linewidth]{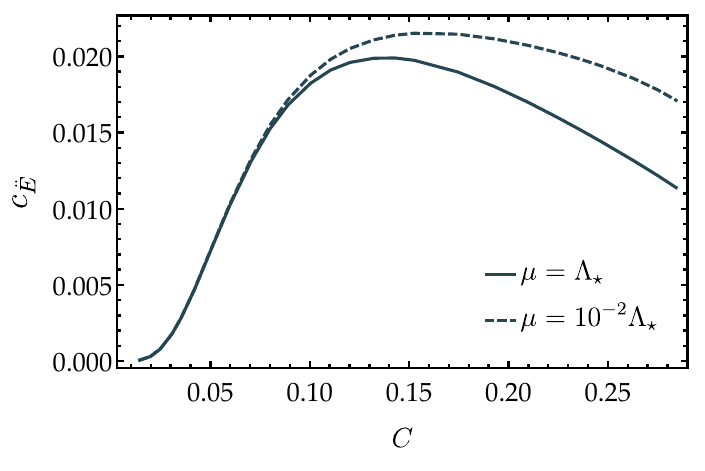}}
            \label{}%
        }
        \subfloat{%
        \includegraphics[width=.49\linewidth]{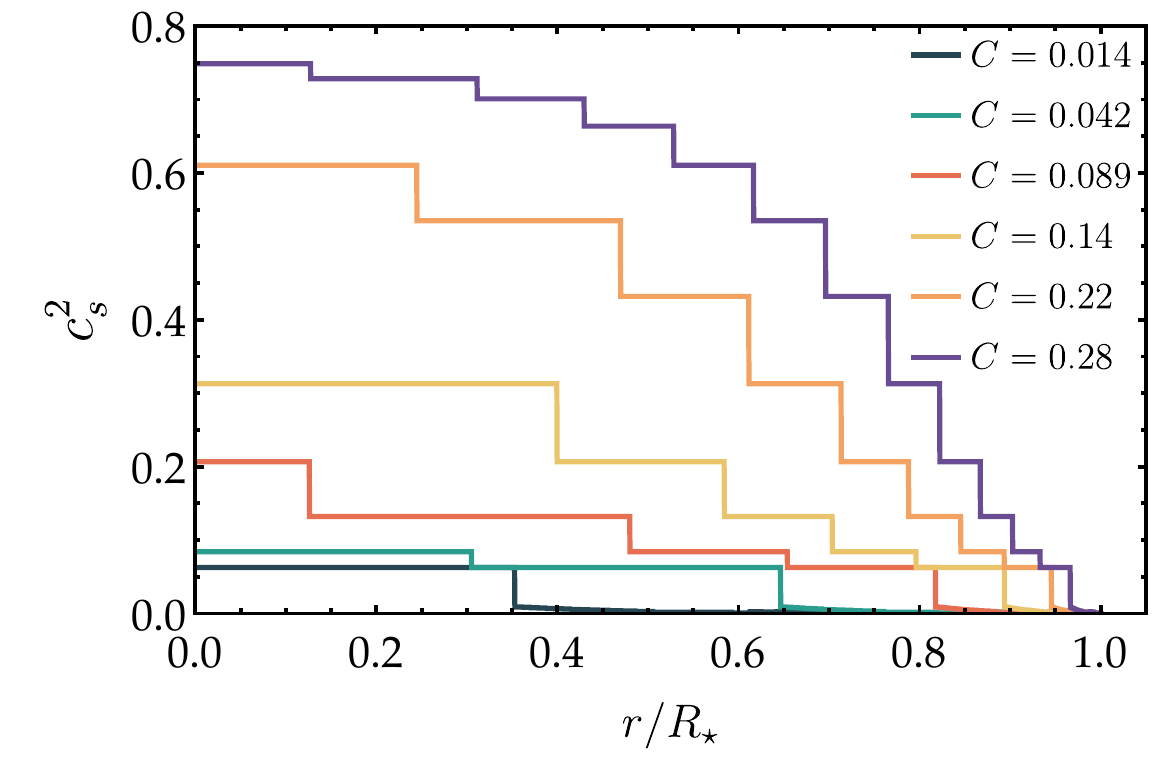}%
            \label{}%
        }\\
        \subfloat{%
        \includegraphics[width=.49\linewidth]{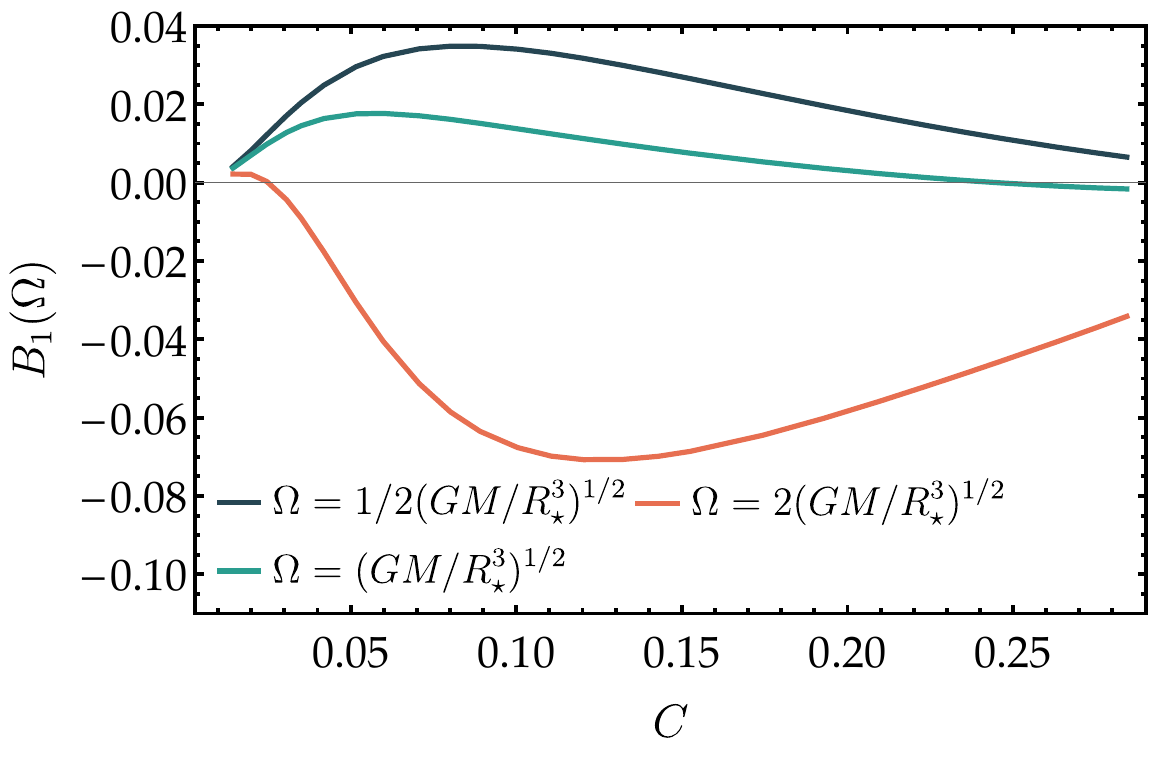}%
            \label{}%
        }\hfill
        \subfloat{%
        \includegraphics[width=.49\linewidth]{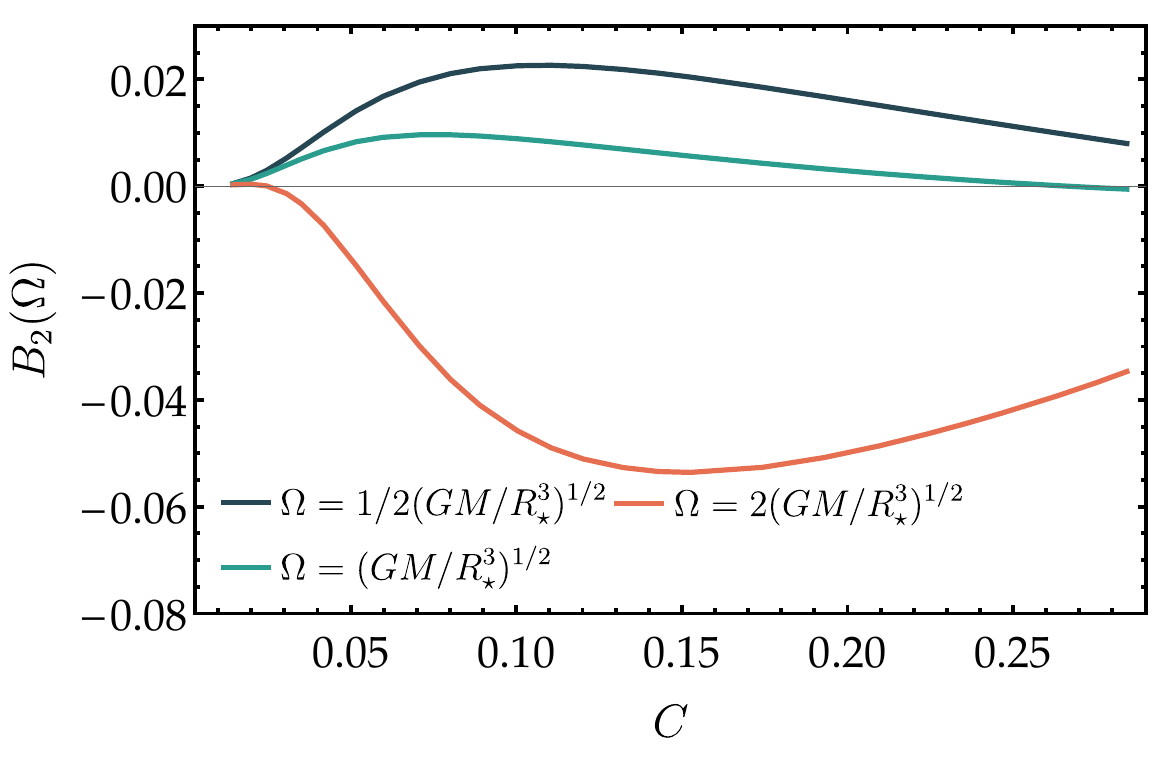}
            \label{}%
        }
         \caption{Static and dynamical Love numbers, sound speed, and arc bounds $B_1$ and $B_2$ for the MS1 equation of state.}
   \label{ms1_plots}
 \end{figure*}

\begin{figure*}[p]
   \subfloat{%
    {\includegraphics[width=.49\linewidth]{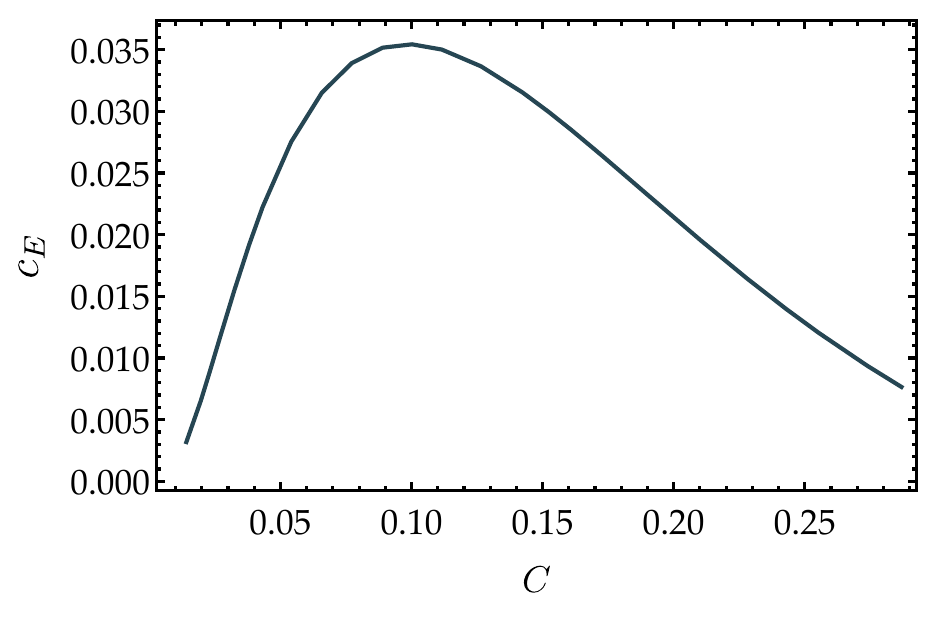}}
            \label{}%
        }
        \subfloat{%
        \includegraphics[width=.49\linewidth]{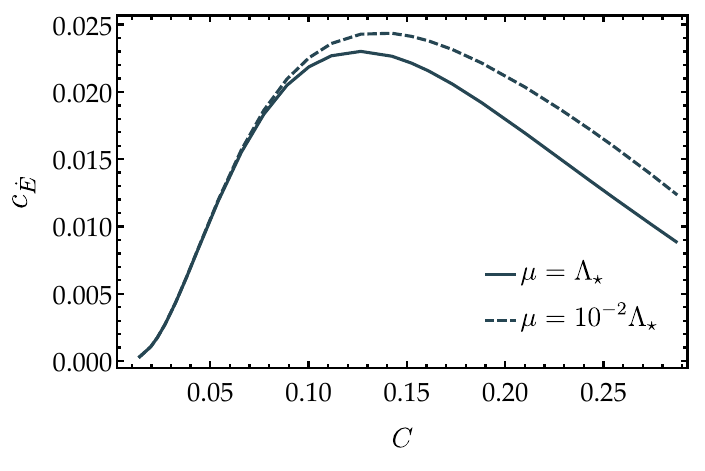}%
            \label{}%
        }\\
        \subfloat{%
    {\includegraphics[width=.49\linewidth]{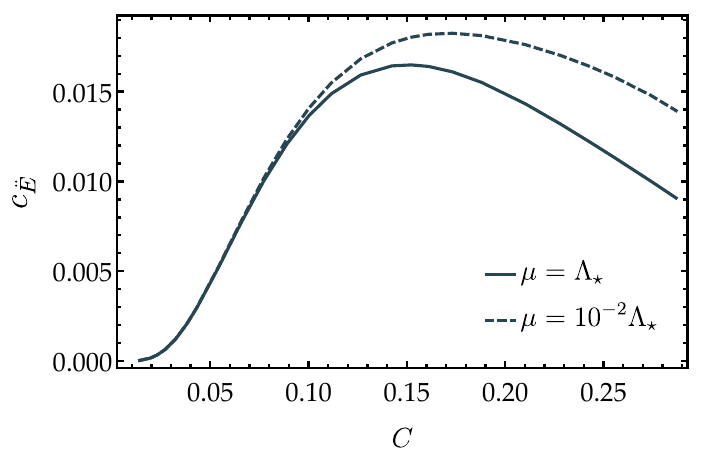}}
            \label{}%
        }
        \subfloat{%
        \includegraphics[width=.49\linewidth]{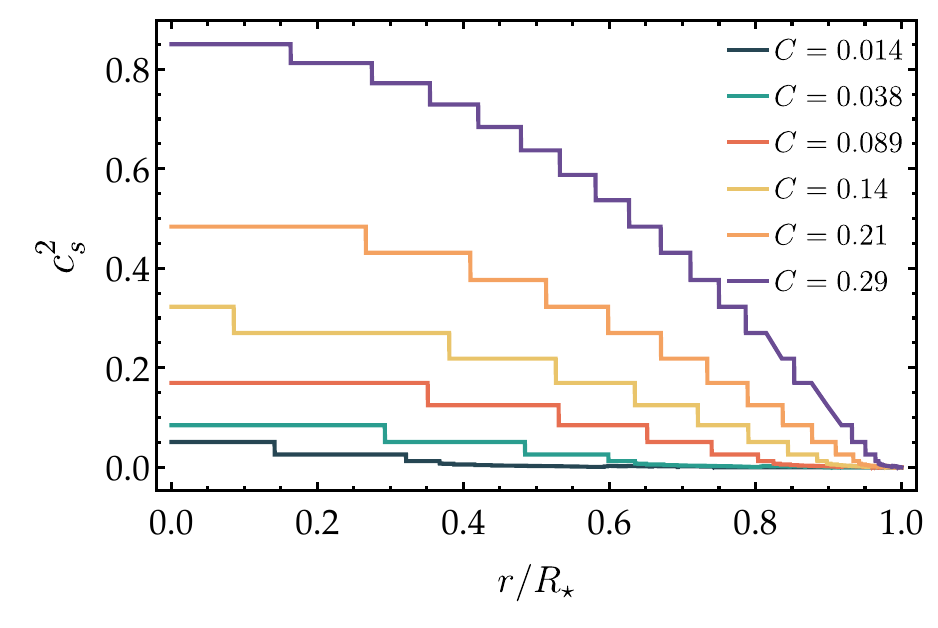}%
            \label{}%
        }\\
        \subfloat{%
        \includegraphics[width=.49\linewidth]{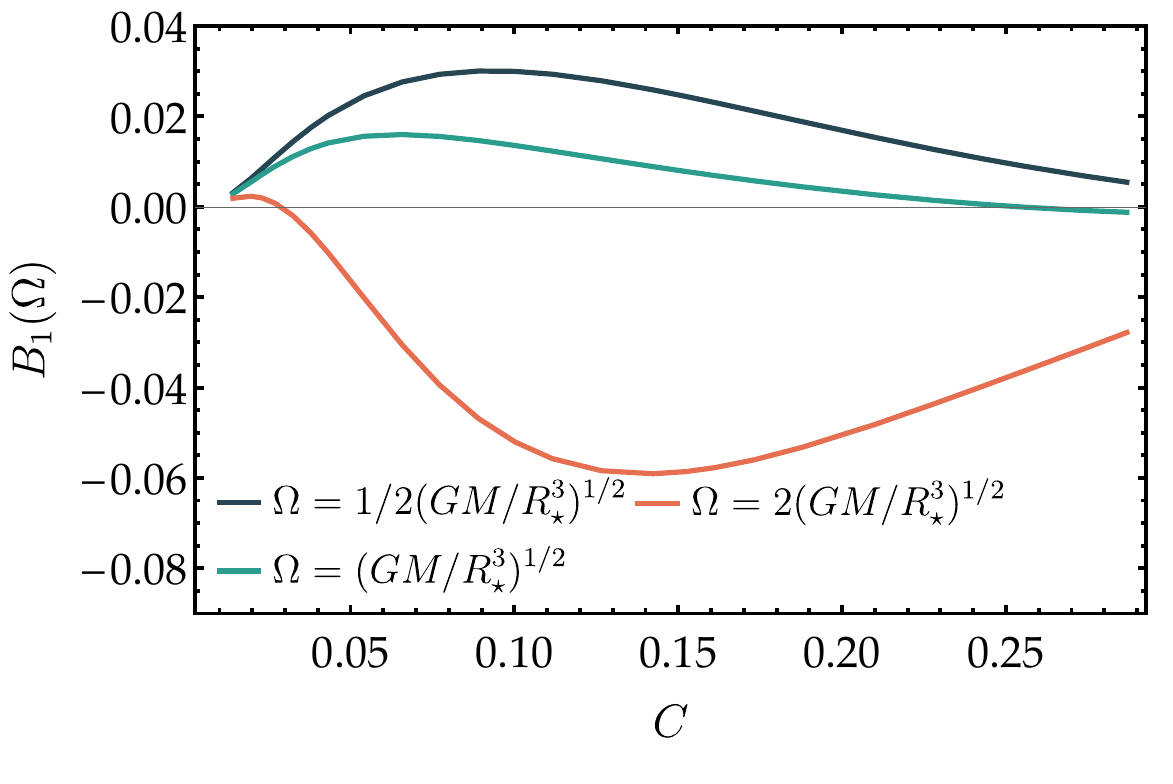}%
            \label{}%
        }\hfill
        \subfloat{%
        \includegraphics[width=.49\linewidth]{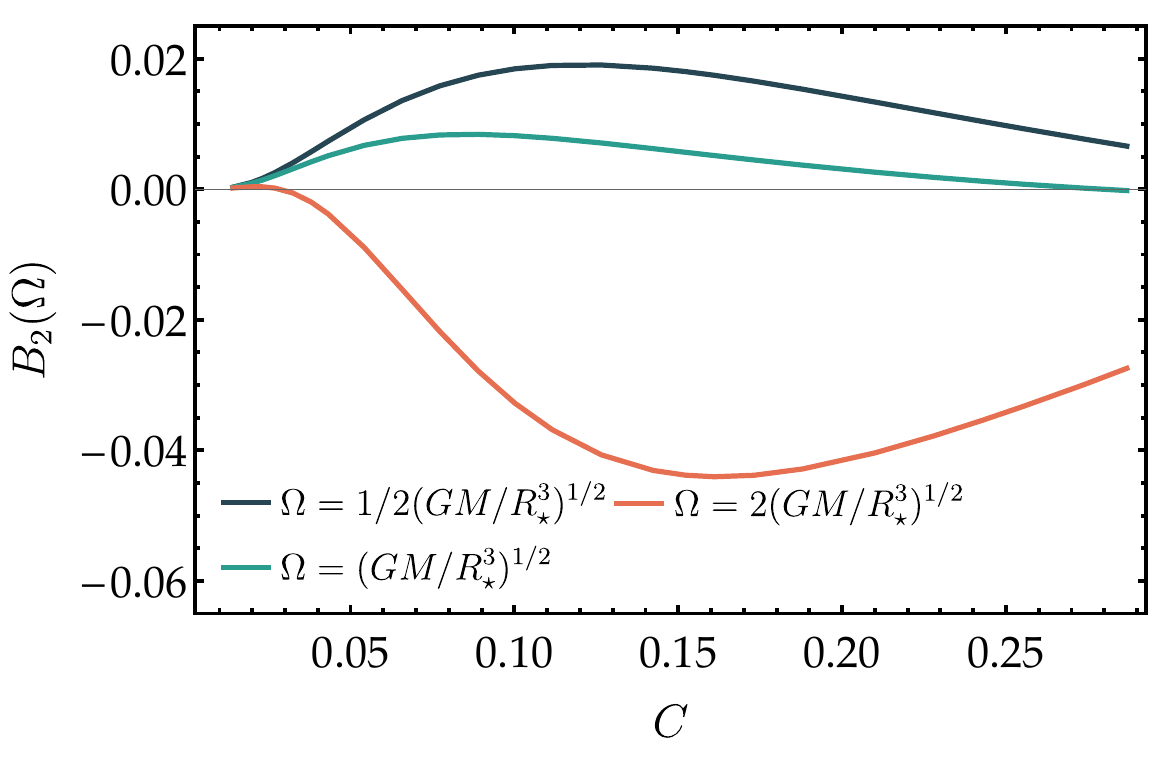}
            \label{}%
        }
         \caption{Static and dynamical Love numbers, sound speed, and arc bounds $B_1$ and $B_2$ for the SLy4 equation of state.}
   \label{sly_plots}
\end{figure*}

\begin{figure*}[p]
        \subfloat{%
    {\includegraphics[width=.49\linewidth]{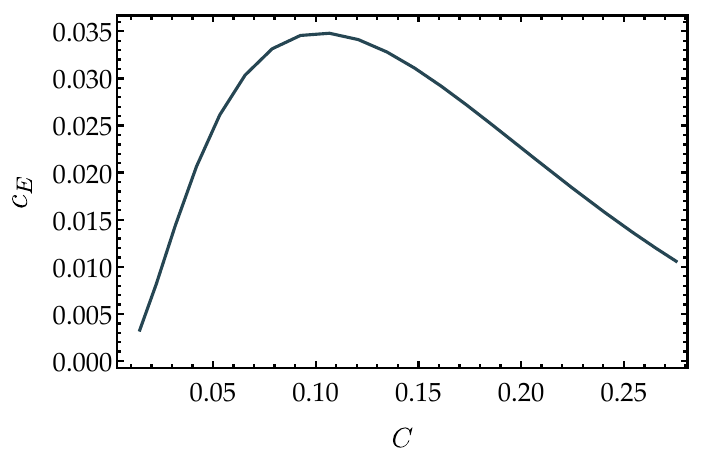}}
            \label{}%
        }
        \subfloat{%
        \includegraphics[width=.49\linewidth]{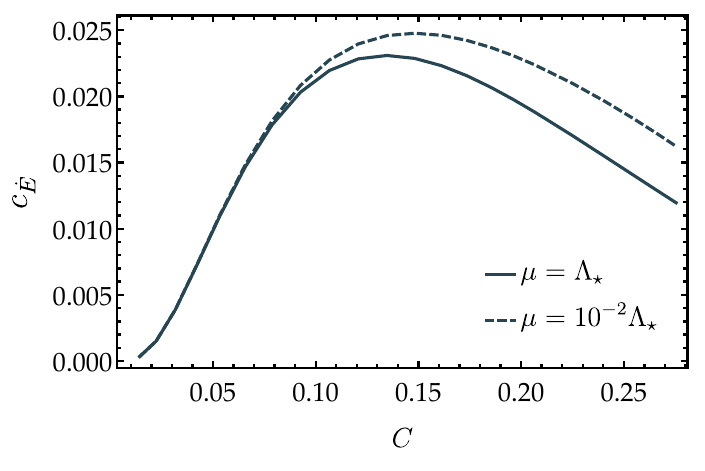}%
            \label{}%
        }\\
        \subfloat{%
    {\includegraphics[width=.49\linewidth]{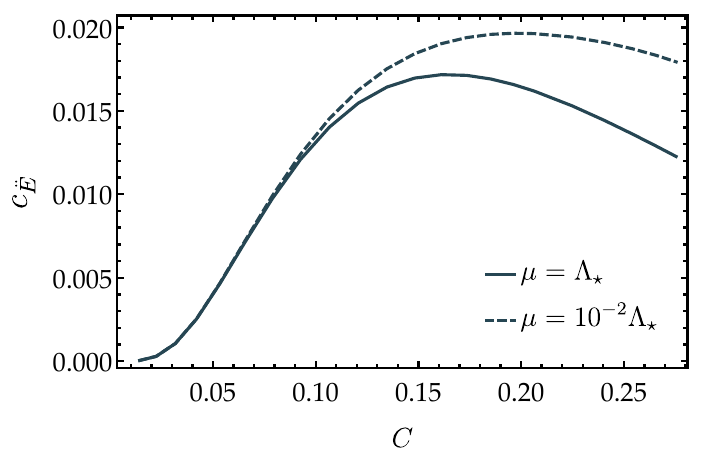}}
            \label{}%
        }
        \subfloat{%
        \includegraphics[width=.49\linewidth]{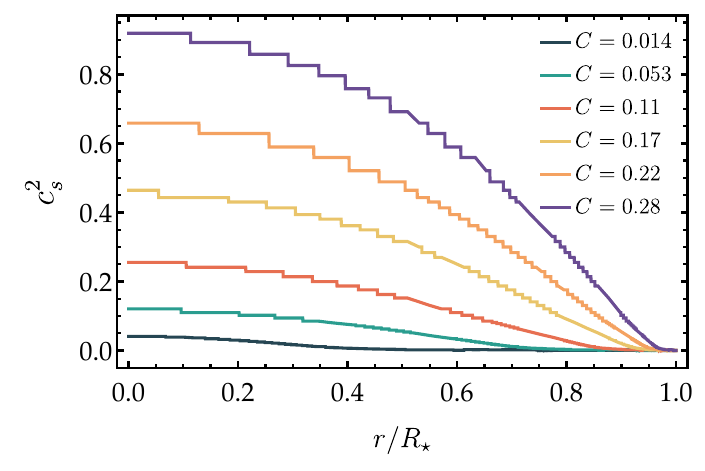}%
            \label{}%
        }\\
        \subfloat{%
        \includegraphics[width=.49\linewidth]{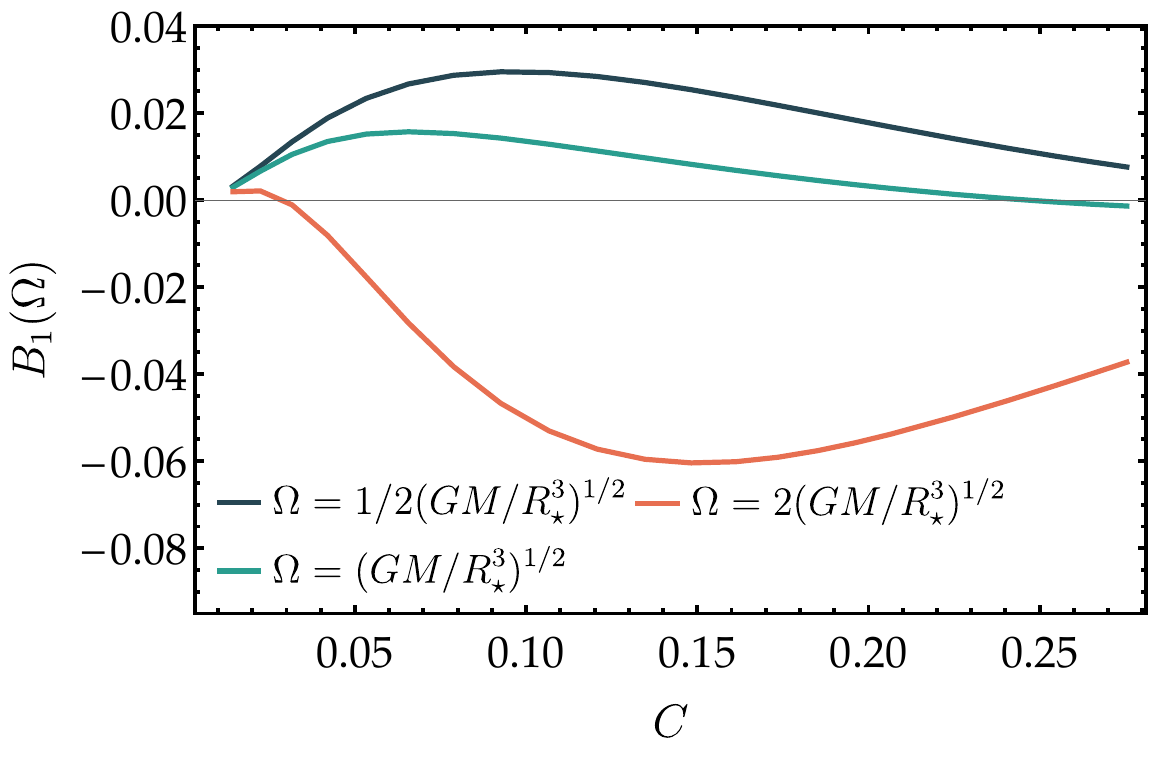}%
            \label{}%
        }\hfill
        \subfloat{%
        \includegraphics[width=.49\linewidth]{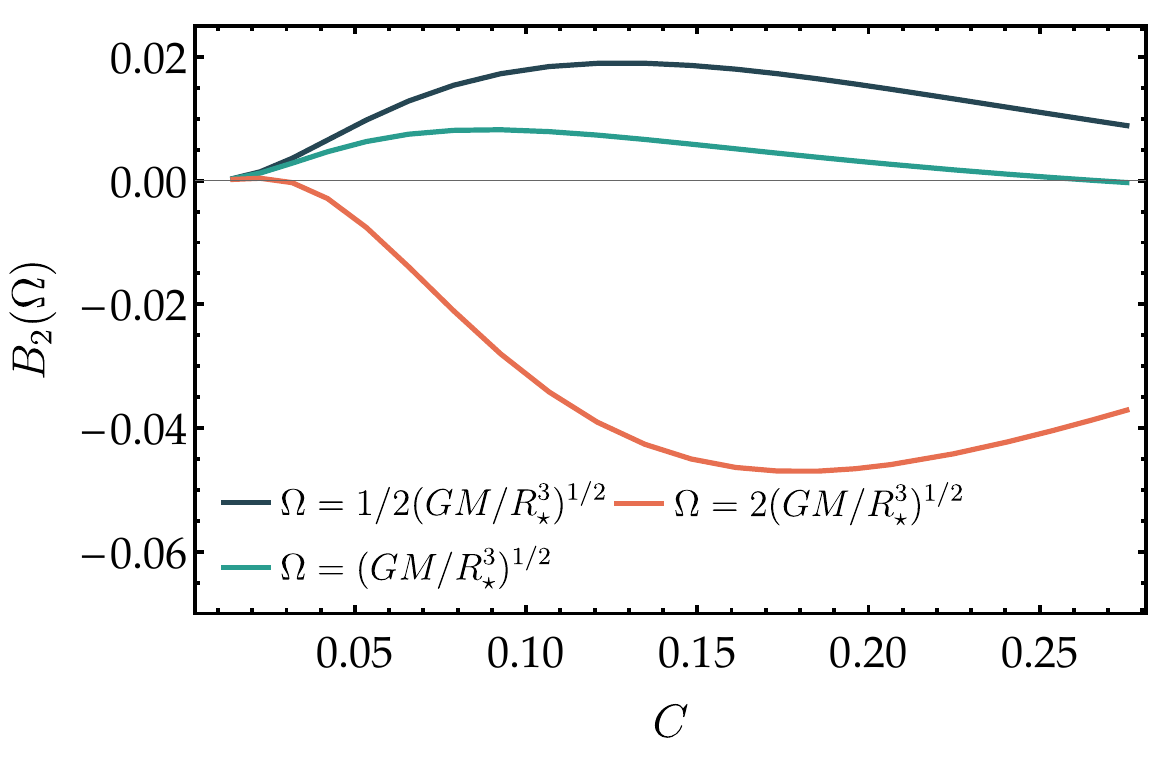}
            \label{}%
        }
         \caption{Static and dynamical Love numbers, sound speed, and arc bounds $B_1$ and $B_2$ for the AP4 equation of state.}
   \label{ap4_plots}
\end{figure*}

\begin{figure}[t]
\centering
    {\includegraphics[width=.55\linewidth]{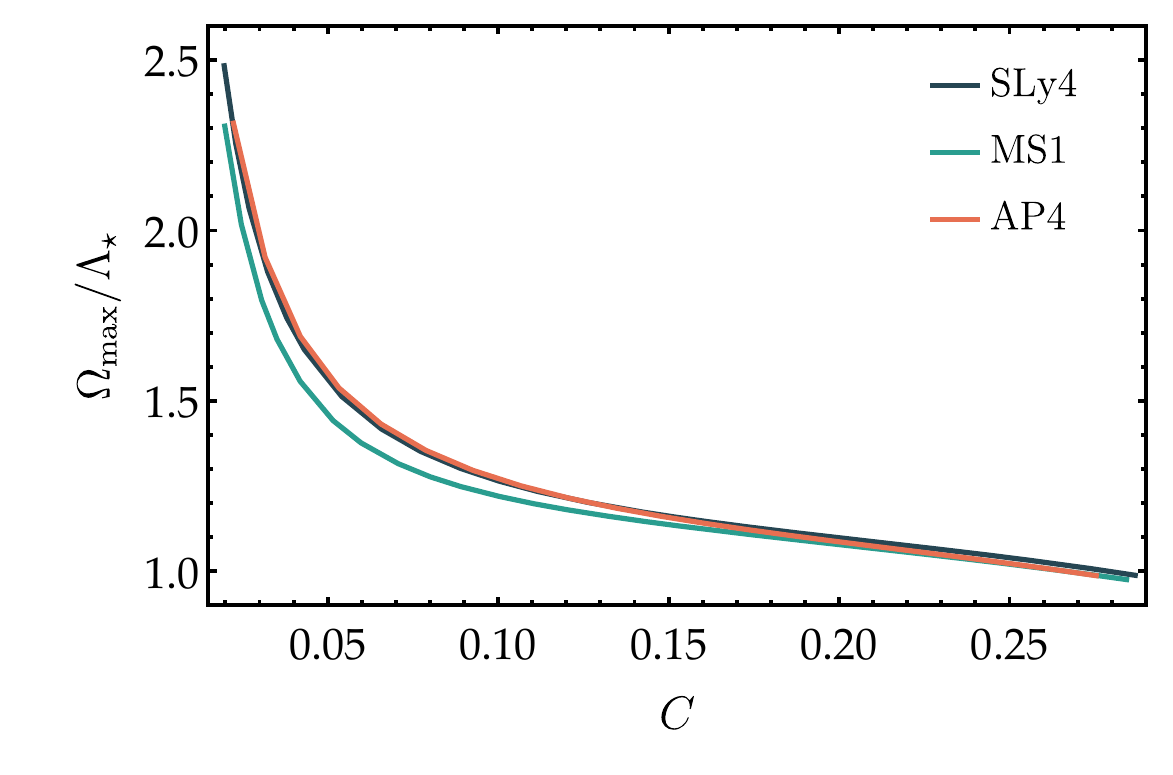}}
         \caption{Comparison of the maximum value of $\Omega$ for different realistic equations of state. In all cases, we fixed $\mu=\Lambda_\star$.}
   \label{Omega_max_comp}
\end{figure}

The results are summarized in three sets of plots in Figures~\ref{ms1_plots}, \ref{sly_plots} and \ref{ap4_plots}, corresponding to the MS1, SLy4, and AP4 tabulated equations of state, respectively. The overall picture is qualitatively similar to that obtained for the polytropic model in Section~\ref{sec:polytropic}. First, we restrict our analysis to the range of compactness for which the sound speed remains subluminal, as commonly required in the literature.\footnote{In the sound-speed plots, the curves appear jagged due to the way the derivatives in Eq.~\eqref{eq:soundspeed} are computed from the tabulated pressure and energy density.} In the plots for the dynamical Love numbers, we take two representative values of $\mu$ for illustration purposes, while we fix $\mu=\Lambda_\star$ in the plots for $B_1$ and $B_2$ (we will discuss the bound $B_3$ further below). 
As in the polytropic case, the plots of $B_1$ and $B_2$ show that the corresponding bounds are not satisfied for arbitrary values of the arc parameter $\Omega$. Requiring both bounds to hold therefore imposes an upper limit on $\Omega$. Figure~\ref{Omega_max_comp} shows the maximum allowed value of $\Omega$ as a function of compactness for the three tabulated equations of state considered. We emphasize that this constraint is independent of the subluminality requirement and translates into a condition on the first mode of the star.
\clearpage
\clearpage

\subsection{On the bound $B_3$ and the $\mu$ dependence in strong gravity}

\begin{figure*}[t!]
        \subfloat{%
    {\includegraphics[width=.49\linewidth]{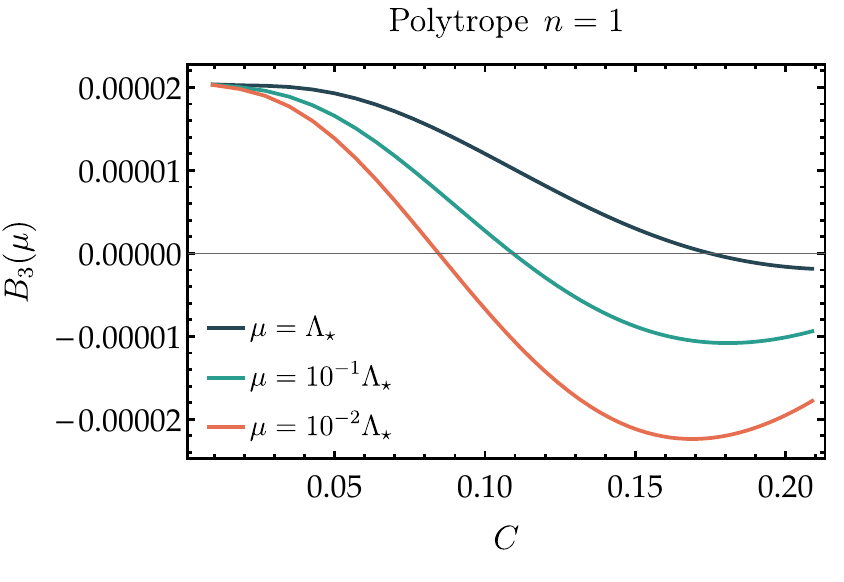}}
            \label{}%
        }
        \subfloat{%
        \includegraphics[width=.49\linewidth]{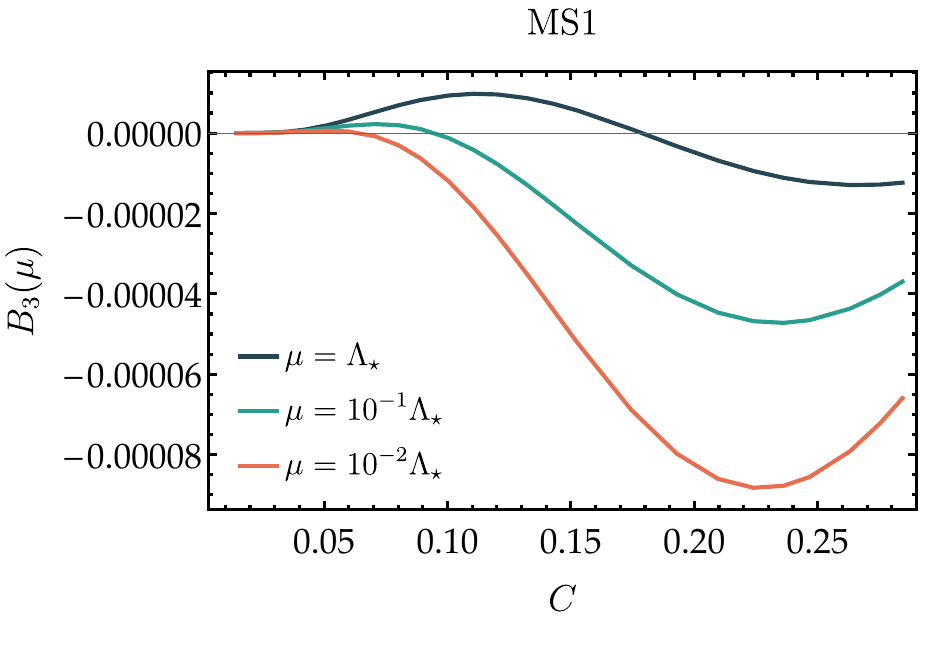}%
            \label{}%
        }\\
        \subfloat{%
    {\includegraphics[width=.49\linewidth]{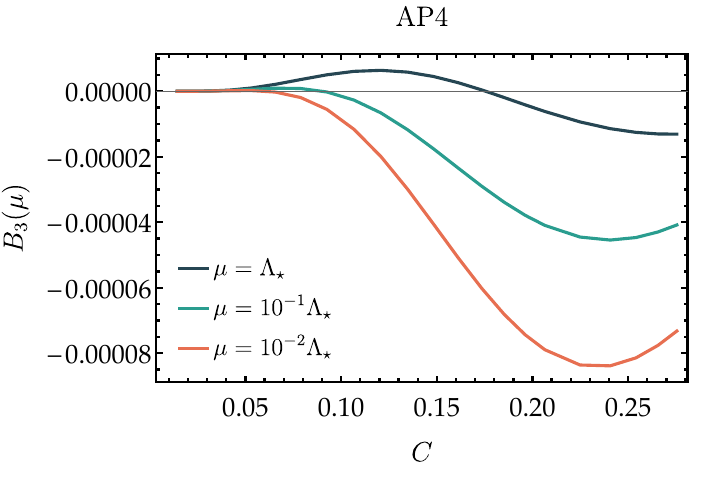}}
            \label{}%
        }
        \subfloat{%
        \includegraphics[width=.49\linewidth]{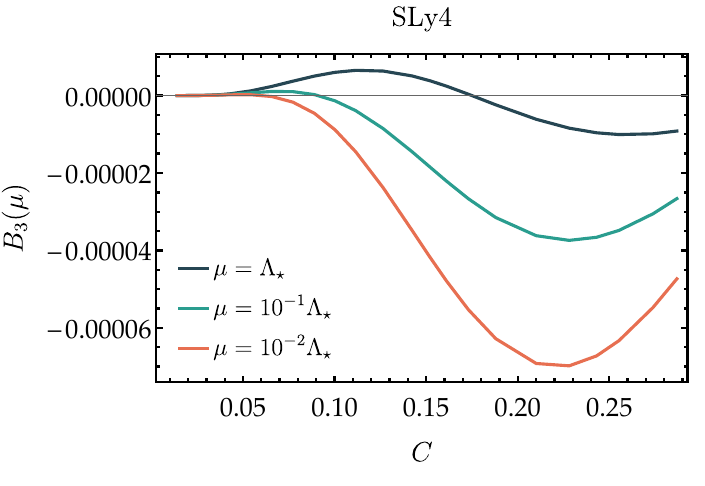}%
            \label{}%
        }\\
         \caption{$B_3$ bound for different equations of state.}
   \label{B3_plots}
\end{figure*}

The third bound, $B_3$ in Eq.~\eqref{CE_bounds2}, deserves separate attention. In Figure~\ref{B3_plots}, we collect the corresponding results for the $n=1$ polytropic model and the MS1, SLy4, and AP4 equations of state as functions of the compactness. As already noted above, unlike $B_1$ and $B_2$, the bound for $B_3$ is independent of $\Omega$ and, for a fixed equation of state, depends only on the compactness and the renormalization scale $\mu$.

In addition, as discussed in Section~\ref{sec:arc_bounds}, $B_3$ is saturated in simple single-mode Green's functions. Therefore, as long as the single-mode approximation provides a good description of the system, at least in the Newtonian regime, one expects a cancellation leading to a small or vanishing combination $c_E c_{\ddot E}-c_{\dot E}^2$ in the low-$C$ limit. This is indeed what we observe in Figure~\ref{B3_plots}. The value of $B_3$ is parametrically suppressed relative to the typical size of the individual Love number coefficients and exhibits a strong dependence on $\mu$. As is apparent from the plots, this suggests that $B_3$ is particularly sensitive to strong-gravity effects. Consequently, better control over the regime of validity of the bounds~\eqref{CE_bounds2} is required before robust conclusions about the equation of state can be drawn from $B_3$. 

From Eq.~\eqref{eq:ccoeffsmain}, we see that while $c_E$ does not run with the renormalization scale $\mu$, $c_{\dot E}\sim-\log(\mu r_s)$ decreases with increasing $\mu$, while $c_{\ddot E}\sim\log(\mu r_s)^2$ grows for both very large and very small $\mu$. Straightforward algebra using the explicit expression for the tidal Love numbers in Eq.~\eqref{eq:ccoeffsmain} gives the renormalization group equation:
\eq{
\mu\partial_\mu\begin{bmatrix}
c_E\\c_{\dot E}\\c_{\ddot E}
\end{bmatrix}=
-\begin{bmatrix}
0&0&0\\4\kappa C^3&0&0\\16\kappa\gamma_{1}C^6&4\kappa C^3&0
\end{bmatrix}
\begin{bmatrix}
c_E\\c_{\dot E}\\c_{\ddot E}
\end{bmatrix}-
\begin{bmatrix}
0\\\frac{32}{45}C^8\\\frac{128}{45}\kappa \gamma_{0}C^{11}
\end{bmatrix}\,,
\label{RG}
}
where the constants are $\kappa=\frac{107}{105}$, $\gamma_0=\frac{150223}{67410}$, and $\gamma_1=\frac{1695233}{4718700}$. Note that the beta function contains both a universal contribution and a contribution that is a linear combination of the Love numbers. Both scale with high powers of $C$. This is expected, since the running is a proxy for general relativistic corrections, and is consistent with the extremely mild $\mu$-dependence of the Love numbers in the $C\to0$ limit.

Moreover, one can show that the following combination of Love numbers is renormalization-group invariant:
\eq{
c_Ec_{\ddot E}-Ac_{\dot E}^2-Bc_{E}c_{\dot E}={\rm const.}\,,
\label{muind}
}
where we have defined
\eq{
A=\frac{1}{2}\frac{c_E}{c_E+\frac{8}{45\kappa}C^5}\,,\qquad B=4\gamma_1C^3\frac{c_E+\frac{8\gamma_0}{45\gamma_1}C^5}{c_E+\frac{8}{45\kappa}C^5}\,.
}
It is then straightforward to verify that the curves defined by Eq.~\eqref{muind}, which are parabolas with curvature $A<1$, always extend beyond the region allowed by the arc bounds, shown in Figure~\ref{arcbound} or \ref{n1_amax}, both for small and for large $\mu$. Indeed, a general feature emerging from Figure~\ref{B3_plots} is that decreasing $\mu$ weakens the bound and lowers the value of $C$ at which $B_3$ becomes negative. In general, decreasing $\mu$ is not necessarily associated with a breakdown of the EFT description. The main lesson we draw from Figure~\ref{B3_plots} is therefore that, while $B_3$ approaches a non-negative value in the limit $C\to 0$, the bound in its present form may not provide the appropriate constraint once general-relativistic effects become dominant.

Violation of the bounds under a change of the arbitrary renormalization scale $\mu$ indicates that at least one of our assumptions about the linear response function (existence/analyticity, positivity, or decay), as discussed in Section~\ref{sec:arc_bounds}, must be violated. A more comprehensive study of this issue is left for future work \cite{unpub}. However, a quick check suggests that modifying the assumed asymptotic decay of the linear response is unlikely to resolve the issue.

Let us relax the decay assumption and allow $G_R(\omega)$ to grow asymptotically, while requiring $G_R(\omega)/\omega^2\to0$ as $\omega\to\infty$. Then, in the definition of the arcs in Eq.~\eqref{ArcEven1}, the integrand is modified as $G_R(z)/z^{2n+1}\to G_R(z)/z^{2n+3}$. This amounts to a shift in the order of the arc variables, $a_{2n}\to a_{2(n+1)}$. For instance, bounds involving the first three moments become
\eq{
a_2(\Omega)\geq\Omega^2a_4(\Omega)\geq\Omega^4a_6(\Omega)\geq0\,,\qquad a_2(\Omega)a_6(\Omega)-a_4(\Omega)^2\geq0\,.
}
In particular, we must have $a_2(\Omega)\propto c_{\dot E}\geq0$. On the other hand, Eq.~\eqref{RG} implies that $c_{\dot E}$ becomes negative for sufficiently large $\mu$. The same conclusion can be reached with four subtractions, i.e.~$G_R(\omega)/\omega^4$, since $a_4(\Omega)\propto c_{\ddot E}$ can become negative for intermediate values of $\mu$.

\newpage
\section{Including dissipation}
\label{diss}

In Section \ref{sec:arc_bounds} we have seen that the requirements of positivity and analyticity (supplied by the assumption of decay at infinity) of the tidal response function can be formulated in terms of a moment problem using the arcs. In particular, in the absence of dissipation, the relation between the arcs and the EFT coefficients is simple, and therefore, the solution to the moment problem translates easily to bounds on these coefficients. On the other hand, when dissipation is taken into account, the arcs involve infinitely many terms, because of the non-vanishing dissipative coefficients (see Eq.~\eqref{eq:arcsdiss}-\eqref{eq:arcsdiss2}). This makes translating the arc bounds into bounds on the tidal response coefficients far more nontrivial. The current section aims at clarifying how the presence of dissipation can obstruct obtaining a bound.

Suppose we are interested in bounding the first $2N+1$ tidal coefficients $\{\lambda_0,\tilde{\lambda}_1,\lambda_2,\dots,\tilde{\lambda}_{2N-1},\lambda_{2N}\}$, including the dissipative coefficients.\footnote{We caution the reader that our notation for $N$ is not entirely consistent with that used in Section~\ref{sec:arc_bounds}. In this section, $N$ denotes the highest order of the EFT coefficient that is considered, namely $\lambda_{2N}$. In Section~\ref{sec:arc_bounds}, by contrast, it specifies the highest considered moment, $m_{2N}$, which corresponds to $\lambda_{4N}$ plus an infinite tower of dissipative coefficients. We hope that this difference in notation will not cause any confusion.} In particular, we pick $N\geq 1$ (since, for $N=0$ the bound collapses to $\lambda_0\geq0$). Let us then decompose the low-energy expansion of the imaginary part of the Green's function (in the $|\omega|<\Lambda$ regime) as follows  
\eq{
{\rm Im}[G_R](\omega)=\sum_{k=0}^{N-1}\tilde{\lambda}_{2k+1}\omega^{2k+1}+\sum_{k=N}^{+\infty}\tilde{\lambda}_{2k+1}\omega^{2k+1}\equiv \sigma^{<}_N(\omega)+\sigma_N^{>}(\omega)\,.
\label{headtail}
} 
The first summation, denoted by $\sigma^{<}_N(\omega)$, involves the dissipative coefficients that belong to the set of coefficients we aim to bound, while the second sum $\sigma_N^{>}(\omega)$ represents the problematic infinite number of terms that we need to somehow get rid of. Since we want to bound the coefficients up to $\lambda_{2N}$, we focus exclusively on the arcs $a_{2n}(\Omega)$, for $n=0,1,\dots,N$. It is straightforward to write the infinite sum for each arc in this set, in terms of $\sigma_N^{>}(\omega)$. From Eq.~\eqref{AL2}, 
\begin{equation}
\label{AL2app}
\spl{
a_{2 n}(\Omega)&=\lambda_{2 n}-\frac{2}{\pi} \sum_{k=0}^{+\infty} \frac{\tilde{\lambda}_{2 k+1} \Omega^{2(k-n)+1}}{2(k-n)+1}\,\\
&=\lambda_{2 n}-\frac{2}{\pi} \sum_{k=0}^{N-1} \frac{\tilde{\lambda}_{2 k+1} \Omega^{2(k-n)+1}}{2(k-n)+1}-\frac{2}{\pi}\int_0^\Omega\dd{z}\,\frac{\sigma_N^{>}(z)}{z^{2n+1}}=\frac{2}{\pi}\int_\Omega^{+\infty}\dd{z}\,\frac{{\rm Im}[G_R](z)}{z^{2n+1}}\,.
}
\end{equation}
In the second line, we have decomposed the infinite sum, as in Eq.~\eqref{headtail}, into the first $N$ dissipative terms, and the infinite tail. The infinite summation is then turned into an integral over $\sigma_N^{>}(z)$, as one can directly check.\footnote{Here we have used that the summation in Eq.~\eqref{headtail} is uniformly convergent for $|z|<\Lambda$, so that we can exchange the sum and the integral.} The last equality is to remind the expression of the arcs in terms of an  integral of ${\rm Im}[G_R](z)$ over frequencies above $\Omega$. 

Although the two integrals in the last line of Eq.~\eqref{AL2app} share some similarities, their contributions are genuinely different. Indeed, there is no guaranteed positivity for $\sigma_N^>(z)$, but only the full imaginary part ${\rm Im}[G_R](z)=\sigma^{<}_N(z)+\sigma_N^{>}(z)$ is sign definite by passivity. It is useful to rewrite the integral over $\sigma_N^{>}(z)$ by  changing variable $z=\Omega\sqrt{x}$
\eq{
\frac{2}{\pi}\int_0^\Omega\dd{z}\,\frac{\sigma_N^>(z)}{z^{2n+1}}=\frac{1}{\Omega^{2n}}\int_0^1\dd{x}\,x^{N-n}\rho(x)\,,
}
which is of the form of $(N-n)$-th order moment of a (signed) measure
\eq{
\frac{1}{\pi}\frac{\dd{x}}{x^{N+1}}\sigma_N^>(\Omega\sqrt{x})\equiv\dd{x}\rho(x)\,.
\label{rhomeasure}
}
Notice that in the limit $x\to0$, the tail goes to zero as $\sigma_N^>(\Omega\sqrt{x})\sim x^{N+1/2}$, therefore $\rho(x)\sim x^{-1/2}$, which is integrable around $x=0$. We rewrite the arcs as
\eq{
\label{newarc}
\Omega^{2n}a_{2 n}(\Omega)=\lambda_{2 n}\Omega^{2n}-\frac{2}{\pi} \sum_{k=0}^{N-1} \frac{\tilde{\lambda}_{2 k+1} \Omega^{2k+1}}{2(k-n)+1}-\int_0^1\dd{x}\rho(x)\,x^{N-n}=\int_{0}^{1} \dd{\mu(x)}\, x^{n}\,,
}
where the last equality is simply the same right-hand side we had in Eq.~\eqref{A3}. As mentioned above, while $\rho(x)$ is not necessarily positive, ${\rm Im}[G_R](\omega)=\sigma^{<}_N(\omega)+\sigma_N^{>}(\omega)$ is positive for $|\omega|\leq\Omega$. Therefore, we define the following new measure over $x\in[0,1]$:
\eq{
\dd{\tilde{\mu}(x)}\equiv\frac{\dd{x}}{\pi x}\bigg(\sigma^{<}_N\big(\Omega\sqrt{x}\big)+\sigma^{>}_N\big(\Omega\sqrt{x}\big)\bigg)=\left[\frac{1}{\pi} \sum_{k=0}^{N-1}\tilde{\lambda}_{2 k+1}\Omega^{2k+1}\,x^{k-1/2}+x^{N}\rho(x)\right]\dd{x}\,.
\label{posIR}
}
The function $\rho(x)$ and the dissipative coefficients $\{\tilde{\lambda}_1,\dots\tilde{\lambda}_{2N-1}\}$ are constrained such that $\tilde{\mu}$, or equivalently its moments, remains positive.

Let us consider the simplest case with $N=1$, corresponding to the set $\{\lambda_0,\tilde{\lambda}_1,\lambda_2\}$. To simplify the notation we define $g_{2k}\equiv\lambda_{2k}\Omega^{2k}$ and $g_{2k+1}\equiv\frac{2}{\pi}\tilde{\lambda}_{2k+1}\Omega^{2k+1}$. Eq.~\eqref{newarc} for $n=0,1$ gives
\begin{align}
g_0-g_1-\int_0^1\dd{x}\rho(x)\,x& =\int_0^1\dd{\mu(x)}\, ,\label{c0N1}\\
g_2+g_1-\int_0^1\dd{x}\rho(x)& =\int_0^1\dd{\mu(x)}x\, . \label{c2N1}
\end{align}
Positivity of Eq.~\eqref{posIR} for $N=1$, gives  $g_1/2+x^{3/2}\rho(x)\geq0$. Let us pick $\rho(x)$ to be a narrow bump, as shown in Figure~\ref{bump}, with width $\sim\delta$ and height $\sim\overline{\rho}/\delta^{3/2}$. As long as $\overline{\rho}\geq-g_1/2$ the positivity condition is satisfied, no matter what $\delta$ is. In particular, since $g_1\geq0$ by the positivity assumption, $\overline{\rho}$ is allowed to take either positive or negative values. We can evaluate the $\rho$-moments in Eqs.~\eqref{c0N1} and \eqref{c2N1} for this choice:
\eq{
\int_0^1\dd{x}\rho(x)\,x\simeq\frac{1}{2}\overline{\rho}\sqrt{\delta}\,,\qquad \int_0^1\dd{x}\rho(x)\simeq\frac{1}{\sqrt{\delta}}\overline{\rho}\,,
}
where we have approximated the functional form of $\rho(x)$ with a top-hat, since we are only interested in the scaling with $\delta$. We see that in the limit $\delta\to0$, the first moment can be made arbitrarily small, while the zeroth moment can be made arbitrarily large. In this limit, it is consistent to keep $g_0$ and $g_1$ fixed in Eq.~\eqref{c0N1}, while arbitrarily increase or decrease $g_2$, in Eq.~\eqref{c2N1}, depending on the sign of $\overline{\rho}$. We conclude that there is neither an upper nor a lower bound on $g_2$.
\fg{
\centering
\includegraphics[width=0.52\textwidth]{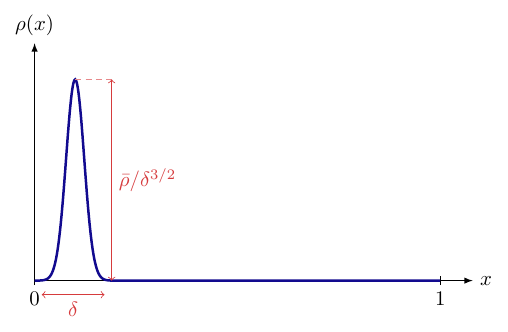}
\caption{The schematic form for the density $\rho(x)$, which parametrizes the infinite sum of ${\rm Im}[G_R](\omega)$ at low energies, used to show that there cannot be an upper bound, nor a lower bound on $g_2$, when no extra constraints are imposed on dissipation.}
\label{bump}
}

The choice of $\rho(x)$ can be made more systematic by parameterizing it in terms of Legendre polynomials
\eq{
\rho(x)=\sum_{j=0}^{+\infty}\alpha_jP_j(2x-1)\,,
}
where we use the Legendre polynomials with a shifted argument. These are orthogonal over the interval $[0,1]$. Notice that the sum is convergent, since $\rho(x)$ is assumed to be finite at $x=1$ and $\rho(x)\sim x^{-1/2}$ around $x=0$. The  arc moments in Eq.~\eqref{newarc} are written as 
\eq{
m_n=g_{2 n}-\sum_{k=0}^{N-1} \frac{g_{2k+1}}{2(k-n)+1}-\sum_{j=0}^{N-n}b_j^{N-n}\alpha_j\,,
\label{newarc2}
}
for $n=0,1,\dots,N$, and $b_{j}^{n}\equiv\int_{0}^{1}\dd{x}\,x^{n}P_{j}(2x-1)$ are moments with respect to Legendre polynomial. Extra constraints are imposed from moments of $\tilde{\mu}$ in Eq.~\eqref{posIR},
\eq{
\tilde{m}_n\equiv\int_0^1\dd{\tilde{\mu}(x)}\,x^n=\sum_{k=0}^{N-1} \frac{g_{2 k+1} }{2(k+n)+1}+\sum_{j=0}^{N+n}b_j^{N+n}\alpha_j\,.
\label{mtild}
}
Note that both in Eqs.~\eqref{newarc2} and \eqref{mtild} the summation over the Legendre polynomials is truncated since $b_j^n=0$ for $j>n$. As a result, both the arc moments $m_n$ and the new moments $\tilde{m}_n$ are expressed in terms of a finite number of parameters, rather than an infinite number. The two sets of moments are constrained to satisfy Eq.~\eqref{ArcHankel}, namely positive semi-definiteness of a suitable combination of the corresponding Hankel matrices. Notice that, in practice, it is sufficient to consider only $\tilde{m}_0\geq0$, together with the positivity constraints on $m_n$ for $n=0,\dots,N$. Any higher-order $\tilde{\mu}$-moment introduces a new parameter beyond $\{\alpha_0,\dots,\alpha_{N}\}$, which appear in Eq.~\eqref{newarc2} for the $m_n$. As a result, they do not impose extra constraints on the EFT coefficients. 

Finding the boundary of the allowed region for the EFT coefficients can then be formulated as an optimization problem, which is in general a semi-definite program (see for instance Refs.~\cite{Caron-Huot:2020cmc,Simmons-Duffin:2015qma}). Considering once again the simple case with $N=1$, we obtain
\eq{
\label{primal}
\begin{aligned}
&\text{max/min} && \big[1,0,0\big]\cdot\big[g_2,\alpha_0,\alpha_1\big]^{T} 
\\
&\text{subject to} && \begin{bmatrix}
1 & -1/2 & 1/6\\
-1 & 1 & 0\\
0 & -1/2 & -1/6
\end{bmatrix}
\begin{bmatrix}
g_2\\\alpha_0\\\alpha_1
\end{bmatrix}\leq
\begin{bmatrix}
g_0-2g_1\\g_1\\g_1
\end{bmatrix}\,,
\end{aligned}
}
where ``$\rm max$'' and ``$\rm min$'' correspond to finding an upper and a lower bound on $g_2$, respectively, while keeping $g_0$ and $g_1$ fixed. The constraints correspond $m_0\geq m_1\geq0$, and $\tilde{m}_0\geq0$. A convenient approach to solve this problem is to look at the \emph{dual} problem (see Refs.~\cite{Caron-Huot:2020cmc,Simmons-Duffin:2015qma,Boyd_Vandenberghe_2004}) obtained as follows
\eq{
\label{dual}
\begin{aligned}
&\text{min/max} && \pm\big[g_0-2g_1,g_1,g_1\big]\cdot\big[y_1,y_2,y_3\big]^T 
\\
&\text{subject to} && \begin{bmatrix}
1 & -1/2 & 1/6\\
-1 & 1 & 0\\
0 & -1/2 & -1/6
\end{bmatrix}^T
\begin{bmatrix}
y_1\\y_2\\y_3
\end{bmatrix}=
\pm\begin{bmatrix}
1\\0\\0
\end{bmatrix}\,,\, y_i\geq0\,,
\end{aligned}
}
where $+$ is for ``$\rm min$'' and $-$ is for ``$\rm max$''. It is straightforward to show that the dual problem in Eq.~\eqref{dual} is infeasible, i.e.~the set of allowed $y_i$ by the constraints is empty. This implies, using duality theory, that the original optimization problem in Eq.~\eqref{primal}, also known as \emph{primal}, is unbounded, which is consistent with our previous discussion. Moreover, notice that there cannot be a disallowed region for $g_2$ in the middle. The reason is that the allowed region must be a convex set and, therefore, cannot have a gap in the bulk. This argument can be extended to all  higher-order coefficients, both conservative and dissipative. 

We conclude that, once nonzero dissipative coefficients are allowed, without any extra information positivity constraints alone are too weak to constrain the EFT coefficients, beyond $g_0,g_1\geq0$. The conservative case represents the opposite extreme, in which all  dissipative coefficients are set to zero and a finite allowed region can be obtained, as discussed in Section~\ref{sec:arc_bounds}. By continuity, one expects that imposing restrictions on the magnitude of the dissipative coefficients should lead to nontrivial bounds. As an example, it is easy to see from Eq.~\eqref{primal} that restricting the magnitude of $\alpha_0$ and $\alpha_1$ implies a bound on $g_2$. However, we are not aware of a physically motivated setting in which such restrictions can be imposed. We leave a more systematic exploration of this question for future work.

\newpage

\section{Conclusions}

In this paper, we have studied the consequences of causality for the dynamical response coefficients in the point-particle effective theory, at leading order in the gradient expansion. We showed that, under the assumptions of positivity (i.e., absorptivity of the system; see Eq.~\eqref{posim}) and suitable high-frequency convergence of the Green's function, classical retardedness of the response implies a series of inequalities among the Wilson coefficients of the EFT. In the absence of dissipation, the first few bounds take the form \eqref{CE_bounds2}.

Despite the analogy with the positive moment problem for scattering amplitudes in unitary and causal quantum field theories in Minkowski space~\cite{Bellazzini:2020cot}, the physical interpretation and consequences of our bounds are significantly different. First, as is common in the context of spontaneous breaking of Lorentz invariance~\cite{Creminelli:2022onn,Creminelli:2024lhd,Creminelli:2025rxj}, we worked with the linear response function, which is a genuinely off-shell quantity, in contrast to the $S$-matrix.\footnote{It is known that the retarded Green's function has a simpler analytic structure than scattering amplitudes~\cite{Correia:2025enx}.}
In addition, unlike in the positive moment problem for the $S$-matrix, the arcs associated with the retarded Green's function receive contributions from an infinite number of dissipative low-energy coefficients. This makes the bounds in the presence of dissipation considerably more involved, as discussed in detail in Section~\ref{diss}. In order to investigate their consequences for gravitating objects, we therefore restricted our analysis to the conservative sector.

In Section~\ref{sec:NeutronStar}, we explicitly computed the dynamical Love numbers for a class of neutron stars modeled as perfect fluids. For the equation of state, we considered both polytropic models and a set of standard tabulated equations of state. The main outcome of our analysis is twofold. First, in the low-compactness regime, where the assumptions underlying the bounds are more robustly justified, the bounds imply an upper limit on the energy scale at which the low-frequency expansion of the response kernel breaks down, which in turn translates into an upper bound on the frequency of the leading mode of the star. This constraint is independent of the usual thermodynamic constraints and subluminality conditions commonly imposed on the equation of state.
We have checked that the resulting upper bound on the mode frequency is consistent with previous explicit calculations of the fundamental mode of neutron stars for specific choices of the equation of state. Additionally, our bound $B_3$ provides a way to quantify the validity of the single-mode approximation.

Second, for more compact stars, where the renormalization-group running of the Love numbers becomes non-negligible, the bounds become strongly sensitive to the renormalization scale and are generally violated as this scale is lowered or increased. This confirms an expectation already suggested by the black-hole example: the bounds \eqref{CE_bounds2} need not hold once strong-gravity effects become important. A more detailed analysis of the assumptions underlying the bounds \eqref{CE_bounds2}, and of their validity in the strong-gravity regime, is presented in a companion work~\cite{unpub}.

We note that, although $C\simeq0.05$--$0.1$ is below the compactness of canonical neutron stars, such values can in principle occur for low- and subsolar-mass neutron-star configurations. Such objects have recently attracted attention as potential gravitational-wave sources, owing in particular to their large tidal deformabilities, and dedicated searches for subsolar-mass neutron-star binaries have already been performed~\cite{Bandopadhyay:2022tbi,Kacanja:2026byy}.

Our results suggest several directions for future work. Natural extensions include generalizing the bounds to the gravito-magnetic response, higher multipoles, the nonlinear regime~\cite{Poisson:2020vap,Riva:2023rcm,Iteanu:2024dvx,Pani:2025qxs}, and rotating solutions.
Throughout this paper, the analyticity properties of the response follow solely from retardedness. It would be interesting to investigate the consequences of full microcausality. For example, in electromagnetism, microcausality constrains combinations of the electric and magnetic susceptibilities of a medium~\cite{Creminelli:2024lhd}. This suggests the possibility of deriving bounds involving both the gravito-electric and gravito-magnetic Love numbers of a compact object. Such an analysis would require going beyond the worldline description and spatially resolving the object.
Another interesting direction would be to constrain or rule out specific families of equations of state. At present, our bounds are not strong enough to rule out equations of state that exhibit explicit pathologies (e.g., superluminality), such as the polytrope with $n=0$, which  saturates the bound in the Newtonian limit. 

In addition, phenomenological universal relations among neutron-star observables are known to hold across broad classes of equations of state. Such relations play an important role in neutron-star phenomenology, as they considerably reduce the dependence of macroscopic observables on the unknown equation of state. Familiar examples include the $I$-Love-$Q$ and $f$-mode-Love relations~\cite{Yagi:2013awa,Yagi:2016bkt}. More recently, this program has been extended to dynamical tides~\cite{Saes:2025jvr} (see also Ref.~\cite{Apostolidis:2026qsg}), revealing a correlation between the static tidal deformability and its leading dynamical correction at second order in frequency for compactnesses $C>0.12$. It would be interesting to investigate to what extent this universality persists at higher orders in frequency and whether it can be combined with our bounds to explore a broader set of equations of state.

A more fundamental, and perhaps conceptually more interesting, open question concerns the strong-gravity regime. Our analysis suggests that a naive extrapolation of the bounds~\eqref{CE_bounds2} to large compactness is unlikely to be valid. Beyond the renormalization-scale dependence discussed above, the assumptions underlying their derivation must be carefully reassessed, as they are no longer necessarily justified in this regime. Understanding this issue is essential to fully exploit the consequences of causality and, more generally, microscopic principles for the point-particle EFT description of compact objects. It may also shed light on how to reconcile the vanishing of black-hole Love numbers with dispersion-relation arguments, a question that has so far remained elusive.

We hope to return to these questions in future work.

\paragraph{Acknowledgements:} We thank Micaela Oertel for helpful discussions. We are particularly grateful to Paolo Creminelli, Valerio De Luca, Leonardo Gualtieri, Takuya Katagiri, Paolo Pani, and Leonardo Senatore for collaborations on topics related to this work. The research of AL and LS has been funded, in part, by the French National Research Agency (ANR) under project ANR-24-CE31-1097-01. This work has received support under the program ``\textit{Investissement d'Avenir}'' launched by the French Government and implemented by ANR, with the reference ANR-18-IdEx-0001 as part of its program ``\textit{Emergence}''.
We acknowledge the \textit{Claude Opus 4.8 model} for useful assistance with technical aspects related to the solution at fourth order in frequency of Appendix~\ref{app:GRcomputation}.

\noindent
\textbf{Note added:} While this work was being finalized, we became aware of Ref.~\cite{Kehagias:2026opd}, which partially overlaps with our analysis. Where the two works overlap, our results are in agreement.

\appendix
\newpage
\section{EFT computation}
\label{app:ppEFT}

\subsection{Graviton one-point response function}

In this appendix, we briefly review the calculation of the graviton one-point function within the point-particle EFT framework, extending previous results~\cite{Combaluzier--Szteinsznaider:2025eoc,Apostolidis:2026qsg,Correia:2026utp} to quartic order in frequency in the even-parity gravitational sector. Let us denote by $h_{\mu\nu}$ the canonically normalized metric perturbation, defined by $\delta g_{\mu\nu}= 2 h_{\mu\nu}/\Mpl^{(D-2)/2}$ in $D$ spacetime dimensions. As is standard, we work within the in-in Schwinger--Keldysh formalism and define  the Keldysh basis,
\begin{equation}
h_+ \equiv \frac{1}{2}\left(h_1 + h_2\right), \qquad
h_- \equiv h_1 - h_2 \,,
\label{eq:keldyshbasispm}
\end{equation}
where $h_1^{\mu\nu}$ and $h_2^{\mu\nu}$ are the fields in the forward and backward branches of the closed-time contour, respectively.

The effective action for $h_\pm$ is obtained by integrating out the object's internal degrees of freedom. Collectively denoting them by $\mathcal{X}$, this amounts to evaluating the path integral
\begin{equation}
\label{eq:PIXX}
    \E^{i\Gamma^\text{in-in}_\text{int}[h_\pm]} = \int\mathcal{D}\mathcal{X}_+\mathcal{D}\mathcal{X}_- \, \E^{iS\left[h_\pm,\mathcal{X}_\pm\right]} ,
\end{equation}
where $S$ is the action \eqref{eq:ppEFT0}.
In Eq.~\eqref{eq:PIXX}, we omit the dependence on the coordinates of the point particle, whose position we assume to be  fixed by working in its instantaneous rest frame.
Using linear response theory, $Q_E$ can be solved for perturbatively in terms of the gravito-electric field $E$ as
\begin{equation}
    \langle Q_{E,I}^{ij}(\tau)\rangle=\int \d \tau'\,K_{IJ}^{(E)}{}^{ij\vert i' j'}(\tau-\tau'){E}^J_{i' j'}(\tau')\,,
\label{eq:lQr}
\end{equation}
where $I,J=\{+,-\}$ are Keldysh indices, and $K_{IJ}^{(E)ij \vert i'j'}$ is the linear-response kernel, which is related to the two-point function of $Q_E^{ij}$ in the Keldysh basis via~\cite{Goldberger:2020fot,Saketh:2022xjb,Saketh:2023bul,Glazer:2024eyi,Saketh:2024juq,Apostolidis:2026qsg,Chakraborty:2026dox}
\begin{equation}
    \langle Q_{E,I}^{ij}(\tau)Q_{E,J}^{i' j'}(\tau')\rangle=-iK_{IJ}^{(E)}{}^{i j\vert i' j'}(\tau-\tau')
    \equiv
    -i K_{IJ}^{(E)}(\tau-\tau')\delta^{( i}_{( i'}\delta^{j)_T}_{j')_T}.
\end{equation}
In the last step, we defined the index-free kernel $K_{IJ}^{(E)}$, exploiting the fact that the (unperturbed) object is spherically symmetric.

From the solution \eqref{eq:lQr}, we obtain an  in-in effective action
\begin{equation}
    \Gamma^\text{in-in}_\text{int}[h_\pm] 
    \supset \int \mathrm{d}\tau_1 \, \mathrm{d}\tau_2  
    K_{{IJ}}^{(E)}(\tau_2-\tau_1) \, E^I_{i j}(\tau_2) \, E^J{}^{i j}(\tau_1) +\cdots \,,
\label{eq:gammainineft}
\end{equation}
where the omitted terms encode higher-multipole contributions, which arise at successive orders in the gradient expansion beyond the quadrupolar sector.
From the effective action \eqref{eq:gammainineft}, the one-point function of the electric component $E_{+a b}$ of the Weyl tensor can be readily computed as~\cite{Combaluzier--Szteinsznaider:2025eoc, Apostolidis:2026qsg}:
\begin{equation}
\label{E_opf}
    \left\langle E_{+a b}(t, \vec{x})\right\rangle_{\mathrm{in}-\mathrm{in}}=i  \int \mathrm{d} \tau_1 \mathrm{d} \tau_2 \, K_{+-}^{(E)}\left(\tau_2-\tau_1\right)\left\langle E_{+a b }(t, \vec{x}) E_{-ik}\left(\tau_2\right)\right\rangle \bar{E}_{+}^{ik}\left(\tau_1\right).
\end{equation}

The expression \eqref{E_opf} contains the following ingredients: $\bar{E}_{+}^{ik}$  is the electric component of the Weyl tensor associated with the external quadrupolar tidal field; the expectation value $\langle E_{+a b }(t, \vec{x}) E_{-ik}\left(\tau_2\right)\rangle$ is the two-point function of the Weyl tensor, which can be computed from the graviton propagator in a suitable gauge; and $K^{({E})}_{+-}(\tau)$ is proportional to the retarded Green's function of the $Q_E$ operators~\cite{Combaluzier--Szteinsznaider:2025eoc}.

Following Refs.~\cite{Combaluzier--Szteinsznaider:2025eoc,Apostolidis:2026qsg}, and working in the de Donder gauge, $\partial^\mu h_{\mu\nu}=\frac{1}{2}\partial_\nu h$, the one-point function can be evaluated straightforwardly, yielding
\begin{equation}
\label{E_opfc}
    \left\langle E_{+a b}(t, \vec{x})\right\rangle_{\mathrm{in}-\mathrm{in}} 
       =  -\frac{\E^{-i\omega t}}{\Mpl^{D-2}} K_{+-}^{(E)}\left(\omega\right) \left[
       \frac{2\Gamma\left(2+\frac{D-3}{2}\right)}{\pi^{\frac{D-1}{2}}}
       \frac{D-3}{D-2} \mathcal{E}_{ik}  \partial_a\partial_b
     \frac{x^ix^k}{\vert\vec{x}\vert^{1+D}} 
     +\mathcal{O}(\omega r)\right]     ,
\end{equation}
where $r\equiv\vert\vec{x}\vert$, 
and $\mathcal{E}_{ij}$ is a symmetric, traceless tensor that characterizes the amplitude of the $t$--$t$ component of the external quadrupolar tidal field. In particular, we defined it as  $\bar h_{00}(\omega)=\Mpl^{(D-2)/2}  \mathcal{E}_{ij}x^i x^j(1+\mathcal{O}(\omega^2 r^2))$. 

While the details of the calculation can be found in Refs.~\cite{Combaluzier--Szteinsznaider:2025eoc,Apostolidis:2026qsg}, a few comments are worth making here.
First, note that it is convenient to keep the number $D$ of spacetime dimensions arbitrary, as we will later resum gravitational nonlinearities. The resummation effectively corresponds to computing a loop diagram in the worldline EFT, which we will regulate using dimensional regularization~\cite{Caron-Huot:2025tlq,Combaluzier--Szteinsznaider:2025eoc,Apostolidis:2026qsg,Correia:2026utp}. Second, in Eq.~\eqref{E_opfc}, we retained only the leading terms in $\omega r$, with all the frequency dependence encoded in the prefactor $K_{+-}^{(E)}(\omega)$. The reason is that subleading terms in $\omega r$ correspond to far-zone corrections to the one-point function. Since we will match this result to a full solution in the near-zone limit, $r\ll\omega^{-1}$ (see Refs.~\cite{Combaluzier--Szteinsznaider:2025eoc,Apostolidis:2026qsg} for details), these terms can be safely neglected. Finally, as assumed from the outset, we focused exclusively on the quadrupolar tidal interaction and disregarded contributions from higher multipoles, as well as from the gravito-magnetic sector.

\subsection{Including gravitational nonlinearities}

In the worldline EFT, the result in Eq.~\eqref{E_opfc} represents the tree-level contribution to the graviton one-point function. To perform a complete matching to a general-relativistic solution, however, it is necessary to account for gravitational nonlinearities within the EFT. In general, this procedure is highly nontrivial and amounts to evaluating divergent classical loop diagrams in the worldline EFT. A convenient approach is to implement a Born-series resummation. This method was applied in Ref.~\cite{Caron-Huot:2025tlq} in the context of a scalar field and was subsequently extended to gravitational perturbations in Refs.~\cite{Combaluzier--Szteinsznaider:2025eoc,Apostolidis:2026qsg,Correia:2026utp}. (See Ref.~\cite{Kosmopoulos:2025rfj} for an alternative approach based on a different regularization scheme.)

In summary, the strategy is to resum the gravitational nonlinearities in the $D$-dimensional Zerilli equation~\cite{Hui:2020xxx}
\begin{equation}
    \left(\frac{\d^2}{\d r^2}-\frac{(\ell-\varepsilon)(\ell-\varepsilon+1)}{r^2} \right)\Psi_{\text{Z}}(r) = V_{\Psi_\text{Z}}(r)\Psi_\text{Z}(r),
    \label{eq:Z_EWE}
\end{equation}
where $\varepsilon=2-D/2$, by means of the perturbative series~\cite{Correia:2024jgr,Caron-Huot:2025tlq,Correia:2026utp}
\begin{equation}
\begin{aligned} \Psi_{\mathrm{Z}}(r)=\Psi_{\mathrm{Z}}^{(h)}(r) & +\int^r \mathrm{~d} r^{\prime} G(r, r^{\prime}) V_{\Psi_{\mathrm{Z}}}(r^{\prime}) \Psi_{\mathrm{Z}}^{(h)}(r^{\prime}) \\ & +\int^r \mathrm{~d} r^{\prime} G(r, r^{\prime}) V_{\Psi_{\mathrm{Z}}}(r^{\prime}) \int^{r^{\prime}} \mathrm{d} r^{\prime \prime} G(r^{\prime}, r^{\prime \prime}) V_{\Psi_{\mathrm{Z}}}(r^{\prime \prime}) \Psi_{\mathrm{Z}}^{(h)}(r^{\prime \prime})+\cdots .
\end{aligned}
\label{eq:Bssolpsi}
\end{equation}
In Eq.~\eqref{eq:Bssolpsi}, $\Psi_{\mathrm{Z}}$ is a generalized Zerilli variable~\cite{Hui:2020xxx}, and $\Psi_{\mathrm{Z}}^{(h)}$ denotes the homogeneous solution of the Zerilli equation~\eqref{eq:Z_EWE},
\begin{equation}
\label{homogen_psi}
    \Psi_{\mathrm{Z}}^{(h)}(r)=\mu^{-\varepsilon} B_{\mathrm{reg}} r^{\ell+1-\varepsilon}+\frac{\mu^{\varepsilon} B_{\mathrm{irr}}}{2 \ell+1-2 \varepsilon} r^{-\ell+\varepsilon},
\end{equation}
where the factors $\mu^{\mp\varepsilon}$ ensure that the field has the correct mass dimension for $D\neq4$, while $B_{\mathrm{reg}}$ and $B_{\mathrm{irr}}$ are integration constants that will subsequently be related to the relativistic vacuum solution in the exterior of the star. Moreover, $G(r,r^{\prime})$ denotes the Green's function of Eq.~\eqref{eq:Z_EWE} and is given by
\begin{equation}
    G(r, r^{\prime})=\frac{r^{\varepsilon-\ell}\left(r^{\prime}\right)^{1+\ell-\varepsilon}-\left(r^{\prime}\right)^{\varepsilon-\ell} r^{1+\ell-\varepsilon}}{2 \varepsilon-2 \ell-1} \theta(r-r^{\prime}),
\end{equation}
while the potential $V_{\Psi_{\mathrm{Z}}}$ on the right-hand side of Eq.~\eqref{eq:Z_EWE} encodes all corrections in $\omega$ and $r_s$ (see Ref.~\cite{Combaluzier--Szteinsznaider:2025eoc} for its explicit expression).
The present calculation differs from those of Refs.~\cite{Combaluzier--Szteinsznaider:2025eoc,Apostolidis:2026qsg} in that we are interested in the solution through fourth order in $\omega$. We therefore require the particular solution to Eq.~\eqref{eq:Z_EWE} through $\mathcal{O}(\omega^4G^9)$.

The relation between the response function $K_{+-}^{(E)}(\omega)$ and the coefficients $B_{\mathrm{reg}}$ and $B_{\mathrm{irr}}$ is obtained by matching the Weyl tensor constructed from the homogeneous solution $\Psi_{\mathrm{Z}}^{(h)}$ to the tree-level result~\eqref{E_opfc}~\cite{Combaluzier--Szteinsznaider:2025eoc,Apostolidis:2026qsg}:
\begin{equation}
\label{Bratio}
    \frac{K_{+-}^{(E)}(\omega)}{\Mpl^{D-2}}\frac{D-3}{D-2}\frac{\Gamma(\frac{D+1}{2})}{\pi^{\frac{D-1}{2}}}D(D-1)(D+1)
    = \frac{B_\text{irr}}{ B_\text{reg}}.
\end{equation}

The resummed particular solution, obtained from the perturbative expansion~\eqref{eq:Bssolpsi}, takes the following form in the $\varepsilon\to0$ limit, up to $\mathcal{O}(\omega^4)$:\footnote{Note that, for a consistent renormalization procedure, one should retain terms up to $\mathcal{O}(\varepsilon^2)$, since the calculation contains $1/\varepsilon^2$ divergences. Here, however, we choose to display only the structure of the divergences.}
\begin{equation}
    \begin{aligned}
    \label{psi_z_bare}
        \Psi_\mathrm{Z}(r)&=r^3 B_{\text{reg}} \Bigg[1-\omega ^2 \left(\frac{2731 \bar{G}_D^2}{9800}+\frac{107 \bar{G}_D^2}{210 \varepsilon }+\frac{1}{70} 107 \bar{G}_D^2 \log (\mu  r)\right)\\
        &+\omega ^4 \left(-\frac{152573716627 \bar{G}_D^4}{15558480000}-\frac{592997 \bar{G}_D^4}{2646000 \varepsilon }+\frac{11449 \bar{G}_D^4}{88200 \varepsilon ^2}+\left(\frac{11449 \bar{G}_D^4}{12600 \varepsilon }-\frac{592997 \bar{G}_D^4}{378000}\right) \log (\mu  r)\right.\\
        &\left.+\frac{11449 \bar{G}_D^4 \log ^2(\mu  r)}{3600}\right)\Bigg]+\frac{B_{\text{irr}}}{r^2}\Bigg[\frac{1}{5}+\omega ^2 \left(\frac{129359 \bar{G}_D^2}{441000}+\frac{107 \bar{G}_D^2}{1050 \varepsilon }+\frac{107}{210} \bar{G}_D^2 \log (\mu  r)\right)\\
        &+\omega ^4 \left(\frac{6347365589 \bar{G}_D^4}{11113200000}+\frac{6553543 \bar{G}_D^4}{30870000 \varepsilon }+\frac{11449 \bar{G}_D^4}{441000 \varepsilon ^2}+\left(\frac{11449 \bar{G}_D^4}{49000 \varepsilon }+\frac{6553543 \bar{G}_D^4}{3430000}\right) \log (\mu  r)\right.\\
        &\left.+\frac{103041 \bar{G}_D^4 \log ^2(\mu  r)}{98000}\right)\Bigg]+\frac{B_{\text{reg}}}{r^2}\Bigg[-\frac{189 \bar{G}_D^5}{32}+\omega ^2 \left(\frac{4937 \bar{G}_D^7}{960 \varepsilon }+\frac{2802329 \bar{G}_D^7}{403200}\right)\\
        &+\frac{64181}{960} \omega^2\bar{G}_D^7 \log (\mu  r)+\omega ^4 \left(\frac{1751056263521 \bar{G}_D^9}{4741632000}+\frac{8153741 \bar{G}_D^9}{756000 \varepsilon }-\frac{528259 \bar{G}_D^9}{403200 \varepsilon ^2}\right.\\
    &\left.+\left(\frac{138613597 \bar{G}_D^9}{756000}-\frac{8980403 \bar{G}_D^9}{403200 \varepsilon }\right) \log (\mu  r)-\frac{152666851 \bar{G}_D^9 \log ^2(\mu  r)}{806400}\right)\Bigg]+\cdots\,,
    \end{aligned}
\end{equation}
where $\bar{G}_D\equiv G M n_D\xrightarrow{D\rightarrow4}\bar{G}\equiv GM$, with $n_D\equiv 4\pi^{\frac{3-D}{2}}\Gamma\left(\frac{D-1}{2}\right)/(D-2)$.
In Eq.~\eqref{psi_z_bare}, we have displayed only the $r^2$ and $r^{-3}$ terms, while all terms with different radial falloffs are omitted in the ellipsis. 

To cancel the $1/\varepsilon$ and $1/\varepsilon^2$ divergences, we introduce counterterms in the effective action that renormalize the Love number couplings and the fields. This amounts to defining the renormalized coefficients $\bar{B}_{\text{reg}}$ and $\bar{B}_{\text{irr}}$, which are related to the bare ones via

\begin{subequations}
\label{B_renorm_12_}
\begin{align}
\label{B_renorm_1_}
B_{\text{reg}}&=\bar{B}_{\text{reg}}\Bigg[1+\omega ^2 \left(\frac{107 \bar{G}^2}{210 \varepsilon }+\frac{107}{105} \bar{G}^2 (\gamma_{\rm E} -1+\log (4\pi))\right)+\omega ^4 \left(\frac{11449 \bar{G}^4}{88200 \varepsilon ^2}+\right.\nonumber\\
&\left.+\frac{1695233 \bar{G}^4}{4630500 \varepsilon }+\frac{\bar{G}^4 (\gamma_{\rm E} -1+\log (4\pi)) (1202145 \gamma_{\rm E} +2188321+1202145 \log (4\pi ))}{2315250}\right)\Bigg]\\
B_{\text{irr}}&=\bar{B}_{\text{irr}}\Bigg[1+\omega ^2 \left(-\frac{107 \bar{G}^2}{210 \varepsilon }-\frac{107}{105} \bar{G}^2 (\gamma_{\rm E} -1+\log (4\pi))\right)+\omega ^4 \left(\frac{11449 \bar{G}^4}{88200 \varepsilon ^2}\right.\nonumber\\
  &\left.-\frac{1695233 \bar{G}^4}{4630500 \varepsilon }+\frac{\bar{G}^4 (\gamma_{\rm E} -1+\log (4\pi)) (1202145 \gamma_{\rm E} -4592611+1202145 \log (4\pi ))}{2315250}\right)\nonumber\\
  &+\bar{B}_{\text{reg}}\Bigg[-\omega ^2 \left(\frac{32 \bar{G}^7}{3 \varepsilon }+\frac{224}{3}  \bar{G}^7 (\gamma_{\rm E} -1+\log (4\pi))\right)+\omega ^4 \left(\frac{856 \bar{G}^9}{315 \varepsilon ^2}\right.\nonumber\\
  &\left.-\frac{8 \bar{G}^9 (11235 \gamma_{\rm E} +289211+11235 \log (4 \pi ))}{33075 \varepsilon }\right.\nonumber\\
  &\left.-\frac{2 \bar{G}^9 \left(67410 \gamma_{\rm E} ^2+3745 \pi ^2-3560412+67410 \left(4 \log ^2(2)+\log (\pi ) \log (16 \pi )\right)\right)}{11025}\right.\nonumber\\
  &-\left.\frac{2 \bar{G}^9 \left(3470532 \log (4\pi )+6 \gamma_{\rm E}  (578422+22470 \log (4\pi ))\right)}{11025}\right)\Bigg]
  \,,
\label{B_renorm_2_}
\end{align}
\end{subequations}
where $\gamma_\text{E}$ is the Euler--Mascheroni constant.

Substituting Eqs.~\eqref{B_renorm_12_} into Eq.~\eqref{psi_z_bare} and taking the limit $\varepsilon\to0$, we finally obtain: 
\begin{align}
    \Psi_\mathrm{Z}^\mathrm{R}(r)= &\, \,  r^3 \bar{B}_{\text{reg}}\Bigg[1- \bar{G}^2 \omega ^2 \left(\frac{214}{105}\log (\mu  r)  + \frac{  2731}{9800} \right) \nonumber \\
    &\qquad +\bar{G}^4 \omega ^4 \left(  \frac{22898}{11025} \log^2(\mu  r) - \frac{10931911}{4630500} \log(\mu  r) - \frac{83304726019}{15558480000}\right)\Bigg] \nonumber \\
    &+\frac{1}{r^2}\Bigg[\frac{\bar{B}_{\text{irr}}}{5}-\frac{189}{32}   \bar{G}^5 \bar{B}_{\text{reg}} \nonumber \\
    &\qquad+\bar{G}^2\omega ^2 \left( \bar{B}_{\text{irr}} \left( \frac{214}{525}\log (\mu  r)+\frac{  111383}{441000}\right)+\bar{G}^5 \bar{B}_{\text{reg}} \left(\frac{3011}{80}\log (\mu  r)+\frac{ 1545209}{57600}\right)\right) \nonumber \\
    &\qquad+\bar{G}^4\omega ^4 \bigg( \bar{B}_{\text{irr}} \left(\frac{22898}{55125}\log^2 (\mu  r)+\frac{5095969}{4630500}\log (\mu  r)-\frac{3413631623}{25930800000}\right) \nonumber \\
    &\qquad\quad-\bar{G}^5 \bar{B}_{\text{reg}} \left(\frac{34347}{2800}\log^2 (\mu  r)-\frac{4815653203}{21168000}\log (\mu  r)-\frac{5999650820213}{23708160000}\right)\bigg)\Bigg].
\label{psi_EFT_renorm}
\end{align}
Note that the expression for $\Psi_\mathrm{Z}^\mathrm{R}(r)$ reproduces previous results~\cite{Combaluzier--Szteinsznaider:2025eoc,Apostolidis:2026qsg} through order $\omega^2$. The terms at order $\omega^4$, by contrast, are new. 
Note also that there is no explicit dependence on odd powers of $\omega$ in Eq.~\eqref{psi_EFT_renorm}, since the frequency enters the differential equation~\eqref{eq:Z_EWE} only through even powers. Odd powers of $\omega$ can nevertheless enter through the coefficients $\bar{B}_{\rm reg}$ and $\bar{B}_{\rm irr}$.

\newpage

\section{General-relativistic computation and matching to EFT}
\label{app:GRcomputation}

The EFT results of Appendix~\ref{app:ppEFT} are general and completely agnostic about the nature of the object. In particular, the coefficients $\bar{B}_{\text{reg}}$ and $\bar{B}_{\text{irr}}$, which are related to the renormalized EFT couplings through Eq.~\eqref{Bratio}, are  so far  generic. They can be determined once an explicit description of the object is provided. This is  the subject of this appendix and of Section~\ref{sec:NeutronStar}.

We work  here in a fully general-relativistic setting for completeness. In particular, we will reproduce the existing results through order $\omega^2$~\cite{Apostolidis:2026qsg} and extend them to order $\omega^4$ for the first time.

\subsection{Exterior solution}

Let us start with the exterior problem. The dynamics of parity-even gravitational perturbations is governed by the Zerilli equation in $D=4$ vacuum general relativity:\footnote{In this section, we no longer need to work in generic spacetime dimensions and set $D=4$ throughout.}
\begin{equation}
\label{Zerilli}
    f(r)\partial_r\left[f(r) \partial_r \Psi_{\mathrm{Z}}(r)\right]+\left[\omega^2-f(r)\left(\frac{r_s}{r^3}+\frac{2 n}{3 r^2}+\frac{8 n^2(2 n+3)}{3\left(2 n r+3 r_s\right)^2}\right)\right] \Psi_{\mathrm{Z}}(r)=0 ,
\end{equation}
where here $f(r)\equiv 1-r_s/r$ and $n\equiv \frac{1}{2}(\ell-1)(\ell+2)$, with $\ell$ denoting the angular momentum quantum number, which we will later set to $\ell=2$. In Eq.~\eqref{Zerilli}, we are adopting the notation of Ref.~\cite{Apostolidis:2026qsg}. For a pedagogical introduction and derivation, see, e.g., Refs.~\cite{Berti:2009kk,Rodriguez:2026iot}.

The goal is to solve Eq.~\eqref{Zerilli} perturbatively in $\omega r_s$ in the range $r_s < r \ll 1/\omega$. This interval is usually referred to as the ``near zone,'' in contrast to the far zone, where $\omega r\gg 1$. Despite its name, the near zone covers a parametrically large range of $r$ and, in particular, overlaps with the region where the EFT is valid. A perturbative solution in $\omega r_s$ is therefore sufficient to perform the matching and determine the effective couplings. In particular, for a star with radius $R_\star \gtrsim \mathcal{O}(1)r_s$, the perturbative expansion is valid throughout the region $R_\star \leq r \ll 1/\omega$.

To this end, it is useful to introduce the combination $g\equiv\omega r_s\ll 1$ and the rescaled field 
\begin{equation}
    u(x(r)) \equiv \left(\frac{r}{r_s}\right)^\ell \Psi_{\mathrm{Z}}(r)\,,
\qquad  x\equiv\frac{r_s}{r} .
\label{eq:rescaledfield}
\end{equation}
Then, Eq.~\eqref{Zerilli} for $\ell=2$ reads~\cite{Combaluzier--Szteinsznaider:2025eoc,Apostolidis:2026qsg}
\begin{equation}
\label{eq:ux}
    \mathcal{L}(u)=g^2 S(u)\,,
\end{equation}
where 
\begin{equation}
    \mathcal{L}(u)\equiv x(1-x) u^{\prime\prime}(x)+(6-7 x) u'(x)-\frac{3 x (27 x+58)+32}{(3 x+4)^2}u(x) \, , \quad S(u)\equiv \frac{1}{(x-1) x^3}u(x),
\end{equation}
which we solve order by order in $g$ for $0 < x < 1$ by expanding $u(x)$ as 
\begin{equation}
    u(x)= u^{(0)}(x)+g u^{(1)}(x)+g^2u^{(2)}(x)+\cdots .
\label{eq:uxexp}
\end{equation}
After expanding in $g$, Eq.~\eqref{eq:ux} becomes
\begin{equation}
    \mathcal{L}\big(u^{(n)}\big)=S\big(u^{(n-2)}\big).
    \label{eq:equn}
\end{equation}
The general solution to Eq.~\eqref{eq:equn} is obtained by adding the homogeneous solution and a particular solution. The homogeneous solution of order $n$ is
\begin{equation}
u_h^{(n)}(x)=a_n\Phi_+(x) + b_n\Phi_-(x)\, ,
\end{equation}
where
\begin{equation}
    \Phi_+(x)=\frac{4+6 x-3 x^3}{x^5 (4+ 3 x)}, \, \quad \Phi_-(x)=- \frac{ x (12+ (24+13 x)x)+3 \left(4+6 x-3 x^3\right) \log (1-x)}{3x^5 (4+3 x)},
\end{equation}
and $a_n$ and $b_n$ denote the integration constants. For $n=0,1$ the source term vanishes, therefore the homogeneous solution is sufficient. Starting at second order, the perturbative equation acquires a nontrivial source term. A particular solution can be obtained from the source term, using variation of parameters, or equivalently, the Green's function method. The solution at order $n$ is then written as: 
\begin{equation}
    u^{(n)}(x)=u_h^{(n)}(x) - \Phi_+(x)\int^x \dd{x}^{\prime}\frac{\Phi_-(x^{\prime})S\big(u^{(n-2)}\big)}{x^{\prime}(1-x^{\prime}) W(x^{\prime})}+\Phi_-(x)\int^x \dd{x}^{\prime}\frac{\Phi_+(x^{\prime})S\big(u^{(n-2)}\big)}{x^{\prime}(1-x^{\prime}) W(x^{\prime})}\,,
    \label{unsol}
\end{equation}
where from the Wronskian we obtain, $x(1-x)W(x)=x(1-x)(\Phi_+\Phi_-^{\prime}-\Phi_-\Phi_+^{\prime})=x^{-5}$. Notice that the tower of the even and odd order solutions are decoupled, and identical. Therefore, it is sufficient to compute $u^{(0)}$, $u^{(2)}$, $u^{(4)}$, etc. The odd order solutions are obtained by shifting the integration constants, i.e.~$a_{2i}\to a_{2i+1}$ and $b_{2i}\to b_{2i+1}$.

It can be shown that the solution~\eqref{unsol} belongs to the class of harmonic polylogarithm functions \cite{Remiddi:1999ew,Maitre:2005uu}. These are defined recursively as successive integrals of logarithm multiplied by rationals. More precisely, a harmonic polylogarithm $H({\vec w},x)$ of weight ${\vec w}=\{a,{\vec w'}\}$ satisfies
\begin{equation}
\frac{\D}{\D x}\, H({\vec w},x) = f_{a}(x)\, H({\vec w'},x)\,,
\end{equation}
in which the components of ${\vec w}$ can take values $0,1$, or $-1$, corresponding to $f_0(x)=1/x$, $f_1(x) = 1/(1-x)$, and $f_{-1}(x) = 1/(1+x)$. A harmonic polylogarithm of zeroth-order weight is defined to be $H(,x)=1$. Then, at weight one, we have $H(0,x)=\log(x)$, $H(1,x)=-\log(1-x)$, and $H(-1,x)=\log(1+x)$, where the integration constants are fixed by convention; see Ref.~\cite{Remiddi:1999ew} for details. Functions $H({\vec w},x)$ of weight less than 4 can be reduced to simpler functions, such as logarithms or polylogarithms, e.g.~$H(\{0,1\},x)={\rm Li}_2(x)$. At higher weights, however, harmonic polylogarithms cannot in general be expressed solely in terms of  simpler functions. We encounter these more complicated functions when solving Eq.~\eqref{eq:equn}, starting at fourth order in $g$.  

Once the solutions to Eqs.~\eqref{eq:equn} have been obtained, we can substitute them into Eqs.~\eqref{eq:uxexp} and \eqref{eq:rescaledfield} to obtain the perturbative solution for the Zerilli variable $\Psi_{\text{Z}}$ in the near zone. The complete solution for the Zerilli variable is rather involved and is reported in the Mathematica code available at \cite{gitTA}. Here, we present only its large-$r$ expansion, which will be needed for matching to the EFT:

\begin{align}
     \Psi_{\mathrm{Z}}^{\ell=2}(r) \xrightarrow{r\rightarrow\infty}
      &\,\, a_0\frac{ r^3}{r_s^3}+\frac{r_s^2}{r^2}\left(\frac{1 }{5}b_0-\frac{189  }{1024}a_0\right) \nonumber 
      \\
      &+\omega  r_s\left[a_1\frac{r^3}{r_s^3}+\frac{r_s^2}{r^2}\left(\frac{1 }{5}b_1-\frac{189 }{1024}a_1\right)\right] \nonumber 
      \\
      &+\omega ^2 r_s^2\left[\frac{r^3}{r_s^3} \left(a_2+a_0\frac{107 }{210 }\log \left(\frac{r_s}{r}\right)\right)+\frac{r_s^2}{r^2}\left(b_2-\left(a_0\frac{3011 }{10240}+b_0\frac{107}{1050}\right) \log \left(\frac{r_s}{r}\right)\right)\right] \nonumber 
      \\
      &+\omega ^3 r_s^3\left[\frac{r^3}{r_s^3} \left(a_3+a_1\frac{107 }{210 }\log \left(\frac{r_s}{r}\right)\right)+\frac{r_s^2}{r^2}\left(b_3-\left(a_1\frac{3011 }{10240}+b_1\frac{107}{1050}\right) \log \left(\frac{r_s}{r}\right)\right)\right] \nonumber 
      \\
      &+\omega ^4 r_s^4\Bigg[\frac{r^3}{r_s^3} \left(a_4+\left(a_0\frac{1695233 }{9261000}+a_2\frac{107}{210}\right) \log \left(\frac{r_s}{r}\right)+a_0\frac{11449}{88200}\log ^2\left(\frac{r_s}{r}\right)\right) \nonumber 
      \\
      &\qquad\quad+\frac{r_s^2}{r^2}\left(b_4- \left(a_0\frac{2634215 }{7225344}+a_2\frac{1987 }{5120}+b_0\frac{1695233 }{46305000}+b_2\frac{107 }{210}\right)\log \left(\frac{r_s}{r}\right)\right. \nonumber 
      \\
      &\qquad\quad\left.-\frac{11449 (945 a_0-1024 b_0) }{451584000}\log ^2\left(\frac{r_s}{r}\right)\right)\Bigg]+\cdots,
\label{Zer_sol_inf_GR}
\end{align}
A word of caution regarding Eq.~\eqref{Zer_sol_inf_GR}: starting at order $n=2$, we have shifted the coefficients $a_n$ and $b_n$ by suitable constants to make the final result more compact. This is immaterial, as it simply amounts to a redefinition of the integration constants of the homogeneous solution.

Comparing Eq.~\eqref{Zer_sol_inf_GR} with the EFT result \eqref{psi_EFT_renorm}, we obtain the following matching conditions:
\begin{subequations}
\begin{align}
r_s^3\bar{B}_\text{reg}  = &\, \,  a_0 + a_1 \omega r_s+ \omega^2 r_s^2\left[a_2+  a_0 \left(\frac{2731}{39200}+\frac{107}{210}\log (\mu  r_s)\right) \right]
\nonumber\\
&+\omega ^3 r_s^3\left[ \left(a_3+\frac{2731}{39200}a_1\right)+\frac{107}{210} a_1 \log (\mu  r_s)\right]
\nonumber\\
&+\omega ^4 r_s^4\left[a_4+a_2 \left(\frac{2731 }{39200}+\frac{107}{210}  \log (\mu  r_s)\right)\right.
\nonumber\\
&\left.+a_0 \left(\frac{84512980501 }{248935680000}+\frac{16191817}{74088000}\log (\mu  r_s)+\frac{11449}{88200}\log ^2(\mu  r_s)\right)\right] ,
\\
r_s^{-2}\bar{B}_\text{irr}  = & \, \,  b_0 + b_1 \omega r_s 
\nonumber\\ 
&+ \omega^2r_s^2 \left[5 b_2 +\frac{945 }{1024}a_2 -\frac{1015283 }{1032192}a_0-\frac{111383}{352800}b_0 -  
\left(a_0+\frac{107}{210}b_0\right) \log (\mu   r_s)
\right]
\nonumber\\
&+\omega ^3r_s^3 \left[5 b_3-\frac{1015283 a_1 }{1032192}+\frac{945 a_3}{1024}-\frac{111383 b_1 }{352800}-\frac{1}{210} (210 a_1+107b_1) \log (\mu  r_s)\right]\nonumber\\
&+\omega ^4 r_s^4\left[5 b_4+\frac{945 a_4 }{1024}+\frac{35053240247 b_0}{248935680000}-b_2 \left(\frac{107}{42}   \log (\mu  r_s)+\frac{111383 }{70560}\right)\right.
\nonumber\\
&\left.-a_2 \left(\frac{3011}{2048}\log (\mu  r_s)+\frac{13160171 }{10321920}\right)- \frac{398716783}{216760320}a_0\log (\mu  r_s)\right.\nonumber\\
&\left.-a_0\frac{6993556173743 }{3641573376000}-\frac{547961}{24696000}b_0\log (\mu  r_s)+\frac{11449}{88200}b_0\log ^2(\mu  r_s)
\right] .
\end{align}
\end{subequations}

Using the condition \eqref{Bratio}, we can then express the renormalized $\ell=2$ electric-type response couplings in $D=4$ in terms of the integration constants $a_n$ and $b_n$ of the vacuum solution:
\begin{multline}
\label{K_pm_ab}
   \frac{45}{8G^4M^5}\bar  K_{+-}^{(E)}(\omega)
    = \frac{b_0}{a_0}
    + \omega r_s \left(\frac{b_1}{a_0} -\frac{b_0a_1}{a_0^2}\right)
    + \omega^2r_s^2\bigg[\!- \left(1+\frac{107}{105}\frac{b_0}{a_0}\right)\log (\mu r_s)
\\
    -\frac{67981 }{176400}\frac{ b_0}{ a_0} +\frac{a_1^2 b_0}{a_0^3}-\frac{a_1b_1}{a_0^2}-\frac{b_0a_2}{a_0^2}+ \frac{945 a_2}{1024 a_0}+\frac{5 b_2}{a_0}-\frac{1015283}{1032192}
    \bigg]
\\
 + \omega^3r_s^3\bigg[\frac{945}{1024}\frac{a_3}{a_0}-\frac{67981}{176400}\frac{b_1}{a_0}+\frac{5 b_3}{a_0}+\frac{a_1^2 b_1+2 a_1 a_2 b_0}{a_0^3}+\frac{a_1^3 b_0}{a_0^4}-\frac{945}{1024 }\frac{a_1a_2}{a_0^2}
 \\ +\frac{67981 }{176400}\frac{a_1 b_0}{a_0^2}-\frac{5 a_1 b_2}{a_0^2}-\frac{a_2 b_1}{a_0^2}-\frac{a_3 b_0}{a_0^2}- \frac{107}{105}\left(\frac{b_1}{a_0}-\frac{ b_0a_1}{a_0^2}\right)\log (\mu  r_s)\bigg] \\
+ \omega^4r_s^4\bigg[\frac{-a_1^3 b_1-3 a_1^2 a_2 b_0}{a_0^4}+\frac{a_1^4 b_0}{a_0^5}-\frac{6744009765173}{3641573376000}+\frac{945 }{1024}\frac{a_1^2 a_2}{a_0^3}-\frac{67981 }{176400}\frac{a_1^2 b_0}{a_0^3}+\frac{5 a_1^2 b_2}{a_0^3}\\
+\frac{2 a_1 a_2 b_1}{a_0^3}+\frac{2 a_1 a_3 b_0}{a_0^3}+\frac{a_2^2 b_0}{a_0^3}-\frac{945 }{1024 }\frac{a_1 a_3}{a_0^2}+\frac{67981}{176400}\frac{a_1 b_1}{a_0^2}-\frac{5 a_1 b_3}{a_0^2}-\frac{945 }{1024 }\frac{a_2^2}{a_0^2}+\frac{67981 }{176400 }\frac{a_2 b_0}{a_0^2}\\
-\frac{5 a_2 b_2}{a_0^2}-\frac{a_3 b_1}{a_0^2}-\frac{a_4 b_0}{a_0^2}-\frac{203943 }{573440 }\frac{a_2}{a_0}+\frac{945 }{1024 }\frac{a_4}{a_0}-\frac{21388060129 }{124467840000 }\frac{b_0}{a_0}-\frac{67981 }{35280 }\frac{b_2}{a_0}+\frac{5 b_4}{a_0}\\
-\frac{107}{105}\log (\mu  r_s) \left(\frac{a_1^2 b_0}{a_0^3}-\frac{a_1b_1}{a_0^2}-\frac{a_2 b_0}{a_0^2}+\frac{945 a_2}{1024 a_0}-\frac{32869 }{1258320 }\frac{b_0}{a_0}+\frac{5 b_2}{a_0}+\frac{76383379}{61358080}\right)\\
+\frac{107}{210}\left(1+\frac{107}{105}\frac{ b_0}{a_0}\right) \log ^2(\mu  r_s)\bigg] .
\end{multline}

The expression \eqref{K_pm_ab} is relatively involved, but it exhibits some simple structural features that are worth highlighting.
First, except at static and linear order in $\omega$, both the conservative and dissipative response coefficients generically run with the renormalization scale. This running arises from classical loop divergences and the associated renormalization in the in-in effective action. In particular, at each order in $\omega$, the $\beta$-function---i.e., the coefficient governing the non-analytic logarithmic running---is uniquely determined by the integration constants entering at lower orders. For instance, once the static Love number, given by the ratio $b_0/a_0$, is specified, the coefficient of the $\omega^2 r_s^2\log(\mu r_s)$ term is completely fixed, including its sign~\cite{Combaluzier--Szteinsznaider:2025eoc,Apostolidis:2026qsg}. The same structure persists at higher orders. This recursive determination follows from the non-renormalization of the static response, which in turn is a consequence of the degeneracies and symmetries of the homogeneous problem~\cite{Kol:2011vg,Hui:2020xxx,Hui:2021vcv,Hui:2022vbh,Charalambous:2021kcz,Charalambous:2022rre,Parra-Martinez:2025bcu}.
In addition, note that some non-analytic terms receive contributions that are completely independent of the integration constants---see, e.g., the $\log(\mu r_s)$ and $\log^2(\mu r_s)$ terms appearing at second and fourth order in $\omega$, respectively. These constitute \emph{universal} contributions: they are independent of the microscopic nature of the object and are entirely determined by the gravitational interactions encoded in the Einstein--Hilbert action. Consequently, they are identically present in the response of black holes as well~\cite{Combaluzier--Szteinsznaider:2025eoc}.

The result \eqref{K_pm_ab} can be expressed more succinctly by introducing two frequency-dependent coefficients, $A$ and $B$, which absorb the integration constants $a_n$ and $b_n$:
\begin{subequations}
\label{A,B}
\begin{align}
    A(\omega)&\equiv a_0+\omega r_s a_1+\omega ^2r_s^2 a_2+\omega ^3r_s^3 a_3+ \omega ^4 r_s^4 a_4, \\
    B(\omega)&\equiv b_0+ \omega r_s b_1+\omega ^2 r_s^2\left(\frac{945}{1024}a_2+5 b_2\right)+ \omega ^3 r_s^3\left(\frac{945 }{1024}a_3+5 b_3\right)+ \omega ^4 r_s^4\left(\frac{945 }{1024}a_4+5 b_4\right).
\end{align}
\end{subequations}
Combining these results, we obtain the following expression for the $\ell=2$ electric-type Love number:
\begin{equation}
\begin{aligned}
    \frac{45}{8G^4M^5} \bar K_{+-}^{(E)}(\omega)
    =&\, \frac{B(\omega)}{A(\omega)}+\omega ^2 r_s^2\Bigg[-\frac{67981 }{176400}\frac{B(\omega) }{A(\omega)}-\frac{1015283 }{1032192}-\left(1+\frac{107}{105}\frac{B(\omega) }{A(\omega)}\right) \log (\mu  r_s)\Bigg]\\
    &+\omega ^4 r_s^4\Bigg[-\frac{21388060129}{124467840000}\frac{B(\omega) }{A(\omega)}-\frac{6744009765173 }{3641573376000}\\
    &+\left(\frac{32869}{1234800}\frac{B(\omega) }{A(\omega)}-\frac{76383379 }{60211200}\right) \log (\mu  r_s)+\left(\frac{11449}{22050}\frac{B(\omega) }{A(\omega)}+\frac{107 }{210}\right) \log ^2(\mu  r_s)\Bigg]\,,
\end{aligned}
\label{eq:Komegabis}
\end{equation}
where we discard terms of $\mathcal{O}(\omega^5)$ and higher. 

The expression \eqref{K_pm_ab}, or equivalently \eqref{eq:Komegabis}, determines the EFT coefficients in terms of the parameters characterizing the exterior solution. These parameters can, in turn, be fully determined from the matter energy-momentum tensor once the interior problem has been solved and the appropriate continuity conditions have been imposed at the surface of the object (see Section~\ref{sec:NeutronStar}).

\newpage
\linespread{.95}
\addcontentsline{toc}{section}{References}
\bibliographystyle{utphys}
{\small
\bibliography{eftbib}
}

\end{document}